\documentclass[sigconf,edbt]{acmart-edbt2022}

\def\BibTeX{{\rm B\kern-.05em{\sc i\kern-.025em b}\kern-.08em
    T\kern-.1667em\lower.7ex\hbox{E}\kern-.125emX}}

\usepackage{booktabs} 

\setcopyright{rightsretained}

\acmDOI{}

\acmISBN{978-3-89318-086-8}

\acmConference[EDBT 2022]{25th International Conference on Extending Database Technology (EDBT)}{29th March-1st April, 2022}{Edinburgh, UK} 
\acmYear{2022}

\usepackage{soul}
\usepackage{listings}
\usepackage{amsmath}
\DeclareMathOperator*{\argmax}{arg\,max}
\usepackage{mathtools}
\DeclarePairedDelimiter\ceil{\lceil}{\rceil}

\usepackage[linesnumbered,ruled]{algorithm2e}
\usepackage{subcaption}

\definecolor{Blue}{rgb}{0, 0, 255}
\definecolor{Yellow}{rgb}{255, 255, 0}
\definecolor{Red}{rgb}{0, 255, 255}

\newcommand{\boldstart}{\vspace{0.5em}\noindent\textbf}

\newcommand{\sysName}{{\sf Maliva}\xspace}

\lstdefinelanguage{SQL}{%
  alsoletter={-_:(\$\\},
  morekeywords={
    select,
    from,
    where,
    and,
    or,
    in,
    as,
    group,
    by,
    join,
    on,
    with,
    update,
    set,
    delete,
    insert,
    limit
  },
  emph={tweetsSample20},  
  emphstyle=\slshape\bfseries\color{blue},
  basicstyle=\sf,
  keywordstyle=\textbf,
  identifierstyle=\texttt,
  morecomment=[s]{/*}{*/},
  commentstyle=\slshape\bfseries\color{blue},
  sensitive=false,
  literate={<=}{{\litleq}}1 {>=}{{\litgeq}}1 {//}{{\litdoubleslash}}1 
}[keywords,comments,strings]

\lstnewenvironment{sql}[1]
   {\lstset{language=SQL,columns=flexible,basicstyle=\footnotesize,basewidth={.3em,.2em},mathescape=true,title=#1}}
   {}
\lstnewenvironment{sqlnotitle}
   {\lstset{language=SQL,columns=flexible,basicstyle=\footnotesize,basewidth={.3em,.2em},mathescape=true}}
   {}

\SetCommentSty{mycommfont}
\SetKwInOut{Parameter}{Parameters}

\begin{document}
\title{\sysName: Using Machine Learning to Rewrite Visualization Queries Under Time Constraints}

\author{Qiushi Bai, Sadeem Alsudais, Chen Li, Shuang Zhao}
\affiliation{%
  \institution{Department of Computer Science, UC Irvine, CA 92697, USA}
}
\email{{qbai1,salsudai}@uci.edu,{chenli,shz}@ics.uci.edu}

\renewcommand{\shortauthors}{}

\begin{abstract}
  We consider data-visualization systems where a middleware layer translates a frontend request to a SQL query to a backend database to compute visual results. We study the problem of answering a visualization request within a limited time constraint due to the responsiveness requirement.  We explore the optimization options of rewriting an original query by adding hints and/or doing approximations so that the total time is within the time constraint.  We develop a novel middleware solution called \sysName based on machine learning (ML) techniques.  It applies the Markov Decision Process (MDP) model to decide how to rewrite queries and uses training instances to learn an agent to make a sequence of decisions judiciously for an online request.  We give a full specification of the technique, including how to construct an MDP model, how to train an agent, and how to use approximating rewrite options.  Our experiments on both real and synthetic datasets show that \sysName performs significantly better than a baseline solution that does not do any rewriting, in terms of both the probability of serving requests interactively and query execution time.
\end{abstract}

\maketitle

\section{Introduction}
\label{sec:intro}

As a powerful way for people to gain insights from data quickly and intuitively,  visualization is becoming increasingly important in the Big Data era.  A common architecture to support data visualization has three tiers: a backend database, a middleware layer, and a user-facing frontend.  The middleware translates a visualization request to a query (typically in SQL) to the database, and sends the query answers to the frontend to display.  This architecture is widely used due to its benefits of supporting in-situ analytics at the data source, and utilizing the database's built-in capabilities of efficient storage, indexing, query processing, and optimization.  {\em Responsiveness} is critical in these systems~\cite{journals/tvcg/LiuH14, conf/sigmod/CrottyGZBK16, journals/tvcg/BattleCNMCS20}, and a request needs to be served within a time budget, e.g., $500ms$.  This requirement is especially challenging when the data volume is large, and the user request has ad-hoc conditions on attributes of various types.  

In this paper, we study the problem of {\em answering visualization requests with a predetermined time constraint}. We focus on middleware-based solutions, with the advantage that they treat the backend database as a black box without changes, and can leverage the computing capabilities to do in-situ analytics.  We consider both rewritings that return exact results and rewritings that return approximate results.  As a motivating example, consider a system that visualizes social media tweets on the US map with a time constraint of $500ms$.  Its backend database has a {\tt tweets} table with attributes {\tt Content}, {\tt Location}, and {\tt CreateAt}.

\begin{figure}[htbp]
  \centering
  \begin{subfigure}[t]{0.45\linewidth}
    \includegraphics[width=\linewidth]{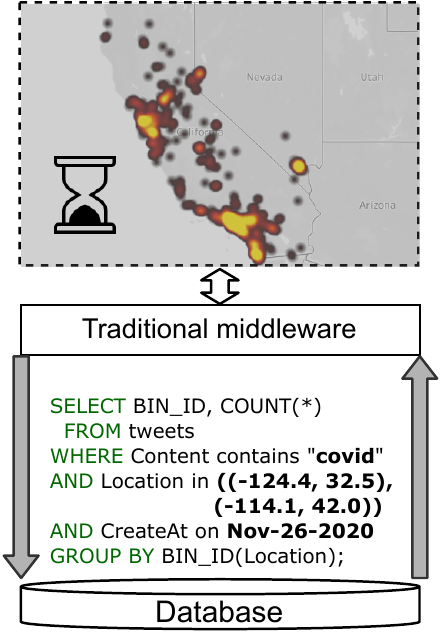}
    \caption{The original SQL query takes $3.35s$.}
  \end{subfigure}
  \hfill
  \begin{subfigure}[t]{0.45\linewidth}
    \includegraphics[width=\linewidth]{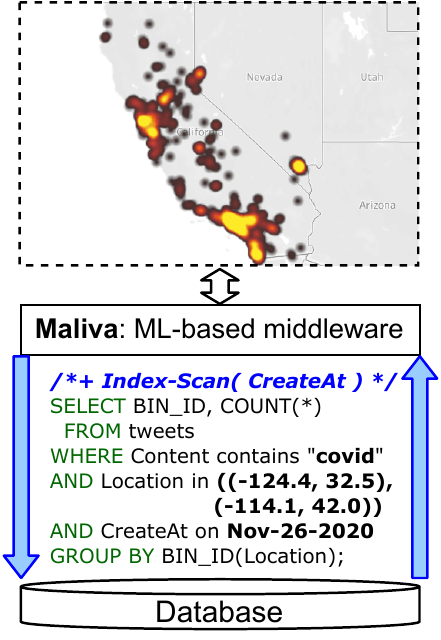}
    \caption{A rewritten query with a hint takes $0.33s$.}
  \end{subfigure}
  \caption{Equivalent rewriting option: adding query hints helps the database compute results with a time constraint ($500ms$).
  \label{fig:option1}}
\end{figure}

\boldstart{Equivalent rewriting options.}  Suppose a user asks for a spatial heatmap of tweets containing the keyword {\tt covid} on the Thanksgiving day of 2020 in a region. The middleware creates a SQL query shown in Figure~\ref{fig:option1}(a), which takes $3.35$ seconds to execute.  
For this query, the physical plan generated by the database uses the keyword to access the inverted index on the {\tt Content} attribute to retrieve candidate records, then filters them using the two other conditions.
If we rewrite the query into an equivalent new query by adding a hint (Figure~\ref{fig:option1}(b)), the rewritten query takes only $0.3$ seconds , as the hint helps the database generate a more efficient physical plan, which uses the temporal filtering condition to access the B+ Tree index on the {\tt CreateAt} attribute. 

\begin{figure}[htbp]
  \centering
  \begin{subfigure}[t]{0.45\linewidth}
    \includegraphics[width=\linewidth]{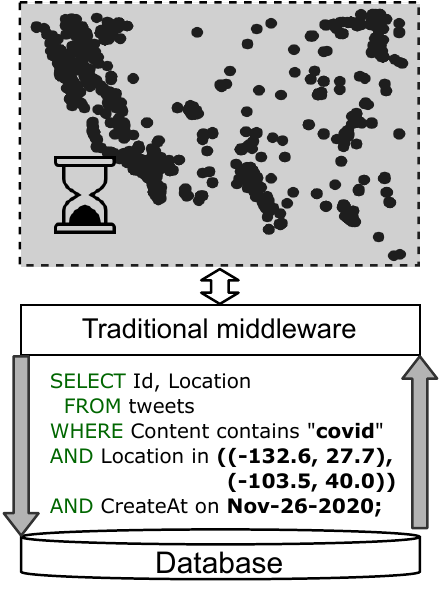}
    \caption{The query takes $4.28s$ (no hints can reduce it).}
  \end{subfigure}
  \hfill
  \begin{subfigure}[t]{0.45\linewidth}
    \includegraphics[width=\linewidth]{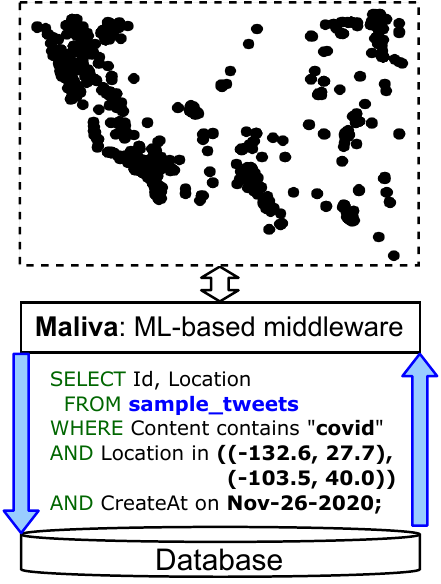}
    \caption{A rewritten query using a sample table takes $0.45s$.}
  \end{subfigure}
  \caption{Approximation rewriting option: rewriting the query to compute an approximate result within the time constraint.\label{fig:option2}}
\end{figure}

\boldstart{Approximation rewriting options.}  Figure~\ref{fig:option2}(a) shows another visualization request on a larger region, which takes at least $4.28s$ for the database to run, no matter what hints we add.  In this case, we rewrite the query by using random sampling, resulting in an approximated query that takes only $0.45s$ to run (see Figure~\ref{fig:option2}(b)).

\boldstart{Why the database fails?}
For the query in Figure~\ref{fig:option1}(a), there are many reasons the database can fail to generate an efficient plan. One is the estimation error of the query cost due to an underestimation of the keyword {\tt covid}'s selectivity.  The cost-estimation problem in optimizers is notoriously hard~\cite{acmblog:is-query-optimization-a-solved-problem}.  For example, in our experiments (Section~\ref{sec:experiments}), out of the $602$ visualization queries that had at least one physical plan that could finish within $500ms$, PostgreSQL failed to choose an efficient plan for $269$ queries due to its cost-estimation errors.  Although there are many higher-accuracy estimators such as~\cite{conf/icde/WuCZTHN13, journals/pvldb/WuWHN14, journals/pvldb/SunL19,  journals/pvldb/MarcusNMZAKPT19, conf/sigmod/ParkZM20, conf/sigmod/HasanTAK020, conf/sigmod/MarcusNMTAK21}, their higher estimation cost prevents them from being adopted by a general-purpose database.  For example, for OLTP queries that need to be finished within milliseconds, spending tens of milliseconds for the cost estimation is unacceptable.  Our key observation is that for visualization applications where requests come with a predetermined time constraint, the middleware can be more opportunistic than the traditional database.  For instance, it can spend tens of milliseconds to estimate the costs of multiple rewritten queries, then choose an efficient one to answer the request within the given time constraint.

\boldstart{Challenges.} We may enumerate all possible rewritten queries by applying different hints to a given original query.  We then use one of the aforementioned query-time estimators (QTE for short) to estimate the execution time of these rewritten queries and choose the most efficient one.  
There are several challenges in using this approach in the context of interactive visualization. \textbf{(C1)} A main challenge is that the cost of estimating the execution time of a rewritten query can be significant given a tight time constraint.  For example, in Bao~\cite{conf/sigmod/MarcusNMTAK21}, estimating the execution time of all rewritten queries for one original query can take up to $320ms$ in their experiments.  \textbf{(C2)} Another challenge is the uncertainty caused by the estimation error of the QTE, and the fact that the backend database may or may not follow the provided hints to generate a physical plan.  \textbf{(C3)} The third challenge is quality.  For queries without equivalent rewritten queries that can meet the time constraint, approximate rewriting options need to be explored.  It is critical to maximizing the quality of the result while ensuring the query time is within the time constraint.

We address these challenges by introducing a novel machine-learning-based technique called \sysName, which stands for ``Machine Learning for Interactive Visualization.''   The technique formulates the middleware task as a Markov Decision Process (MDP). For a given time budget, we train an MDP agent to balance the planning time and the execution time of the rewritten queries. By learning from previous experiences, the MDP agent judiciously explores different rewriting options, so that the total time (including planning and query execution) is within the time limit. (We address challenge \textbf{C1} in Section~\ref{sec:mdp-model}.)  Applying reinforcement learning, which inherently handles uncertainty, \sysName solves the issues of inaccurate time estimation and database ignoring the query hints when rewriting a query. (We address challenge \textbf{C2} in Section~\ref{sec:mdp-training}.)  By considering visualization qualities of rewritten queries in the reward design of the MDP model, \sysName makes the best effort to maximize the result’s quality while ensuring the query time is within the time limit. (We address challenge \textbf{C3} in Section~\ref{sec:generalization}.)  Our experiments show that  \sysName has a much higher chance (70$\times$ more than the original query) to generate an execution plan such that the total time is within a time limit. Interestingly, it can also reduce query execution time.  Both improvements show the significant benefits of adding learning capabilities to the middleware to support responsive visualization.

The rest of the paper is organized as follows.
After formulating the middleware query-generation problem in Section~\ref{sec:problem-formulation}, we give an overview of \sysName in Section~\ref{sec:system-overview}. We present the details of this MDP-based solution, including its states, actions, transitions, and rewards (Section~\ref{sec:mdp-model}). We present how \sysName trains an MDP agent offline and uses it to generate a rewritten query online (Section~\ref{sec:mdp-training}). We generalize the MDP-based solution to be quality-aware by considering approximation rewriting (Section~\ref{sec:generalization}). Lastly, we report the results of a thorough experimental evaluation of \sysName  to show its performance and benefits (Section~\ref{sec:experiments}).

\subsection{Related Work} \label{sec:related-work}

Visualization is a broad topic studied in many communities, and here we focus on efficiency-related works. A survey~\cite{journals/tkde/GodfreyGL16} summarized studies on interactive data analytics and visualization, and there are several recent studies on this topic~\cite{conf/sigmod/JiangR018,conf/sigmod/Psallidas018a,journals/debu/LeeP18,journals/pvldb/Kraska18}.


{\em Approximate Query Processing (AQP).}
There are many techniques for computing approximate answers to queries~\cite{journals/pvldb/ZhangWYJ16,conf/sigmod/ZengADAS15,conf/cidr/MozafariRMMCBB17,conf/sigmod/ParkMSW18,journals/dase/LiL18,conf/sigmod/GuoFCB18,conf/icde/ParkCM16,journals/pvldb/WangCLY15,conf/sigmod/DingHCC016,journals/pvldb/RahmanAKBKPR17,conf/sigmod/PengZWP18,journals/pvldb/BudiuGSWKA19}.
These approaches focus on developing approximation solutions to compute high-quality visualization. Existing solutions can be adopted as approximation rules in \sysName, such as Sample+Seek~\cite{conf/sigmod/DingHCC016}, which generates error-bounded visualization results by running queries on a small sample table.

{\em Datacube-based approaches.}
Related studies include~\cite{journals/tvcg/LinsKS13,journals/tvcg/PahinsSSC17,journals/tvcg/WangFWBS17,journals/cgf/LiuJH13,conf/icde/YuS20,conf/sigmod/CrottyGZBK16,conf/icde/KamatJTN14,conf/chi/MoritzHH19}.
In these approaches, the predefined cube intervals cannot support visualization queries with arbitrary numerical range conditions. The proposed \sysName system efficiently computes results for visualization queries with arbitrary conjunctive selection conditions.

{\em Progressive visualization.}
There are solutions to show visualization results progressively~\cite{conf/chi/MoritzFD017,conf/bigdataconf/ImVM13,journals/pvldb/CrottyGZBK15,conf/chi/FisherPDs12,conf/sigmod/CrottyGZBK16,journals/pvldb/BudiuGSWKA19}.
For instance, DICE~\cite{conf/sigmod/CrottyGZBK16} uses random and stratified samples to present an approximate result and then incrementally updates the result.
These progressive visualization systems can adopt the proposed \sysName middleware to further optimize the intermediate queries to increase their efficiency.

{\em Prefetching-based approaches.}
Techniques including \cite{conf/sigmod/BattleCS16,conf/icde/YuMS17,conf/sigmod/RundensteinerWXCWYH07,conf/ieeevast/ChanXGH08} accelerate visualization queries by prefetching or caching their results. For example, ForeCache~\cite{conf/sigmod/BattleCS16} divides visualizations into tiles and prefetches them based on predicted user behaviors. \sysName is orthogonal to these techniques, and it can be adopted by them to further optimize the database queries.

{\em Visualization using big data systems.}
These techniques use Hadoop, Spark, and Hive to support visualizations~\cite{conf/ssdbm/YuZS18,journals/pvldb/BudiuGSWKA19,journals/cgf/TaoLWBDCS19,conf/icde/EldawyMJ16, conf/bigdataconf/ChengSKBW13}.
For instance, HadoopViz~\cite{conf/icde/EldawyMJ16} and GeoSparkViz~\cite{conf/ssdbm/YuZS18} use Hadoop and Spark to generate high-resolution visualizations. Their focus is on offline construction, not on an interactive visualization for queries with ad-hoc conditions. 
The proposed \sysName middleware technique is complementary to these solutions.

{\em ML for visualization.} The survey in~\cite{journals/corr/abs-2012-00467} summarized studies of applying ML techniques to different stages during the whole visualization pipeline. Examples are \cite{conf/icde/LuoCQ0020, journals/tvcg/WangFCZFSYC18} for data cleaning and preparation and \cite{conf/chi/HuBLKH19, journals/corr/abs-2009-12316, conf/sigmod/LuoQ00W18} for visualization recommendation. Our proposed system focuses on applying ML techniques to solve performance issues at the middleware.

{\em ML-based query optimization.} ML has recently used in database optimizers~\cite{conf/sigmod/MarcusNMTAK21, conf/sigmod/MarcusP18, conf/icde/Yu0C020, journals/pvldb/MarcusNMZAKPT19, conf/sigmod/SikdarJ20, journals/corr/abs-1808-03196, conf/sigmod/TrummerWMMJA19}, selectivity estimation~\cite{conf/sigmod/ParkZM20, conf/sigmod/HasanTAK020}, and cost estimation~\cite{journals/pvldb/SunL19}.  {\em Comparison with Bao}: The recent Bao technique~\cite{conf/sigmod/MarcusNMTAK21} uses hints to generate optimized queries by modeling the optimization as a multi-armed bandit problem. In the training phase, Bao applies Thompson sampling to minimize the training time and maximize the accuracy of its neural-network-based query time estimator (QTE). In the online query processing phase, Bao uses a brute-force strategy that enumerates all the candidate query-hint sets, estimates the query time of all rewritten queries, and chooses the fastest query-hint set as the output.  Bao assumes that the cost of estimating query time is negligible. This assumption is not valid in \sysName as our problem is responsive visualization with a stringent time constraint (e.g., $500ms$).   \sysName focuses on the online planning problem when a brute-force strategy is not acceptable.  We compared \sysName with Bao in the experiments and the results show \sysName outperformed Bao in various metrics.


\section{Problem Formulation}
\label{sec:problem-formulation}

\boldstart{Visualization architecture.}  We consider a typical three-tier data-visualization system that consists of a backend database, a middleware layer, and a frontend.  For each frontend visualization request, let $Q$ be the {\em original} SQL query for the request. Let $\tau$ be a time limit that quantifies the expected responsiveness of the system.  Ideally, we want the total delay, from the time the user submits a request to the time the result is shown on the frontend, to be within $\tau$.  The original query $Q$ may not meet the time-limit constraint when the backend database cannot generate a physical plan that is fast enough. To solve this problem, {\sysName} rewrites $Q$ with two kinds of options: {\em query hints} and {\em approximation rules}. By adding a query hint to $Q$, {\sysName} can help the backend database generate an efficient physical plan that computes the result within the time limit. For expensive queries where no physical plan can meet the time limit, {\sysName} can add an approximation rule to the original query such that the backend database computes an approximate result to trade the visualization quality for responsiveness.  Note that the proposed approach also works in a more general setting of approximate query processing (AQP) where a time constraint is given.

\boldstart{Query hints.}  A {\em query hint} in a database is an addition to the SQL standard that instructs the database engine on how to execute the query. For example, a hint may tell the engine to use or not to use an index (even if the query optimizer would decide otherwise)~\cite{wiki:query-hints}.  A query hint does not change the semantic meaning of the query, i.e., the result computed by the database engine with the hint remains the same.  Databases such as AsterixDB~\cite{asterixdb:query-hints}, MySQL~\cite{mysql:optimizer-hints}, Oracle~\cite{oracle:optimizer-hints}, PostgreSQL~\cite{postgresql:query-hints}, and SQL Server~\cite{sqlserver:query-hints} support a variety of query hints.  For example, in Figure~\ref{fig:example-queries}(b), \sysName adds two hints {\em + Index-scan(t CreateAt)} and {\em Nest-Loop-Join(t u)} to the original query. They suggest the engine to use the index on the \texttt{CreateAt} attribute to scan the table \texttt{t}, and do a nest-loop join on tables \texttt{t} and \texttt{u}.

\begin{figure}[htbp]
\begin{minipage}[t]{.45\columnwidth}
\begin{sql}{(a) Original Query (Q)}

SELECT BIN_ID, COUNT(*)
    FROM tweets t, users u
WHERE t.Content contains "covid"
        AND t.Location in ((-124.4, 32.5),
                                                    (-114.1, 42.0))
        AND t.CreateAt on 'Nov-26-2020'
        AND u.TweetCnt in [100, 5000]
        AND t.user_id = u.id
    GROUP BY BIN_ID(t.Location);
\end{sql}
\end{minipage}
\hfill
\begin{minipage}[t]{.45\columnwidth}
\begin{sql}{(b) Rewritten Query (RQ)}
/*+ Index-scan(t CreateAt),  
        Nest-Loop-Join(t u) */
SELECT BIN_ID, COUNT(*)
    FROM tweetsSample20 t, users u
WHERE t.Content contains "covid"
        AND t.Location in ((-124.4, 32.5),
                                                    (-114.1, 42.0))
        AND t.CreateAt on 'Nov-26-2020'
        AND u.TweetCnt in [100, 5000]
        AND t.user_id = u.id
    GROUP BY BIN_ID(t.Location);
\end{sql}
\end{minipage}
\caption{A original query and a rewritten query. \label{fig:example-queries}}
\end{figure}

\boldstart{Approximation rules.}  An {\em approximation rule} is a method to rewrite the original SQL query to compute an approximate result, and the new query takes less time.  There are various approximation rules available in database systems, such as adding a {\em ``Limit''} clause, applying a SQL-standard {\em ``TableSample''} operator on a table, or substituting a table with a smaller table randomly sampled from the original table.  For example, in Figure~\ref{fig:example-queries}(b), \sysName rewrites the original query by substituting the table \texttt{tweets} with a sample table \texttt{tweetsSample20} with $20\%$ randomly selected records.

Now we formally define rewriting options, rewritten queries, and the query-rewriting problem.

\begin{definition}{(Rewriting Option) \label{def:rewrite-option}}
  Let $H$ be a set of query-hint sets and $A$ be a set of approximation-rule sets. A rewriting option (``RO'' for short) is a tuple $(h, a)$, where $h \in H$ and $a \in A$. Note that both $h$ and $a$ can be the empty set $\emptyset$.
\end{definition}

For instance, the rewriting option in Figure~\ref{fig:example-queries}(b) is a tuple with a query-hint set of ``use the index on \texttt{CreateAt} and do a nest-loop join on \texttt{t} and \texttt{u}'' and an approximation-rule set of ``substituting the table \texttt{tweets} with the sample table \texttt{tweetsSample20}''.

\begin{definition}{(Rewritten Query) \label{def:rewritten-query}}
  Given an original SQL query $Q$ and a rewriting option $RO$, a rewritten query (``RQ'' for short) is a new SQL query generated by applying $RO$ onto $Q$. If $RO = (\emptyset, \emptyset)$, then $RQ = Q$.
\end{definition}

For example, Figure~\ref{fig:example-queries}(b) is a rewritten query for the original query in Figure~\ref{fig:example-queries}(a).

\boldstart{Query-rewriting problem.} Given a visualization request's original SQL query $Q$, and a time limit $\tau$, we want to generate a rewriting option, such that the total time of the corresponding rewritten $RQ$, including planning and query execution, is within $\tau$ and the quality of $RQ$'s result is maximized. To quantify the quality, we assume a given visualization quality function $F$. Let $r(Q)$ be the result of the original query $Q$ and $r(RQ)$ be the result of the rewritten query $RQ$. Then $F(r(Q), r(RQ))$ computes the quality of $r(RQ)$.

In Sections~\ref{sec:system-overview},~\ref{sec:mdp-model}, and~\ref{sec:mdp-training}, we study the case of using query hints only (i.e., without changing query results). In Section~\ref{sec:generalization}, we study the case where approximation rules are also used.

\section{\sysName: ML-based Query Rewriting}
\label{sec:system-overview}

\begin{figure*}[tb]
  \centering
  \includegraphics[width=0.95\linewidth]{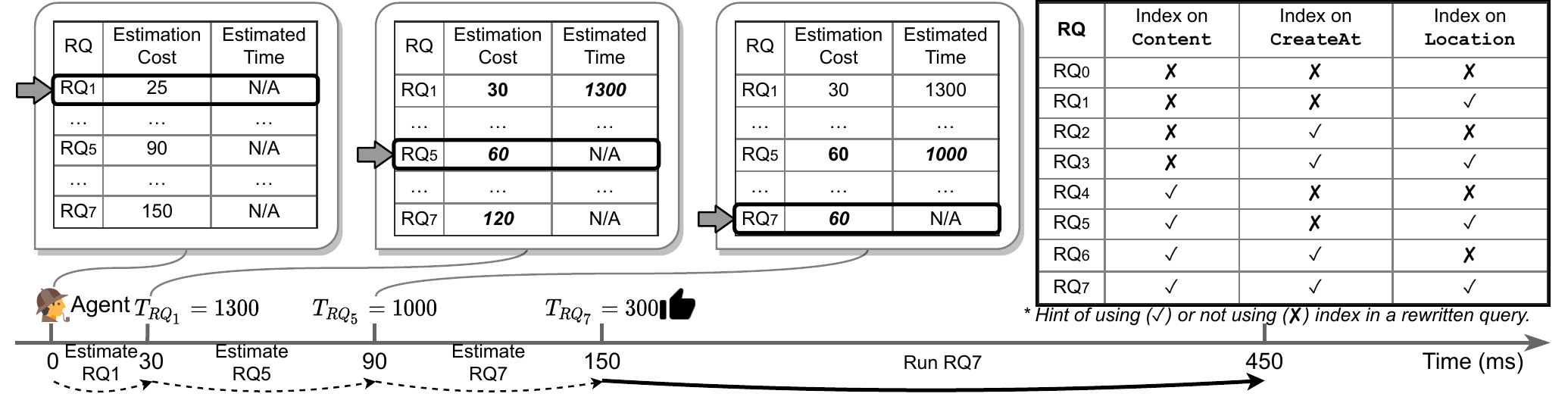}
  \caption{The Query Rewriter acts like an agent who makes a sequence of decisions to generate a rewritten query (with a total time $\leq 500 ms$). At time $0$, the agent considers rewritten query $RQ_1$ due to its low estimation cost (estimated $25ms$, the actual $30ms$ is updated once $RQ_1$ is explored). After estimating its execution time ($1,300ms$), the agent knows that $RQ_1$ is not viable since the total time is longer than $500 ms$. The estimation of $RQ_1$ affects the costs for estimating $RQ_5$ and $RQ_7$. The agent explores $RQ_5$ and then $RQ_7$.  With the estimated execution time being $300ms$ and the elapsed time being $150ms$, $RQ_7$ is decided as a viable rewritten query because the total time ($450ms$) is within $500 ms$.}
  \label{fig:sequential-decsision-making}
\end{figure*}

We now introduce the middleware technique called ``\sysName'' to solve the aforementioned query-rewriting problem.  We first give an overview of the technique, then use an example to explain the details.

\begin{figure}[htb]
  \centering
  \includegraphics[width=0.95\linewidth]{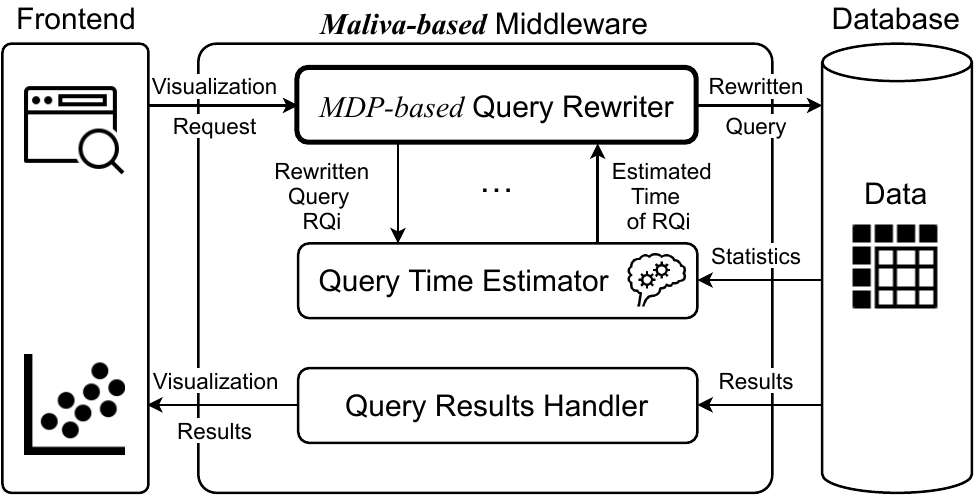}
  \caption{Overview of \sysName.}
  \label{fig:architecture}
\end{figure}

\boldstart{Overview.}  As illustrated in Figure~\ref{fig:architecture}, \sysName rewrites the original SQL query to answer a visualization request within a time budget.  It considers a predefined set of rewriting options, which we denote as $\Omega=\{RO_1,\ldots\}$. For each $RO_i$, the rewritten query is denoted as $RQ_i$. The set of candidate rewritten queries is $\Phi=\{RQ_1,\ldots\}$.

\sysName has a {\em Query Rewriter} that enumerates possible RQs 
and uses a {\em Query Time Estimator} (QTE) to estimate the execution time of each of them. The {\em Query Rewriter} uses the best effort to choose an RQ such that the total time, including the planning process and query execution, is within the time budget $\tau$.  Such an RQ is called {\em viable}.  The middleware then sends the rewritten query to the database. The {\em Query Result Handler} sends the retrieved result to the frontend to visualize.

Na\"ively enumerating all available RQs in $\Phi$ is computationally prohibitive due to two reasons. First, the cost of {\em Query Time Estimator} to estimate the execution time of a rewritten query is not negligible.  For instance, in some cases it could take up to $70ms$~\cite{journals/pvldb/SunL19} or $300ms$~\cite{conf/icde/WuCZTHN13}. Second, the number of RQs increases exponentially when the number of applicable indexing choices increases. For example, consider a selection query on a table with filtering conditions on $m$ attributes, and the database has an index on each attribute. The number of query-hint sets in $H$ would be $2^{m}$, since the database can use any subset of the $m$ indexes to do filtering and then intersect the record lists to compute the final result.  Therefore, the {\em Query Rewriter} needs to balance the exploration time for query estimation and the execution time of each chosen RQ to find a viable RQ. 

\boldstart{An example.} \sysName views query rewriting as a Markov decision process (MDP)~\cite{books/lib/SuttonB98} and adopts machine learning (ML) to solve this problem.
We use the running example in Section~\ref{sec:intro} to illustrate how \sysName uses an MDP agent to make a sequence of decisions to find a viable RQ.  For simplicity, we assume 
the rewrite-options (RO) set to involve query hints only.
We will generalize the technique to consider approximation rules in Section~\ref{sec:generalization}.  As shown in Figure~\ref{fig:sequential-decsision-making}, a query has three selection conditions on three attributes, and each of which has an index.  Suppose in the set $H$ of query-hint sets, each attribute has a query hint of using or not using the index. Thus, we have $2^3 = 8$ query-hint sets to choose from.
The agent makes a sequence of decisions to estimate the execution times of several rewritten queries and find a viable one $RQ_7$ (that uses the indexes on all three attributes).  Next, we present the details of this MDP-based technique.

\section{MDP Model for Adding Query Hints}
\label{sec:mdp-model}

In this section, we present the details of using an MDP model in \sysName to solve the query-rewriting problem, and discuss how to implement the Query Time Estimator (QTE).

\subsection{MDP Model for Query Rewriting}
\label{subsec:model-def}

\boldstart{MDP}~\cite{books/lib/SuttonB98} is a formalization of sequential decision-making problems where an agent learns to achieve a goal from interaction with an environment. At each time step, the agent is in a state $s$, and chooses an action $a$ available in state $s$. The environment transits the agent to a new state $s'$, and gives the agent a corresponding reward $R(s, a)$.  To train an MDP agent is to find a good policy ${\pi}_{*}$ such that if the agent follows the policy to choose an action for each state, it maximizes the total reward in the end.

We use the MDP model to solve the query-rewriting problem. For simplicity, we first focus on the case where rewriting options do not contain any approximation rules, which means no rewritten queries have quality loss. We will generalize the technique to consider approximation rules in Section~\ref{sec:generalization}. Without considering quality loss, the MDP agent learns to maximize the chance of finding a viable rewritten query for a given visualization request. The agent takes a sequence of actions, and each action chooses an RQ to explore. That is, it asks the query time estimator ({\sf QTE}) to estimate the execution time of the RQ.  The agent chooses an RQ based on the current state, and considers the future cost it needs to pay and the execution time of RQs already explored.  The agent gets a bonus if it finds a viable RQ, or a penalty if it runs out of time. In the offline phase, by analyzing queries in the training workload, the agent learns to maximize the chance to receive a bonus. In the online processing phase, given a new query, the agent decides which RQ to explore in each step to receive a bonus in the end. Now we present the details of how to use MDP to model the process of choosing RQs.

\begin{figure}[ht]
  \centering
  \includegraphics[width=\linewidth]{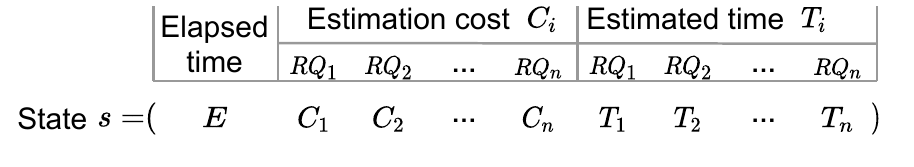}
  \caption{An MDP state in \sysName.}
  \label{fig:state}
\end{figure}

\boldstart{States.} A state represents the past decisions, based on which the agent decides an RQ to consider next. Suppose we are given a predefined set of $n$ ROs, i.e., $\Omega =\{RO_1,\ldots,RO_n\}$. Correspondingly, we have $n$ candidate RQs, denoted as $\Phi = \{RQ_1,\ldots,RQ_n\}$.  A state is a vector
$$s = (E, C_1, C_2, \ldots, C_n, T_1, T_2, \ldots, T_n),$$
which includes three pieces of information, as shown in Figure~\ref{fig:state}.  (1) The elapsed time ($E$) captures how much time we have spent.   (2) The estimation cost ($C_i$) for each possible rewritten query $RQ_i$ captures how much time is needed for the agent to estimate its running time. Each $C_i$ is initialized with a rough estimation collected offline and updated during the online planning phase.  Note that the MDP state does not require the initial $C_i$ values to be accurate, and a rough estimation from history statistics suffices. The actual estimation costs will be collected while the MDP agent processes a query, as will be described soon in the definition of {\em Transitions}.  (3) $T_i$ is the estimated time for each already explored $RQ_i$. Each $T_i$ is initialized with a $0$ value until it is filled with an estimated execution time.

We assume for each rewritten query $RQ_i$, collecting the physical plan and its statistics (e.g., cardinality and cost estimations of each operator) is done by the QTE, and its time is captured by the MDP model's estimation cost ($C_i$). We assume the rewritten queries' physical plans and statistics are not available to the MDP model.  Thus, the proposed MDP model is general, and can be applied to any query shape with any predefined query-hint set.  A natural question is that without the statistics of the explored RQs stored in the state, how can the MDP model make a good decision on which RQ to choose next?  Our answer is that the execution time of a rewritten query implicitly captures the statistics of the physical operators (e.g., the cost of doing an R-Tree index scan on the \texttt{Location} attribute).  By keeping the estimated execution time of each explored RQ in its state, the MDP model can learn the correlations of the execution times between different rewritten queries and make good decisions.  

\boldstart{Actions.} An action, denoted as $a$, is to explore an RQ next. For each RQ, the agent asks the {\sf QTE} to estimate its execution time. Meanwhile, the agent needs to pay cost as it takes time for the QTE to extract query features, possibly by collecting online statistics from the database, and running the estimation model to do the estimation. In the running example, at time $0$, the agent decides to explore $RQ_1$. It asks the {\sf QTE} to estimate $RQ_1$'s execution time.  

\begin{figure}[ht]
  \centering
  \includegraphics[width=0.75\linewidth]{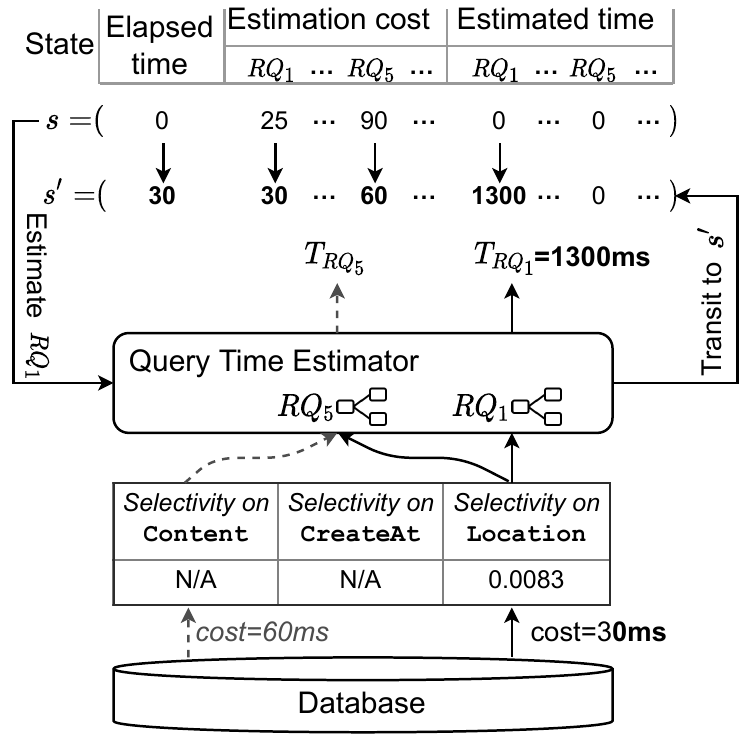}
  \caption{Transition after estimating execution time of $RQ_1$.}
  \label{fig:transition}
\end{figure}

\boldstart{Transitions.} A transition function defines how the environment computes the next state, given the agent's action in the current state. Let the RQ considered by action $a$ in state $s$ be $RQ_i$. We define the transition function $\mathcal{T}$ as follows. First, the {\sf QTE} estimates the time of $RQ_i$, and we add the estimated time $T_i$ to the state.  Second, the estimation costs for other RQs could change. In the running example, to estimate $RQ_1$ that uses the R-Tree index on the \texttt{Location} attribute, we need to collect the spatial filtering condition's selectivity on the \texttt{Location} attribute. To estimate $RQ_5$ that uses both the inverted index on the \texttt{Content} attribute and the R-Tree index on the \text{Location} attribute, we need to collect the selectivity values of the filtering conditions on both attributes. After the agent takes the $RQ_1$ action, we update the estimation cost of $RQ_1$ to be the actual time it costs and update the estimated estimation cost of $RQ_5$ by excluding the cost to collect the selectivity value of the spatial filtering condition. As shown in Figure~\ref{fig:transition}, after estimating the time of $RQ_1$, we add the estimated time $1,300ms$ of $RQ_1$ to the state, update the estimation cost for $RQ_1$ from the estimated $25ms$ to the actual $30ms$, and update the estimation cost for $RQ_5$ from the previous estimated $90ms$ to the new estimated $60ms$.  Lastly, the estimation takes time $\hat{C_i}$, and we add it to the elapsed time so far to indicate how much time the agent has spent to explore different RQs. Note that the $\hat{C_i}$ is the actual cost of estimating $RQ_i$, which could be different from $C_i$ because $C_i$ is an estimated cost for estimating $RQ_i$.

\boldstart{Rewards.} A reward function defines the agent's immediate gain when it takes a particular action $a$ in a given state $s$. In our setting, consider two cases to compute the reward function. (1) The first case is when the agent is at an intermediate state where it still has time for planning but has not yet found a viable rewritten query.  In this case, the agent should not be awarded or punished since it has not made a decision yet. Thus the reward value is $0$.  (2) The second case is when the agent is at a termination state where it decides the rewritten query $\hat{RQ}$, runs it against the database, and collects the execution time $\hat{T}$ of $\hat{RQ}$.  In this case, the agent should be awarded if the total time (including both the planning time and rewritten query execution time) is less than the time budget, or punished if the total time is more than the budget.

The agent decides a rewritten query by considering three situations. The first one is that the agent finds an RQ to be viable based on the estimation of the QTE before running out of time. For example, in Figure~\ref{fig:sequential-decsision-making}, after spending $150ms$ for planning, the agent decides $RQ_7$ as the chosen rewritten query, since the predicted total time of $RQ_7$ is $450ms$, which is within the $500ms$ budget.  The second situation is when the agent uses up the time budget and has to stop planning.  The third situation is when the agent has exhausted all candidate rewritten queries and has to decide which RQ to choose.  In the latter two situations, the agent chooses the fastest RQ known so far as the final decision.

Formally, suppose the generated rewritten query by the agent is $\hat{RQ}$ and the actual running time of query $\hat{RQ}$ is $\hat{T}$. Then the reward function $\mathcal{R}(s, a)$ is defined as follows, 

\begin{equation}
  \mathcal{R}(s, a) = 
       \frac{(\tau - s.E - \hat{T})}{\tau},
\end{equation}
where $s.E$ denotes the elapsed time so far in state $s$. If the total time $s.E + \hat{T}$ is less than the time budget $\tau$, which makes $\mathcal{R}(s, a)$ positive, then the agent receives a reward.  The faster the rewritten query is, the larger the reward will be.  On the other hand, if the total time exceeds the time budget, which makes $\mathcal{R}(s, a)$ negative, then the agent receives a penalty. The slower the rewritten query is, the larger the penalty will be.  Thus, guided by the reward function, the MDP model will learn to find an efficient rewritten query as soon as possible.

\subsection{Query Time Estimator (QTE)}
\label{ssec:approximiate-qte}


Take the sampling-based QTE described in~\cite{conf/icde/WuCZTHN13} as an example.  It first builds an analytical cost model (e.g., linear regression model), and uses it to estimate the execution time of a rewritten query by collecting its statistics online. Specifically, it estimates the selectivity values of the query conditions by running {\tt count(*)} queries on a small sample table, provides the values as input features to the cost model, and uses the model's prediction as the query's execution-time estimation. There are also other possible solutions in the literature~\cite{journals/pvldb/WuWHN14, journals/pvldb/SunL19, journals/pvldb/MarcusNMZAKPT19} that can be used by \sysName.  Note that QTEs are the focus of this paper, and \sysName leverages a given QTE intelligently to balance the planning time and the query execution time.

\section{Training and Using the MDP Agent}
\label{sec:mdp-training}

In this section we discuss how to train the MDP agent offline in \sysName on a workload of visualization requests and use it to generate a viable rewritten query online.

\subsection{Training the MDP Agent}

Suppose we have a workload of queries $W=[q_1, q_2, \ldots, q_m]$. Our goal is to find an optimal policy $\pi^*$ such that for any query $q_i \in W$, the agent following policy $\pi^*$ maximizes the chance to generate a viable rewritten query. We adopt the {\em deep Q-learning} algorithm~\cite{journals/corr/MnihKSGAWR13} for finding an optimal policy for the MDP agent. Its basic idea is to use a neural network (called {\em Q-network}) to represent a policy $\pi$. Given an input of a state vector, the q-network outputs a {\em Q-value}~\cite{journals/ml/WatkinsD92} for each possible RQ in the state. A higher q-value means that the rewritten query is more likely to be viable given the current information. Its training process includes two main steps. The first step is to generate a set of experiences by exploring different planning sequences for queries in the workload repeatedly. Each experience is a 4-tuple $(s, a, s', r')$, meaning that taking action $a$ from state $s$ results in state $s'$ with reward $r'$. The second is to replay those experiences to update the q-network's weights gradually such that the q-network can approximate the q-values of the optimal policy for each state-action pair.

To apply the q-learning algorithm to solve our problem, we first design a q-network architecture to represent the target policy $\pi^*$, then introduce an algorithm for training the agent to maximize the chance to generate viable RQs for the given workload $W$.  Next we provide more details in both steps.

\begin{figure}[ht]
  \centering
  \includegraphics[width=\linewidth]{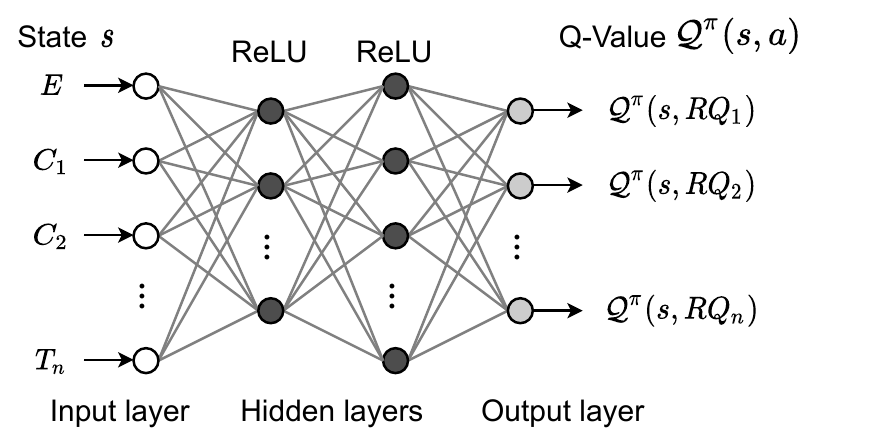}
  \caption{Q-network in \sysName for query rewriting.}
  \label{fig:dqn}
\end{figure}

\boldstart{Q-network for the query-rewriting problem.} A q-network approximates a policy $\pi$ using a neural network, and different weights of the network reflect different policy instances. As shown in Figure~\ref{fig:dqn}, the q-network adopts a simple layered architecture that includes one input layer, two fully connected linear hidden layers, and a linear output layer. The two hidden layers have sizes similar to the input layer. The number of nodes in the output layer is the number of possible actions in the model, i.e., possible RQs. All the hidden layers adopt ``ReLU''~\cite{relu} as the activation function. Given a state $s$, the input of the network is a vector of $s=(E, C_1, C_2, \ldots, C_n, T_1, T_2, \ldots, T_n)$. The output is the q-value $\mathcal{Q}^{\pi}(s, a)$ for each action $a$ in state $s$, where action $a$ can be any rewritten query $RQ_i$ in $RQSet = \{RQ_1, RQ_2, \ldots, RQ_n\}$. 

\begin{algorithm}[htb]
\caption{Training an MDP agent \label{alg:train}}
    \DontPrintSemicolon
    \KwIn{A query workload $W=[q_1, q_2, \ldots, q_m]$ \newline 
          A transition function $\mathcal{T}$ \newline
          A reward function $\mathcal{R}$ \newline
          A time budget $\tau$
    }
    \KwOut{An agent's policy $\pi$}
    
    Replay memory $M$ $\leftarrow$ $\{\}$ with capacity $\mathcal{C}$;\;
    Initialize policy $\pi$ with random weights;\;
    \While{$\pi$ does not converge}{
        $W$ $\leftarrow$ \text{shuffle the queries in $W$};\;
        \For{each query $q$ in $W$}{ 
            State $s$ $\leftarrow$ $(0, C_1, C_2, \ldots, C_n, 0, 0, \ldots, 0)$;\;
            Remaining set $\rho$ $\leftarrow$ query $q$'s all possible RQs $\{RQ_1, RQ_2, \ldots, RQ_n\}$;\;
            Reward $r$ $\leftarrow$ $0$;\;
            \While{($s$, $\tau$, $\rho$) is not at a termination state \label{alg:train:termination}}{
                $f$ $\leftarrow$ generate a random number from [0,1];\;
                \eIf{$f$ $<$ $\epsilon$}{
                    $a$ $\leftarrow$ a random RQ from $\rho$; \label{alg:train:explore}\;
                }{
                    $a$ $\leftarrow$ $\argmax_{RQ_i \in \rho}{\mathcal{Q}^{\pi}(s, RQ_i)}$;  \label{alg:train:exploit}\;
                }
                \tcp*[h]{Estimate query $a$ and transit to state $s'$}\;
                $s'$ $\leftarrow$ $\mathcal{T}(s, a)$; \label{alg:train:transit} \; 
                \tcp*[h]{Compute the immediate reward}\;
                $r'$ $\leftarrow$ $\mathcal{R}(s, a)$; \label{alg:train:reward} \;
                Store experience tuple $(s, a, s', r')$ in $M$; \label{alg:train:store_exp} \;
                \tcp*[h]{Remove query $a$ from the remaining set $\rho$}\;
                $\rho$ $\leftarrow$ $\rho - \{a\}$;                 $s$ $\leftarrow$ $s'$; $r$ $\leftarrow$ $r'$;\;
            }
            Update $\pi$ using a random sample from $M$; \label{alg:train:update}\;
        }\label{alg:train:end}
    }
\end{algorithm}

\boldstart{Training an MDP agent for query rewriting.} Algorithm~\ref{alg:train} details how we train the MDP agent. To apply deep q-learning, we generate the replay memory $M$ of experiences. For a given visualization query workload $W=[q_1, q_2, \ldots, q_m]$, we generate a set of experiences. Each experience is a 4-tuple 
$$(s, a, s', r'),$$ 
where the agent in a state $s$ estimates the time of the hinted query represented by an action $a$ and observes the next state $s'$ with a reward $r'$. Note that different queries can have different optimal policies. Our goal is to learn an optimal policy for the whole workload. We let the agent explore all the queries in the workload $W$ in multiple iterations until the policy converges or the number of runs exceeds a maximum threshold. In each iteration, we shuffle the order of queries to reduce the bias caused by earlier queries on the exploration direction of later queries. For each query $q$ in $W$, we let the agent complete a sequence of decisions. At each step, it selects an RQ to estimate. It pays the cost to estimate the rewritten query's execution time, transits to the next state, and receives an immediate reward. The agent repeats the process until it reaches a termination state (line~\ref{alg:train:termination}) in one of the three cases. The first case is when the estimated time $T(a)$ of the rewritten query in action $a$ suggests it is potentially viable, i.e., $s.E + T(a) \leq \tau$. The second case is when the agent runs out of time, i.e, $s.E \geq \tau$. The third case is when the agent has exhausted all possible RQs, i.e., $\rho = \emptyset$.

When the agent decides which RQ to explore at each step (lines~\ref{alg:train:explore} to~\ref{alg:train:exploit}), we adopt an $\epsilon$-greedy strategy~\cite{journals/corr/MnihKSGAWR13, conf/ki/Tokic10} to balance between the exploration of RQs with uncertain values and the exploitation of RQs known with high values. With an $\epsilon$ probability, the agent chooses a random RQ that it has not been considered before (line~\ref{alg:train:explore}). Otherwise, it selects an RQ that has not been explored with the highest q-value based on the current policy weights (line~\ref{alg:train:exploit}). We start with a high probability ($\epsilon$) of exploration and gradually decrease it to favor exploitation with the training progress.

Once an RQ is decided by the agent as an action $a$, we call the transition function $\mathcal{T}$ (Section~\ref{subsec:model-def}) to transit the agent to the next state $s'$ (line~\ref{alg:train:transit}). We estimate the query $a$'s running time and update the new state $s'$ by adding the estimated time for $a$, adding the cost to the elapsed time, and modifying the costs of affected RQs. We then call the reward function $\mathcal{R}$ (Section~\ref{subsec:model-def}) to compute an immediate reward $r'$ for estimating the RQ in $a$ (line~\ref{alg:train:reward}). To this end, we have generated a new experience tuple $(s, a, s', r')$, and store it in the replay memory $M$ (line~\ref{alg:train:store_exp}). When $M$ reaches its capacity $\mathcal{C}$, we replace existing experiences in a FIFO manner.

After processing a query, we update the policy $\pi$ following the original deep q-learning algorithm~\cite{journals/corr/MnihKSGAWR13} (line~\ref{alg:train:update}). 
We sample a random subset $M'$ of experiences from $M$. For each experience tuple $(s, a, s', r')$ in $M'$, we first compute the target q-value $y$ of the state-action pair $(s, a)$ using the Bellman equation~\cite{journals/ml/WatkinsD92}.
We then update the weights in policy $\pi$ by minimizing the loss value $L$ between the target q-value $y$ and the current q-value, where $L$ is defined as:
$$
L = (\mathcal{Q}^{\pi}(s, a) - y)^2.
$$
We keep updating the policy $\pi$ until it converges, i.e., the total accumulated reward of the training workload $w$ does not improve much in new iterations (e.g., less than $1\%$). 

\boldstart{Accommodating estimation inaccuracy using MDP.}
One advantage of using the MDP framework where an approximate QTE may give inaccurate estimations is its tolerance of the inaccuracy. The MDP model captures the uncertainty in two places. One is the transitions between states that store the estimated times of explored RQs. Although estimated times can have errors, statistically, after learning from the historical queries, the agent understands which action has the highest expected total reward. Another place is the reward definition, where the penalty for making a wrong decision will lead the agent to understand the QTE's mistakes and avoid them in the future.

\subsection{Using MDP to Rewrite Queries Online }

After we train an MDP agent, the query rewriter utilizes the agent to generate a rewritten query for a new visualization query $q$ online. Algorithm~\ref{alg:process} shows the pseudo-code. Starting from an initial state $s$, we use the trained policy $\pi$ to compute the q-values for all the RQs and select the one with the highest q-value as the action $a$ (line~\ref{alg:process:exploit}). We then estimate the running time of query $a$ and transit to state $s'$ (line~\ref{alg:process:transit}). We compute the immediate reward $r'$ for estimating RQ in $a$ (line~\ref{alg:process:reward}). If the action $a$ is a potentially-viable RQ (line~\ref{alg:process:viable}), we output the query $\hat{RQ_i}$ in $a$ as the generated rewritten query. Otherwise, we run out of time for the remaining RQs (line~\ref{alg:process:outoftime}). Then we select the rewritten query $RQ_j$ with the minimum execution time estimated so far and output it. If neither cases happen, we repeat the above process.

\begin{algorithm}[htb]
\caption{Generating an RQ online \label{alg:process}}
    \DontPrintSemicolon
    \KwIn{A new query $q$ \newline 
          A trained policy $\pi$ \newline
          A transition function $\mathcal{T}$ \newline
          A reward function $\mathcal{R}$ \newline
          A time budget $\tau$
    }
    \KwOut{An RQ}
    
    State $s$ $\leftarrow$ $(0, C_1, C_2, \ldots, C_n, 0, 0, \ldots, 0)$;\;
    Remaining set $\rho$ $\leftarrow$ query $q$'s all possible RQs $\{RQ_1, RQ_2, \ldots, RQ_n\}$ ;\;
    Reward $r$ $\leftarrow$ $0$;\;
    \While{$True$}{
        \tcp*[h]{Select a query with the highest q-value predicted by $\pi$} \;
        $a$ $\leftarrow$ $\argmax_{RQ_i \in \rho}{\mathcal{Q}^{\pi}(s, RQ_i)}$; \label{alg:process:exploit}\;
        \tcp*[h]{Estimate query $a$ and transit to state $s'$}\;
        $s'$ $\leftarrow$ $\mathcal{T}(s, a)$; \label{alg:process:transit} \; 
        \tcp*[h]{Compute the immediate reward}\;
        $r'$ $\leftarrow$ $\mathcal{R}(s, a)$; \label{alg:process:reward} \;
        \tcp*[h]{Remove query $a$ from the remaining set $\rho$}\;
        $\rho$ $\leftarrow$ $\rho - \{a\}$; $\leftarrow$ $s'$; $r$ $\leftarrow$ $r'$;\;
        \uIf{$s.E + T(a) \leq \tau$ \label{alg:process:viable}}{
            \Return $\hat{RQ_i}$ represented by $a$; \;
        }
        \uIf{$s.E \geq \tau$ \label{alg:process:outoftime} or $\rho$ $=$ $\emptyset$}{
            \Return $RQ_j$ with the minimum execution time estimated in $s$;\;
        }
    }
\end{algorithm}

\section{Approximation Rewriting Options \label{sec:generalization}}

In this section, we generalize \sysName by considering rewriting options with approximation rules.  Recall that using a query-hint set to rewrite an original query $Q$ into an $RQ$ can help the  database generate an efficient physical plan that computes the actual result without any approximation. However, for expensive queries where no physical plan can meet the time constraint, by applying an approximation-rule set to $Q$, \sysName can generate an $RQ$ that efficiently computes an approximate result within the time budget.  We first extend the MDP model in Section~\ref{sec:mdp-model} to consider approximation rules.  We then discuss two approaches to applying the MDP model to implement a quality-aware query rewriter. The quality-aware query rewriter makes the best effort to generate a viable rewritten query and maximize the result's quality.  In the end, we discuss the trade-offs between the two approaches.

\subsection{Quality-Aware MDP Model}

\begin{figure}[htb]
  \centering
  \includegraphics[width=0.9\linewidth]{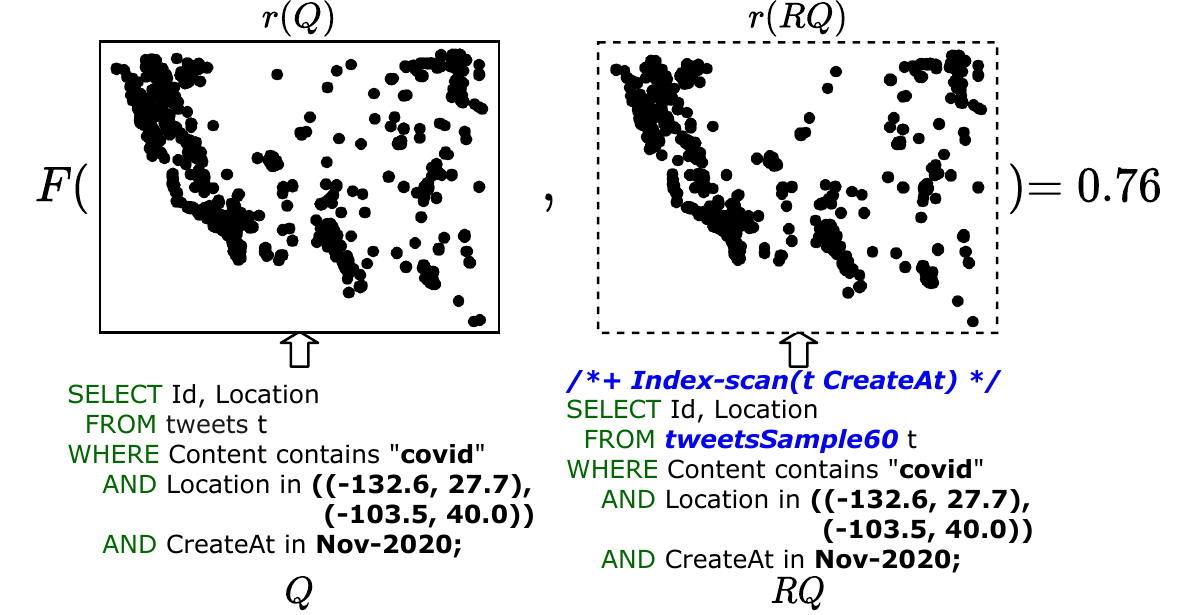}
  \caption{The quality of $RQ$ compared to $Q$ using a Jaccard-based quality function as an example.}
  \label{fig:quality}
\end{figure}

Consider the case where the rewriting options contain both query hints and approximation rules. A rewritten query can return an approximate result with quality loss. We need to let the MDP agent learn to maximize the chance to generate a viable rewritten query and maximize the quality of the query result simultaneously. To quantify the quality of a rewritten query, we assume a given visualization quality function $F$. Let $r(Q)$ be the result of the original query $Q$, and $r(RQ)$ be the result of the rewritten query $RQ$. Then $F(r(Q), r(RQ))$ computes the quality of $r(RQ)$. 
For example, suppose we use the Jaccard similarity function to measure the quality of an approximate result. Figure~\ref{fig:quality} shows that the quality of the scatterplot visualization result of an approximate rewritten query $RQ$ compared to the original query $Q$ is $0.76$.  Note that \sysName does not have restrictions on quality functions, and many functions can be used, such as VAS in~\cite{conf/icde/ParkCM16} for scatterplots and the function of distribution precision in~\cite{conf/sigmod/DingHCC016} for pie charts.

\boldstart{Reward function for a quality-aware MDP model.} 
To achieve the goal of guiding the MDP agent to learn to maximize the chance to generate a viable rewritten query and maximize the quality of the query result simultaneously, we extend the definition of the reward function in Section~\ref{sec:mdp-model}.  Recall that the learning goal of an MDP agent is to maximize the accumulative reward.  In Section~\ref{sec:mdp-model}, once the agent decides a rewritten query, it receives a reward that reflects the query performance in terms of the total running time. Guided by the reward, the agent learns to generate a viable rewritten query quickly.  Similarly, the MDP agent can also learn to quickly generate a viable rewritten query with a high result quality if the final reward reflects both the decided rewritten query's efficiency and quality. The main idea is to combine the efficiency defined in Section~\ref{sec:mdp-model} and the quality.  The new reward function is a weighted summation of both.  Formally, suppose the generated rewritten query by the agent is $\hat{RQ}$ and the actual running time of query $\hat{RQ}$ is $\hat{T}$. Then the new reward function $\mathcal{R}(s, a)$ is defined as follows:

\begin{equation}
  \mathcal{R}(s, a) = 
       \beta \frac{\tau - s.E - \hat{T}}{\tau} + (1 - \beta) F\bigl(r(Q), r(\hat{RQ})\bigr).
\end{equation}

The term $\frac{\tau - s.E - \hat{T}}{\tau}$ represents the efficiency of the rewritten query in terms of running time compared to the time budget.  The function $F(r(Q), r(\hat{RQ}))$ represents the quality of the RQ's result. 
Note that computing $F$ could be expensive since the actual result $r(Q)$ of the original query is required. However, we only need to pay the cost in the offline training phase once. In the online phase, we don't need to compute the $F$ value for a new query when we use the MDP model to explore different RQs.  Since the MDP model learns from the final reward values only, we do not require every query to use the same quality function. In particular, different quality functions can be applied for different training queries to evaluate their visualization qualities, e.g., some queries are visualized as scatterplots and others as heat-maps. 
$\beta \in [0, 1]$ is a parameter that indicates how important the running time is compared to the result quality. 

\subsection{Quality-Aware Query Rewriter}

Now we discuss how to apply the extended MDP model to implement a quality-aware query rewriter.  We present the technical details of two approaches and discuss their pros and cons. We will show the evaluation results in Section~\ref{sec:experiments}.

\begin{figure}[htb]
  \centering
  \includegraphics[width=0.9\linewidth]{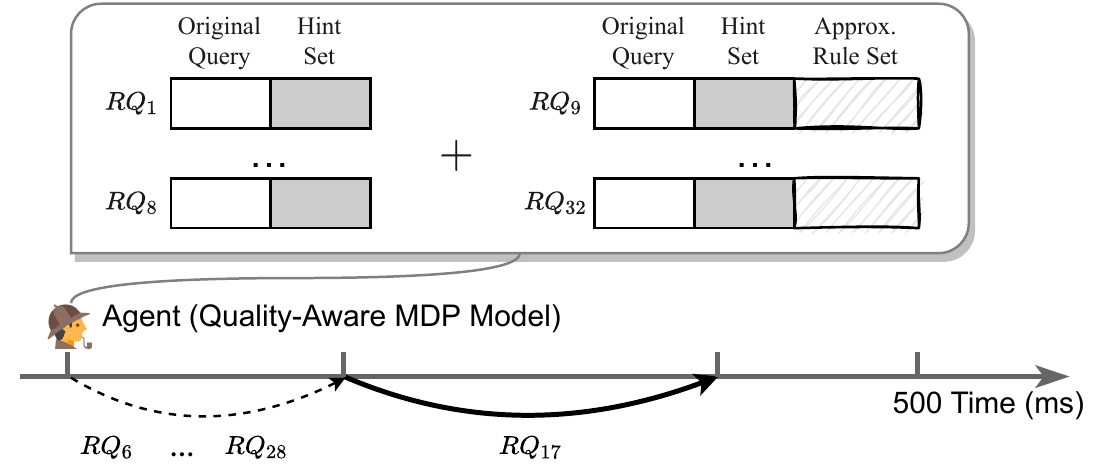}
  \caption{One-stage MDP approach.}
  \label{fig:query-rewriter-one-stage}
\end{figure}

\boldstart{One-stage approach.}
A natural idea is to replace the MDP model in Section~\ref{sec:mdp-model} with the quality-aware MDP model. We  let the MDP agent simultaneously consider query hints and approximation rules as rewriting options. By applying the new reward function combining both the efficiency of the rewritten query and the result's quality, the MDP agent learns to maximize the chance of generating viable rewritten queries and maximize the quality.

\begin{figure}[htb]
  \centering
  \includegraphics[width=0.9\linewidth]{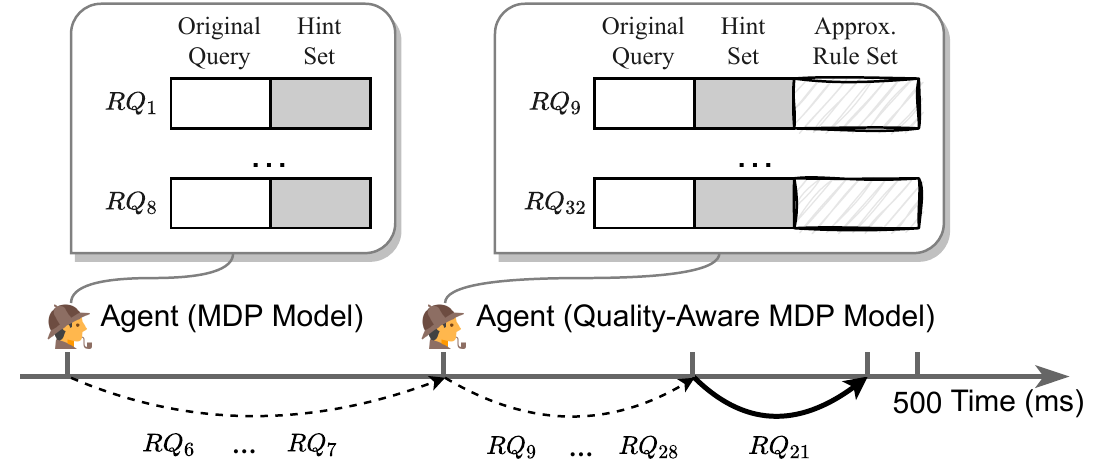}
  \caption{Two-stage MDP approach.  After running the original agent that considers the $8$ query-hint sets defined in Figure~\ref{fig:sequential-decsision-making} without approximation rules, we cannot find a viable RQ. We then run the new agent with the quality-aware MDP model that considers all $8$ query-hint sets combined with $3$ approximation-rule sets (e.g., substituting the \texttt{tweets} table with $20\%$, $40\%$, or $80\%$ sample tables), resulting in $24$ rewritten queries in total. After spending extra time exploring a few RQs, the quality-aware agent chooses $RQ_{21}$ as the final decision.}
  \label{fig:query-rewriter-two-stage}
\end{figure}

\boldstart{Two-stage approach.} A drawback of the previous approach is that the agent might miss a non-approximate viable rewritten query.  To solve this problem, we consider a two-stage approach, with a main idea to let the MDP agent exhaust all candidate query hints first and then explore those approximation rules.  In the two-stage approach, \sysName first runs the original MDP model, excluding the approximation rules. If the agent finds a viable rewritten query, it outputs the $RQ$ as before.  If the agent exhausts all candidate $RQ$s without finding a viable one, and the elapsed time has not exceeded the time budget $\tau$, then we run the new quality-aware MDP model that considers the approximation rules to find a viable $RQ$.  

When the planning time for the original agent is longer than the time budget, the two-stage approach reduces to the case described in Section~\ref{sec:mdp-model}. In this case, the one-stage approach is preferred since it can increase the chance of generating a viable rewritten query considering approximation rules.  When the planning time for the original agent is relatively small compared to the time budget, the two-stage approach has the advantage of not missing any non-approximate viable rewritten queries.

\section{Experiments}
\label{sec:experiments}

We conducted experiments to evaluate \sysName. In particular, we want to answer the following questions: (1) How well does it rewrite queries to support visualization requests? (2) How well does it generalize to different numbers of rewriting options? (3) How well does it perform for different types of queries (e.g. single-table selection queries and multiple-table joining queries)? (4) How well does it generalize to different time budgets, unseen queries and other databases? (5) How does it compare with related solutions? and (6) How much is its training overhead?

\begin{figure*}[htb]
    \centering
    \begin{subfigure}[t]{0.33\linewidth}
        \includegraphics[width=\linewidth]{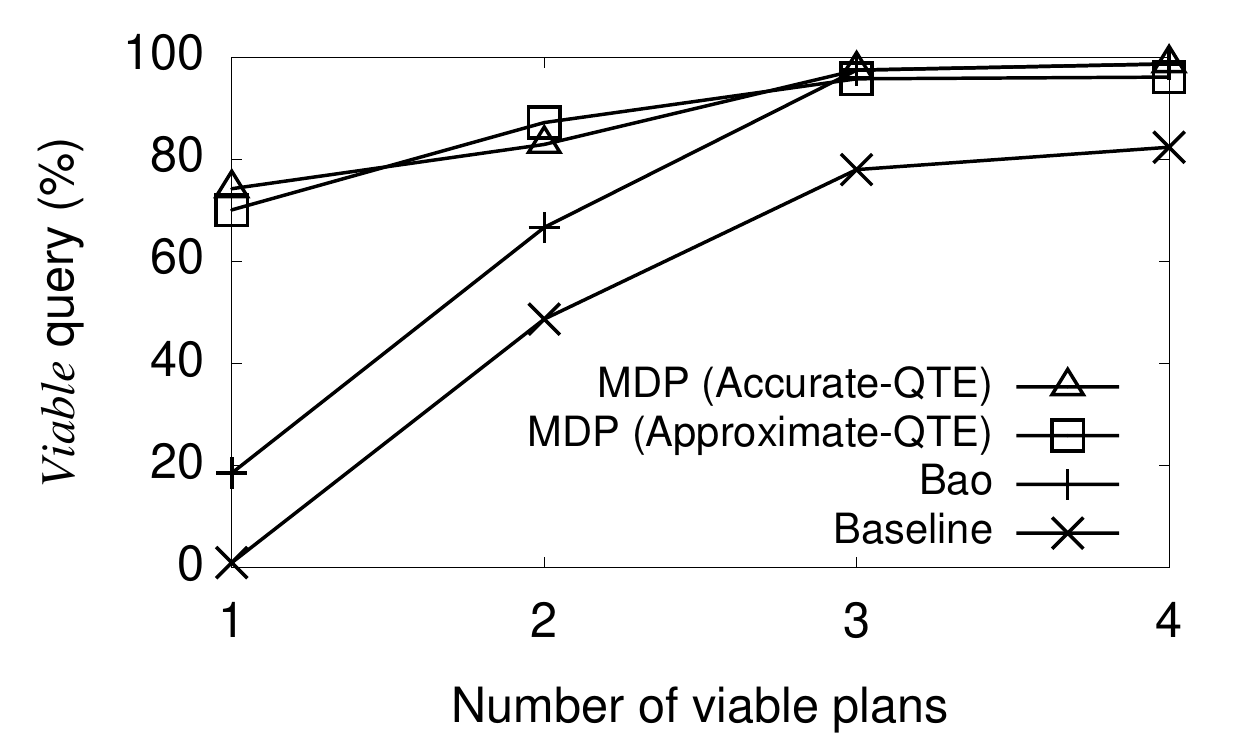}
        \caption{Twitter ($\tau=500ms$).}
    \end{subfigure}
    \hfill
    \begin{subfigure}[t]{0.33\linewidth}
        \includegraphics[width=\linewidth]{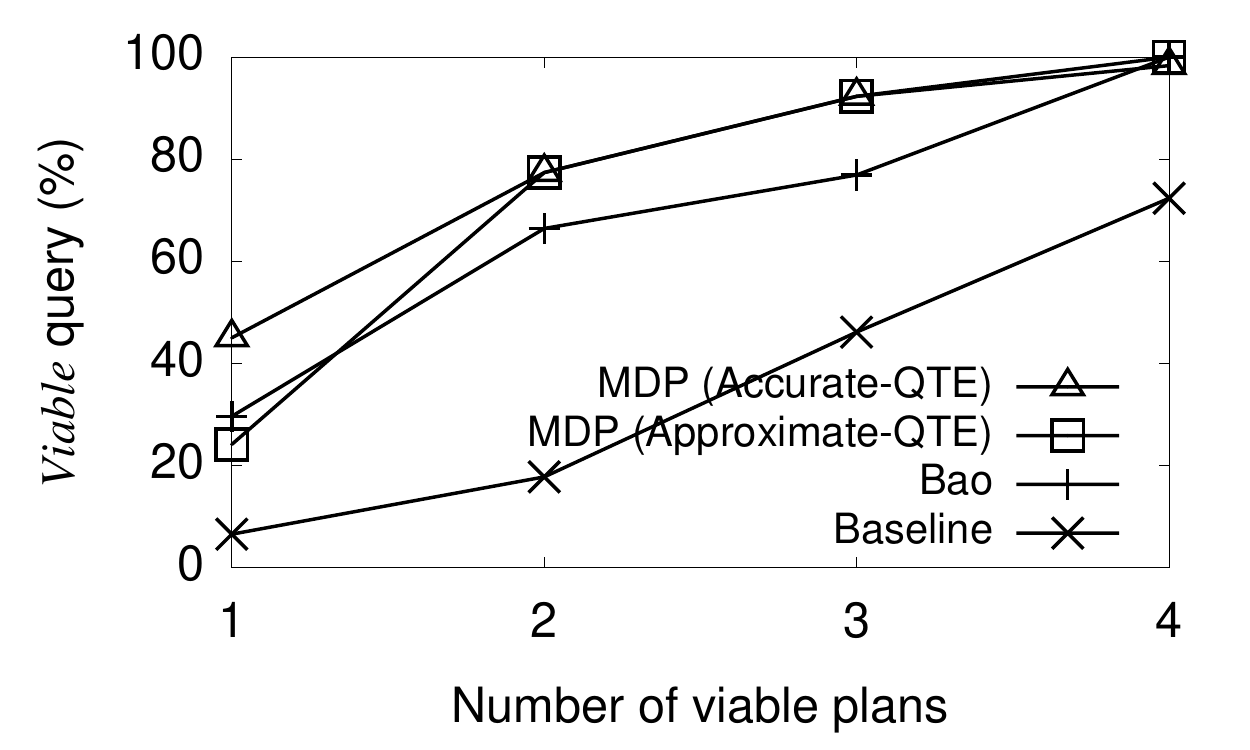}
        \caption{NYC Taxi ($\tau=1s$).}
    \end{subfigure}
    \hfill
    \begin{subfigure}[t]{0.33\linewidth}
        \includegraphics[width=\linewidth]{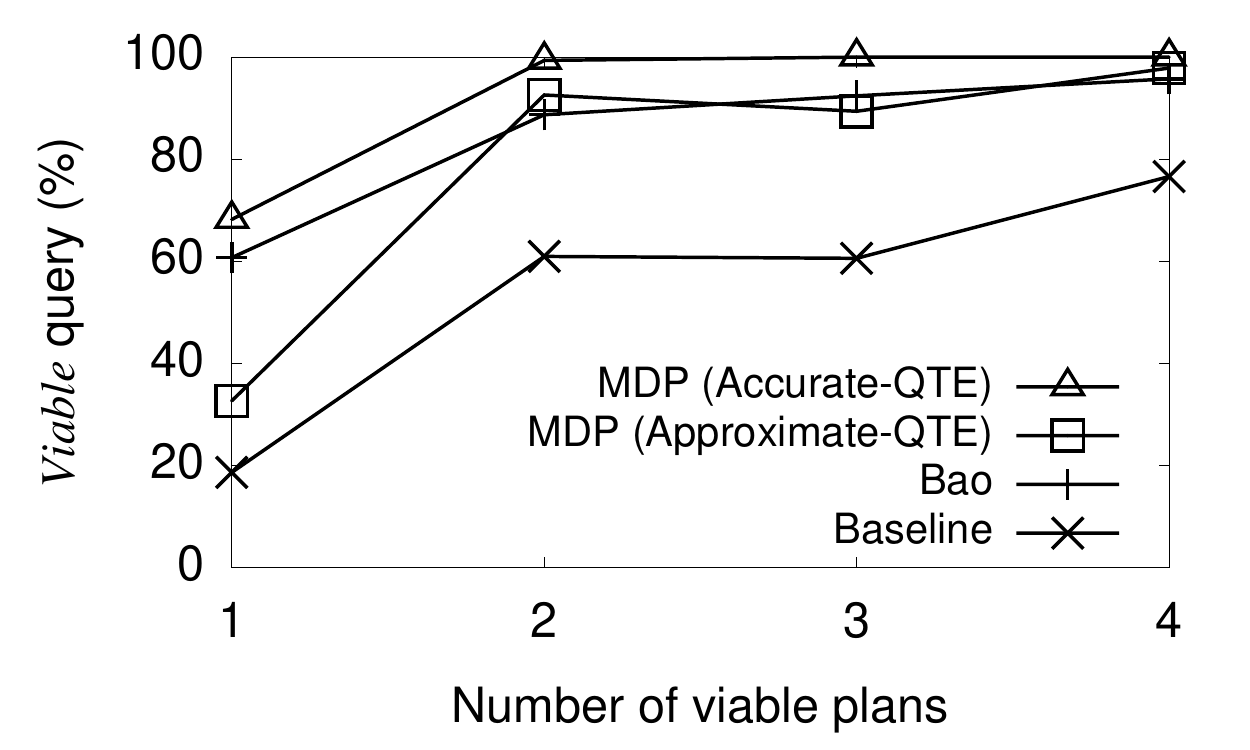}
        \caption{TPC-H ($\tau=500ms$).}
    \end{subfigure}
    \caption{Viable query percentage using PostgreSQL on {\sf Twitter}, {\sf NYC Taxi} and {\sf TPC-H} datasets. \label{fig:viable-query-percentage}}
\end{figure*}

\subsection{Setup}

\boldstart{Datasets.} We used two real datasets and a synthetic one as shown in Table~\ref{table:datasets}. The {\sf Twitter} dataset included 100 million geo-located tweets in the US from November 2015 to January 2017. We kept the timestamp, geo-coordinate, text message, and several user attributes for each tweet in a \texttt{tweets} table. For the experiment on join queries, we used the \texttt{tweets} table and a \texttt{users} table. The former had a foreign key of ``user\_id'' referencing the ``id'' in the latter. We used the geo-coordinate attribute as the output for visualization (e.g., choropleth map, heatmap, or scatterplot). The {\sf NYC Taxi} dataset~\cite{nyc-taxi-data:website} included taxi trip records within three years from 2010 to 2012. The third dataset was generated from the {\sf TPC-H} benchmark~\cite{TPC-H}. We used the line-item table as the fact table. The attributes we used for query selection conditions are shown in Table~\ref{table:datasets}.

\begin{table}[htbp]
\footnotesize
\caption{Datasets.}
\label{table:datasets}
\begin{tabular}{|l|r|r|l|l|}
\hline
\bf{Dataset} & \bf{\shortstack[l]{Record \# \\ (millions)}} & \bf{Size } & \bf{\shortstack[l]{Filtering \\ Attributes}} & \bf{\shortstack[l]{Output \\ Attributes}} \\ \hline
Twitter & 100 & 57GB & \shortstack[l]{text, \\ created\_at, \\  coordinates, \\ users\_statues\_count, \\ users\_followers\_count} & \shortstack[l]{id, \\ coordinates}    \\ \hline
NYC Taxi & 500 & 146GB & \shortstack[l]{pickup\_datetime, \\ trip\_distance, \\ pickup\_coordinates} & \shortstack[l]{id, \\ pickup\_coordinates}    \\ \hline
TPC-H & 300 & 65GB & \shortstack[l]{extended\_price, \\ ship\_date, \\ receipt\_date} & \shortstack[l]{quantity, \\ discount}   \\ \hline
\end{tabular}
\end{table}

\begin{figure*}[htb]
    \centering
    \begin{subfigure}[t]{0.33\linewidth}
        \includegraphics[width=\linewidth]{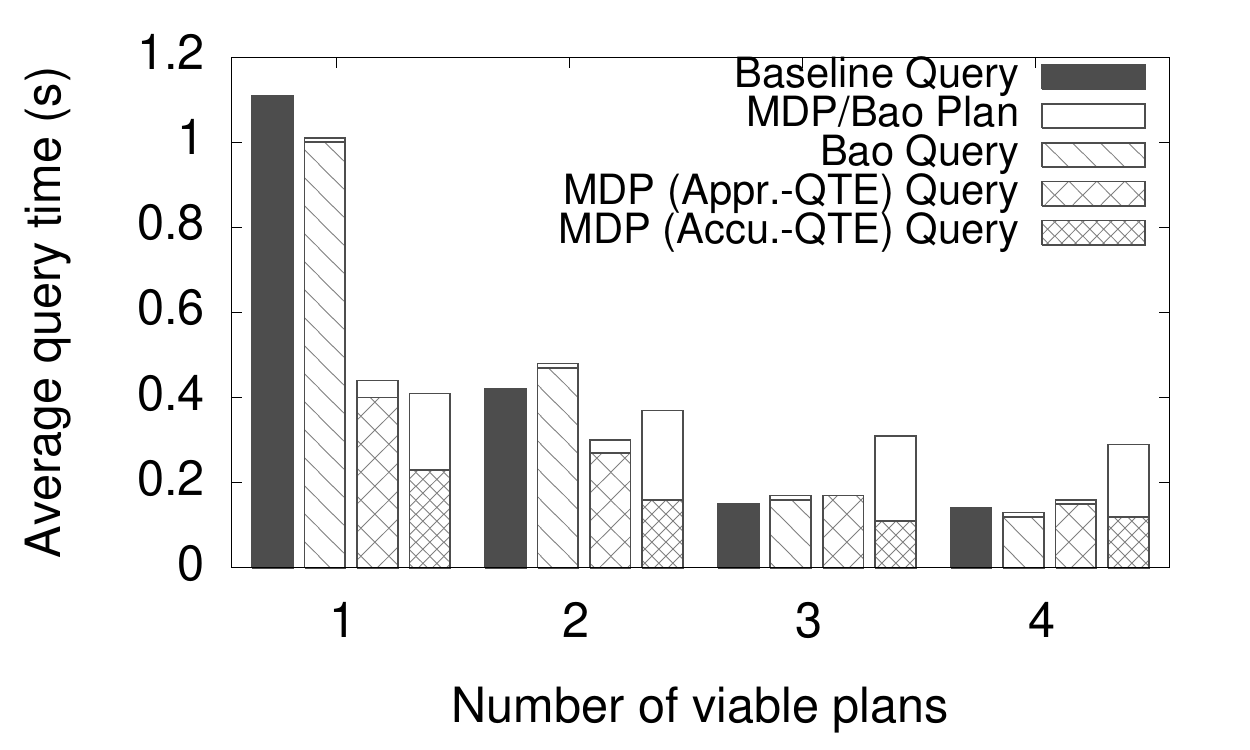}
        \caption{Twitter ($\tau=500ms$).}
    \end{subfigure}
    \hfill
    \begin{subfigure}[t]{0.33\linewidth}
        \includegraphics[width=\linewidth]{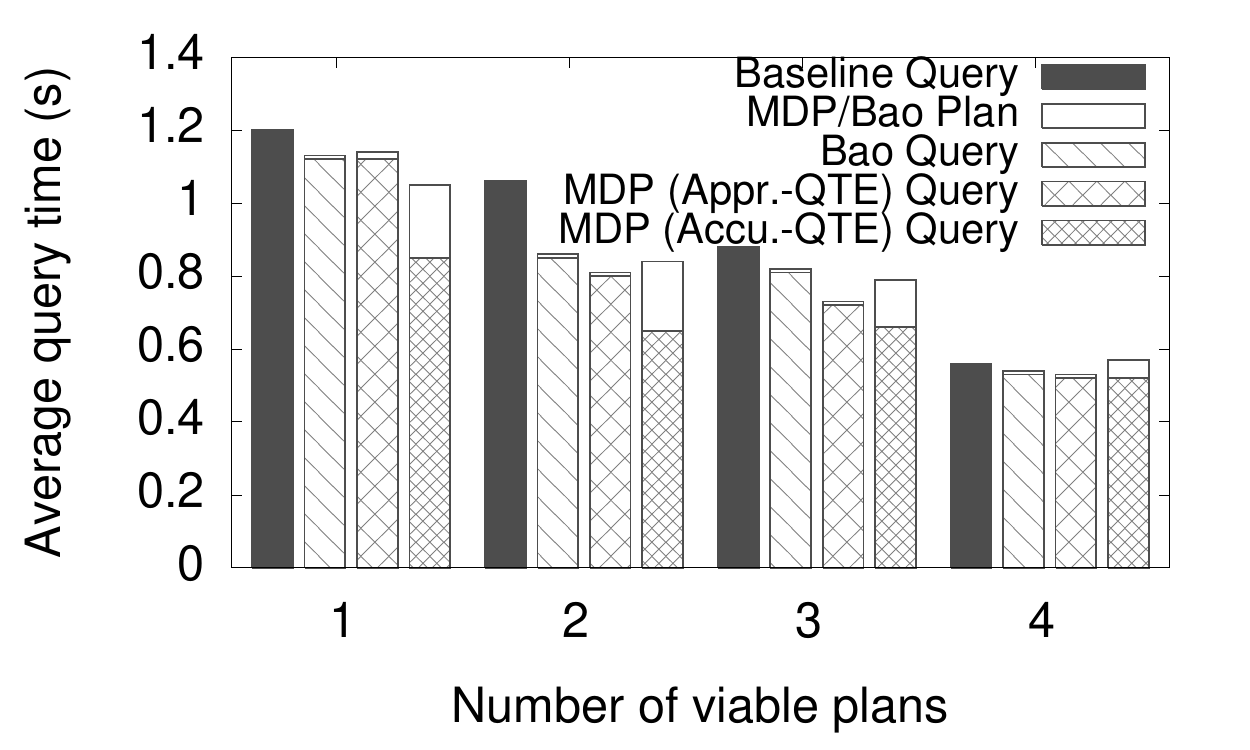}
        \caption{NYC Taxi ($\tau=1s$).}
    \end{subfigure}
    \hfill
    \begin{subfigure}[t]{0.33\linewidth}
        \includegraphics[width=\linewidth]{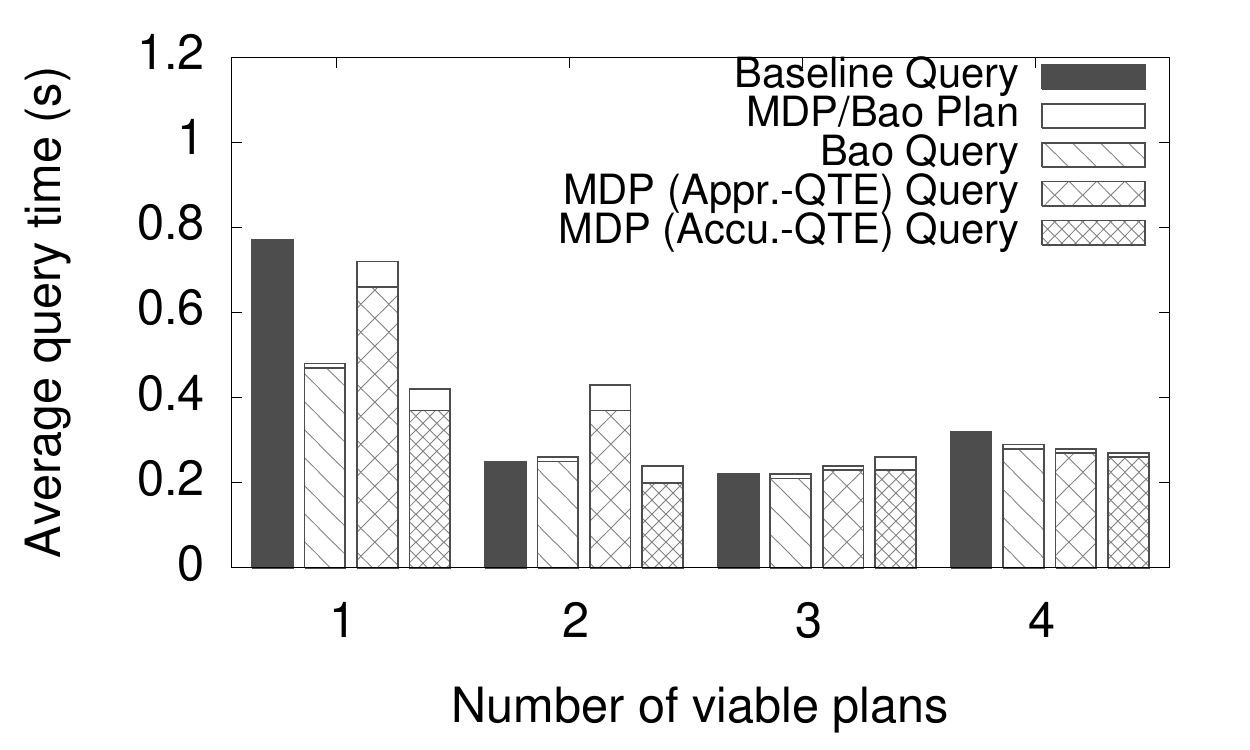}
        \caption{TPC-H ($\tau=500ms$).}
    \end{subfigure}
    \caption{Average query response time using PostgreSQL on {\sf Twitter}, {\sf NYC Taxi} and {\sf TPC-H} datasets. \label{fig:avg-query-time}}
\end{figure*}

\boldstart{Query workloads.} 
We generated random queries on each dataset for training and evaluation. Take {\sf Twitter} dataset as an example. We first randomly sampled a set of tweets from the base table. For each tweet, we generated a query as follows.  We chose the {\tt text}, {\tt created\_at}, and {\tt coordinates} attributes for the selection conditions in the query. We generated three conditions based on the values in the sampled tweet. For {\tt text}, we randomly selected a non-stop word in the original tweet's text message as the keyword condition. For {\tt created\_at}, we generated a temporal range condition with the value in the original tweet as the left boundary. We divided the maximum range on the {\tt created\_at} attribute in the base table into multiple zoom levels, and randomly selected a level to generate the length of the range condition. Suppose the maximum range on {\tt created\_at} had $L$ days. We computed the maximum zoom level on {\tt created\_at} as $Z = \ceil*{log_2(L)}$. If we randomly chose a zoom level from range $[0, Z]$ as $z$, we computed the length of the query condition range as $l = max(L / 2^{z}, 1)$.  Similarly, for the {\tt coordinates} attribute, we used the exact coordinates in the sampled tweet as the center. We randomly chose a zoom level and generated a spatial bounding box as the spatial range condition for the query. 

In the experiments, we evenly divided generated queries into a training workload and an evaluation workload. We further divided the training workload into two parts, one with two thirds of queries for training the agent and the other one with one third of queries for validation. For all experiments, we used a hold-out validation strategy, i.e., we used a workload to train multiple MDP agents, and used a validation workload to choose a best agent to be evaluated on the evaluation workload.  We report the performance on the evaluation workload in the each section.  When evaluating different approaches, the ``difficulty'' of the queries in the evaluation workload played an important role. For example, if none of the physical plans of a query were viable, then no approach can generate a viable plan without approximation. On the contrary, if a query had many (e.g., five) viable physical plans, it was relatively easy for a method to find one. We further divided the evaluation workload into subsets of queries based on their difficulty. To measure the difficulty of a query, we define the following metric. Given a time budget $\tau$, a query's number of viable plans is $\sum_{i=1}^n[T(P_i) \leq \tau]$, where the set $\{P_1, P_2, \ldots, P_n\}$ contains an original query $Q$'s all possible physical plans generated by applying different candidate query hints, and $T(P_i)$ is the execution time of plan $P_i$. We changed the number of viable plans and evaluated different approaches on the same query workload with the same number of viable plans.

\boldstart{QTE implementations}. We implemented two QTEs to evaluate the \sysName's performance.   1) {\em Accurate-QTE}. To isolate the effect of estimation errors on the \sysName' performance, we used the actual execution time of the hinted queries as the estimation, and set up a unit cost parameter to represent the time of collecting the selectivity value of one filtering condition in a given rewritten query. Unless otherwise stated, we used $40 ms$ as the unit cost of collecting one selectivity value for the {\em Accurate-QTE}. 2)  We also implemented the ML-based {\em approximate-QTE} as presented in Section~\ref{ssec:approximiate-qte}.  We used a random sample table~\cite{conf/icde/WuCZTHN13} to estimate the selectivity values of query conditions. The selectivity values were used by the approximate-QTE's ML model to estimate the execution time of queries.

\boldstart{Performance metrics.} We used two metrics to evaluate the performance of different approaches. Recall that a generated rewritten query is ``{\em viable}'' if its total response time (including both the planning time and the querying time) is within a given time budget. The ``viable query percentage'' (VQP) of a solution was the ratio of viable queries over all the queries in the workload. The other metric was called ``{\em Average Query Response Time}'' (AQRT), which was the average total response time of all the queries in the workload.

\boldstart{Query-rewriting Approaches.}  We compared the proposed MDP-based approaches with three related methods, i.e., baseline, naive, and Bao~\cite{journals/corr/abs-2004-03814}.  MDP-based approaches included an MDP agent using an approximate-QTE, i.e., MDP (Approximate-QTE), and an MDP agent using an accurate QTE, i.e., MDP (Accurate-QTE).  In the baseline approach, the middleware relies on the database optimizer to generate a physical plan for the original query. In the naive approach, we used the same approximate QTE as the MDP-based approach, but enumerated all possible RQs in a brute-force way, then chose the best RQ as the output.  The third approach was Bao~\cite{journals/corr/abs-2004-03814}. We used its open-source release~\cite{BaoForPostgreSQL} as the server, which provided interfaces for training the model and using the model to choose the best plan for a given set of query plans. Its original client, which was a PostgreSQL plug-in, did not support query hints for using a specific index, which were required by our visualization queries. To solve this problem, we implemented a new client in Python to support such query hints while keeping their server implementation.

In the experiments, we ran both the database and the middleware on the same AWS t2.xlarge instance with four vCPUs, $16$GB RAM, and a $500$GB SSD drive. We implemented the middleware in Python $3.6$ and the neural network using Pytorch $1.7$. We evaluated \sysName on both PostgreSQL and a commercial database.

\subsection{Performance on Using Query Hints \label{subsec:non-approximate}}

We evaluated the performance of \sysName for only considering query hints in rewriting options (i.e., no approximations). For each dataset, we generated queries with three filtering conditions and set up the rewrite-option set with $8$ query-hint sets, i.e., using or not using the index on each attribute.  We varied the evaluation workloads with different numbers of viable plans, and collected the VQP and AQRT metrics for each approach. Table~\ref{table:workloads} shows the number of queries in the evaluation workloads.

\begin{table}[htbp]
\caption{Number of queries in evaluation workloads.}
\label{table:workloads}
\begin{tabular}{|r|r|r|r|r|r|r|}
\hline
\multicolumn{1}{|l|}{{\bf{\# of viable plans}}} & 0 & 1 & 2 & 3 & 4 & $\geq$ 5 \\ \hline
\multicolumn{1}{|l|}{{\bf{Twitter}}} & 518 & 97 & 234 & 118 & 153 & 69 \\ \hline
\multicolumn{1}{|l|}{{\bf{NYC Taxi}}} & 408 & 91 & 146 & 13 & 181 & 3 \\ \hline
\multicolumn{1}{|l|}{{\bf{TPC-H}}} & 381 & 107 & 310 & 66 & 47 & 0 \\ \hline
\end{tabular}
\end{table}

Figure~\ref{fig:viable-query-percentage} shows viable-query percentages (VQP) on the three datasets.  The MDP-based approaches and Bao outperformed the baseline approach significantly, with MDP (Accurate-QTE) as the best.  For example, on the {\sf Twitter} dataset, for the queries with a single viable plan, both MDP-based approaches increased the VQP from the baseline's $1\%$ and Bao's $20\%$ to more than $70\%$.  In most cases, MDP (Approximate-QTE) performed better than or comparable to Bao.
In one case of the {\sf TPC-H} dataset, Bao performed better than MDP (Approximate-QTE) mainly because Bao's QTE had a much higher accuracy than the approximate QTE for {\sf TPC-H}.  When the number of viable plans increased from $1$ to $4$, the VQP of all approaches increased because the more viable plans existed for a query, the easier it was for each approach to find a viable plan in a short amount of time.

Figure~\ref{fig:avg-query-time} shows the results of the average query-response time (AQRT) of different approaches.  On the {\sf Twitter} dataset, Bao had a comparable AQRT to the baseline, while MDP (Approximate-QTE) had much lower time than the baseline and Bao for queries with one or two viable plans. For example, MDP (Approximate-QTE) reduced the average response time from the baseline's $1.11$ seconds and Bao's $1.01$ seconds to $0.4$ seconds.  On the {\sf NYC Taxi} dataset, Bao and MDP-based approaches had comparable performance and were slightly better than the baseline. On the {\sf TPC-H} dataset, Bao was better than or comparable to the baseline. In two cases, Bao performed better than MDP (Approximate-QTE) because Bao's QTE had a much higher accuracy than the approximate QTE on {\sf TPC-H}.  However, in all cases, MDP (Accurate-QTE) always had a lower query time than Bao and the baseline, which means it generated a more efficient plan.  In cases where MDP (Accurate-QTE) had a longer response time, the extra planning time was the main reason.  At the same time, the high VQP of MDP (Accurate-QTE) proved the ability of the MDP model balancing the planning time and the query-execution time to maximize the chance of generating a viable rewritten query.

\subsection{Effect of Rewrite-Option Number}

The number of rewrite options affects both the state and action space of the MDP model, which will further affect the performance of the MDP agent. We evaluated the effect of the number of rewrite options on the {\sf Twitter} dataset. We set up two workloads of queries with filtering conditions on $4$ attributes and $5$ attributes of the \texttt{tweets} table, resulting in a rewrite-option set with $16$ query-hint sets as rewrite options and the other rewrite-option set with $32$ rewrite options.  We compared the MDP-based approaches with Bao and the baseline approach. We varied the evaluation workloads with different groups of viable-plan numbers and collected the VQP and AQRT metrics of each approach. Table~\ref{table:workloads-dimensions} shows the number of queries for the workloads.

\begin{table}[htbp]
\caption{Workloads with 16 and 32 rewrite options.}
\label{table:workloads-dimensions}
\begin{tabular}{|r|r|r|r|r|r|r|}
\hline
\multicolumn{7}{|c|}{\bf{Workloads with $16$ rewrite options.}} \\
\hline
\multicolumn{1}{|l|}{{\bf{\# of viable plans}}} & 0 & 1-2 & 3-4 & 5-6 & 7-8 & $\geq$ 9 \\ \hline
\multicolumn{1}{|l|}{{\bf{\# of queries}}} & 485 & 150 & 241 & 90 & 132 & 93 \\ \hline
\multicolumn{7}{|c|}{\bf{Workloads with $32$ rewrite options.}} \\
\hline
\multicolumn{1}{|l|}{{\bf{\# of viable plans}}} & 0 & 1-4 & 5-8 & 9-12 & 13-16 & $\geq$ 17 \\ \hline
\multicolumn{1}{|l|}{{\bf{\# of queries}}} & 412 & 141 & 197 & 159 & 151 & 145 \\ \hline
\end{tabular}
\end{table}

\begin{figure}[htb]
  \centering
  \begin{subfigure}[t]{0.65\linewidth}
    \includegraphics[width=\linewidth]{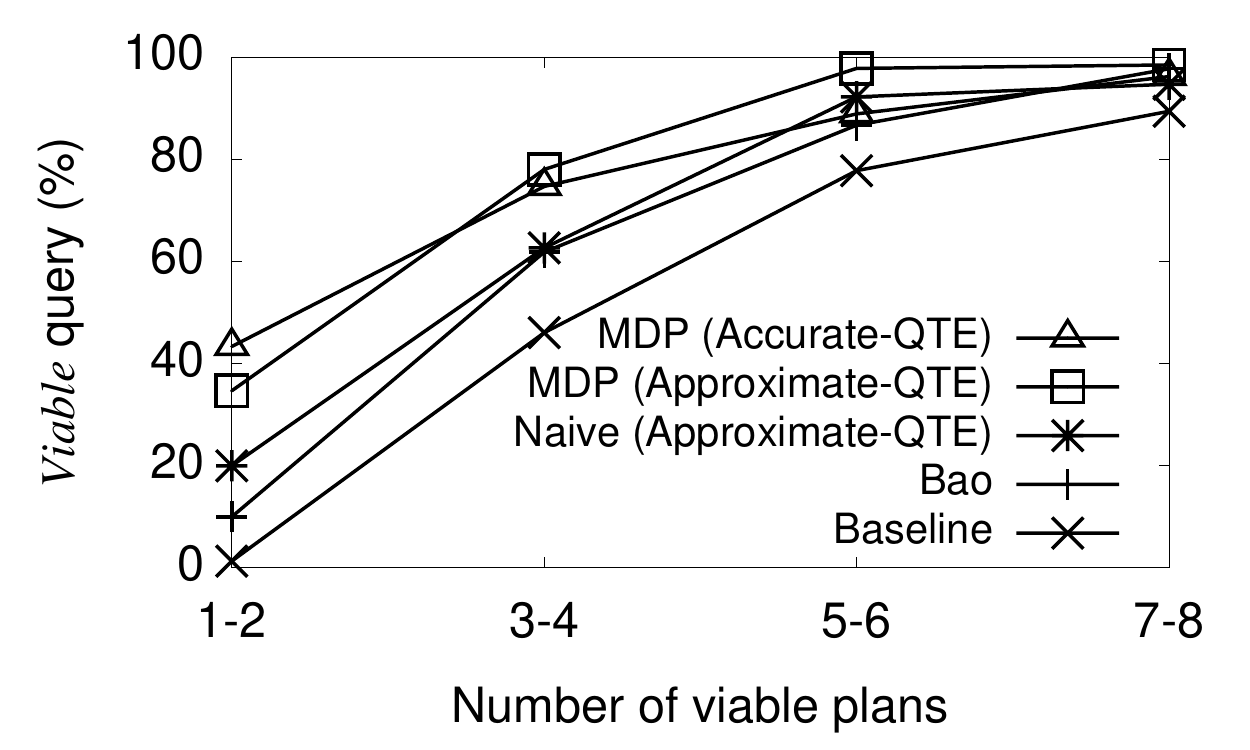}
    \caption{$16$ rewrite options.}
  \end{subfigure}
  \hfill
  \begin{subfigure}[t]{0.65\linewidth}
    \includegraphics[width=\linewidth]{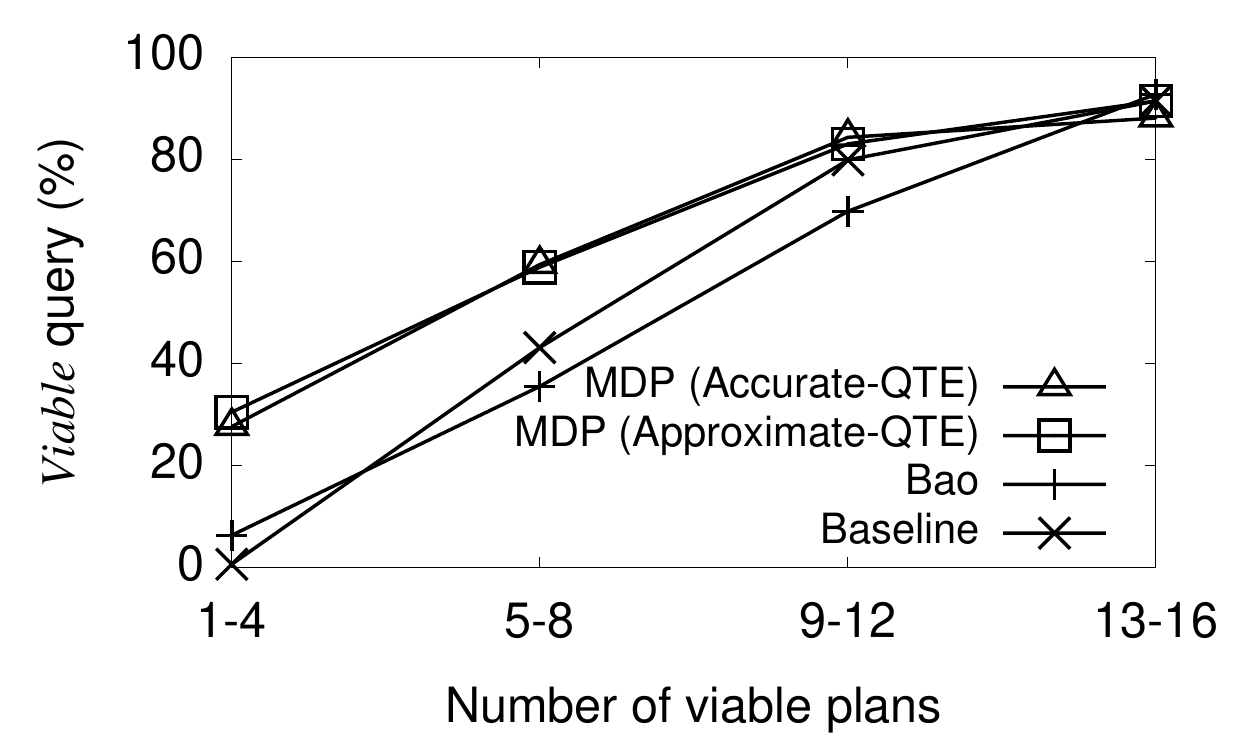}
    \caption{$32$ rewrite options.}
  \end{subfigure}
  \caption{Viable query percentages for different numbers of rewrite options using PostgreSQL on {\sf Twitter} dataset. (Time budget $\tau=500 ms$.) \label{fig:viable-query-percentage-dimensions}}
\end{figure}

Figure~\ref{fig:viable-query-percentage-dimensions} shows the VQP results of different approaches for workloads with $16$ and $32$ rewrite options.  The MDP approaches (similar between using an Approximate-QTE and using an Accurate-QTE) performed the best in both cases, generating up to $40\times$ more viable queries than both Bao and the baseline approach for the workload with $16$ rewrite options on queries with one or two viable plans.  As expected, the advantages of MDP-based approaches over the baseline approach became smaller when the number of rewrite options increased to $32$.  Estimating the query time became more expensive for queries with more rewrite options, and it increased the planning time of MDP-based approaches significantly compared to queries with fewer rewrite options.

\begin{figure}[htb]
  \centering
  \begin{subfigure}[t]{0.65\linewidth}
    \includegraphics[width=\linewidth]{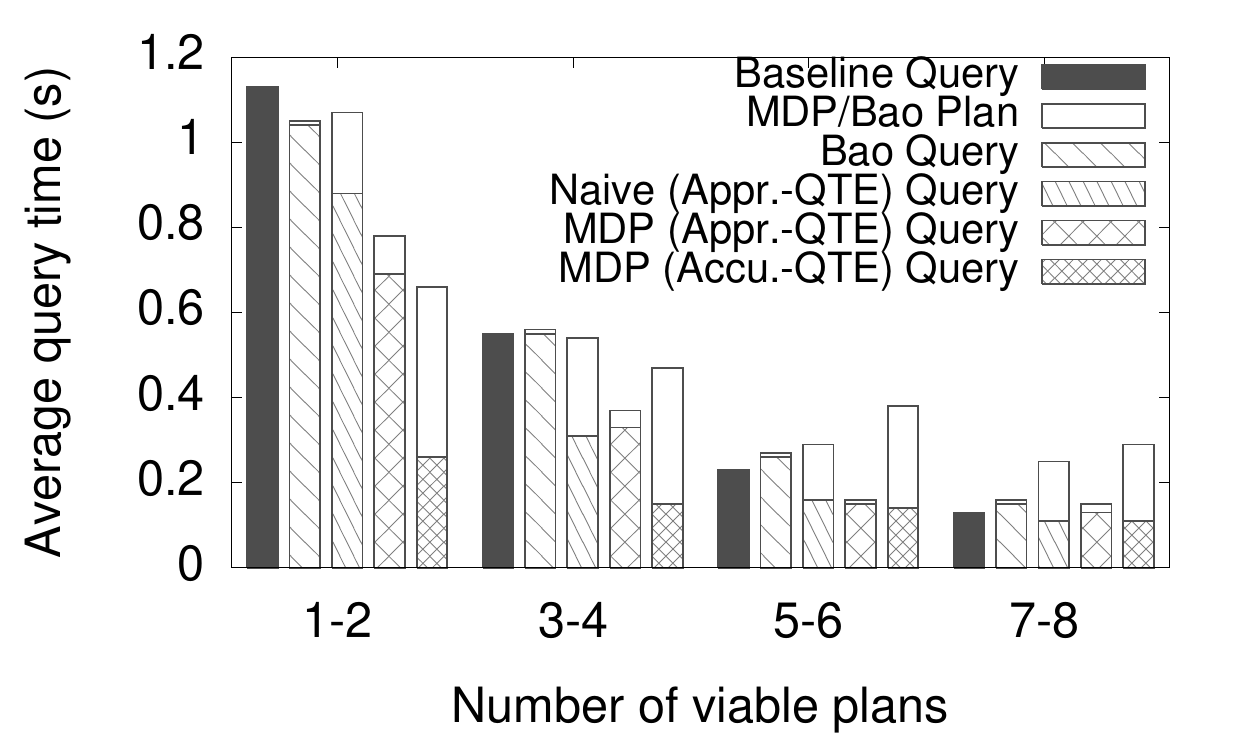}
    \caption{$16$ rewrite options.}
  \end{subfigure}
  \begin{subfigure}[t]{0.65\linewidth}
    \includegraphics[width=\linewidth]{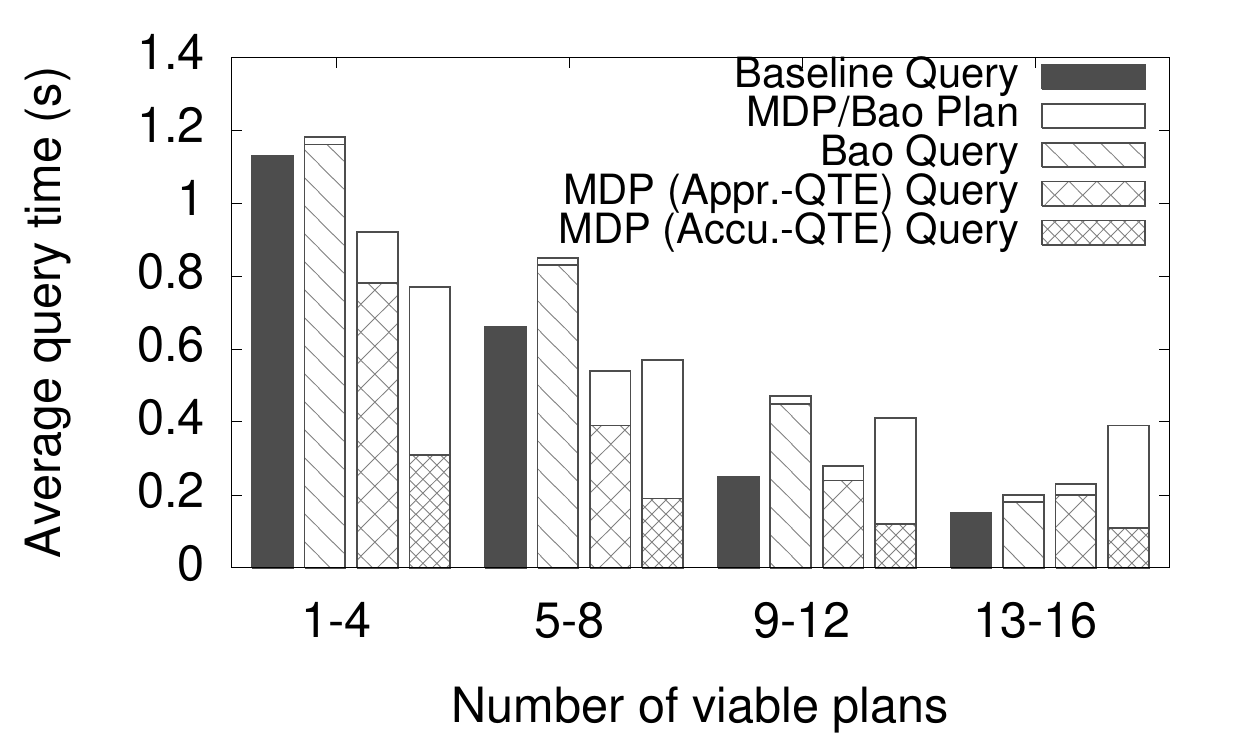}
    \caption{$32$ rewrite options.}
  \end{subfigure}
  \caption{Average query response time for different numbers of rewrite options using PostgreSQL on {\sf Twitter} dataset. (Time budget $\tau=500 ms$.) \label{fig:avg-query-time-dimensions}}
\end{figure}

\begin{figure*}[htb]
    \centering
    \begin{subfigure}[t]{0.33\linewidth}
        \includegraphics[width=\linewidth]{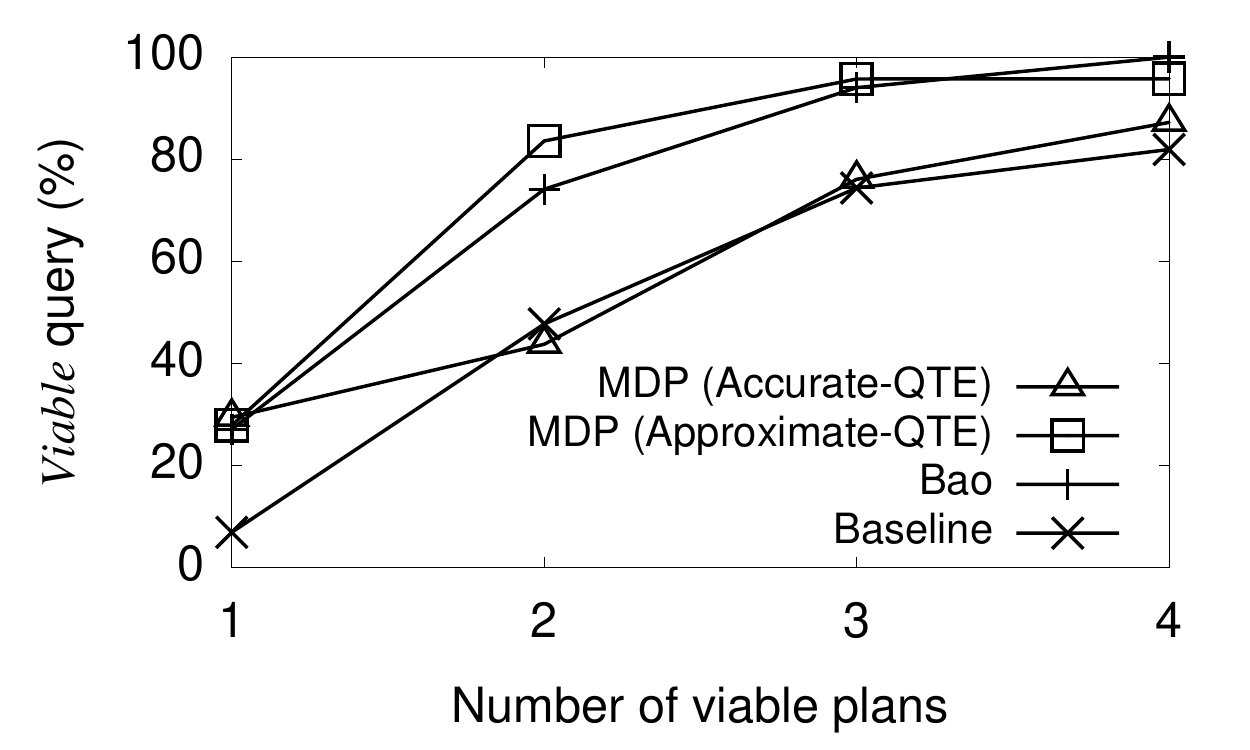}
        \caption{Time budget $\tau = 0.25$ second.}
    \end{subfigure}
    \hfill
    \begin{subfigure}[t]{0.33\linewidth}
        \includegraphics[width=\linewidth]{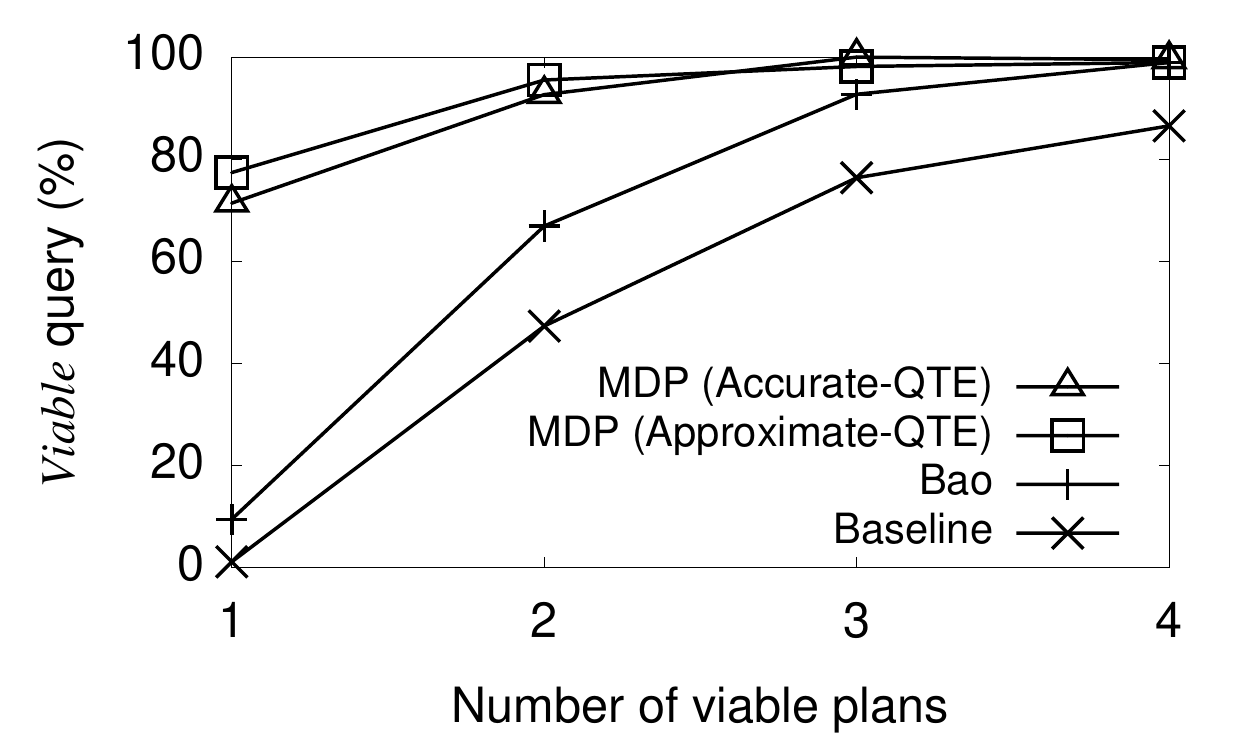}
        \caption{Time budget $\tau = 0.75$ second.}
    \end{subfigure}
    \hfill
    \begin{subfigure}[t]{0.33\linewidth}
        \includegraphics[width=\linewidth]{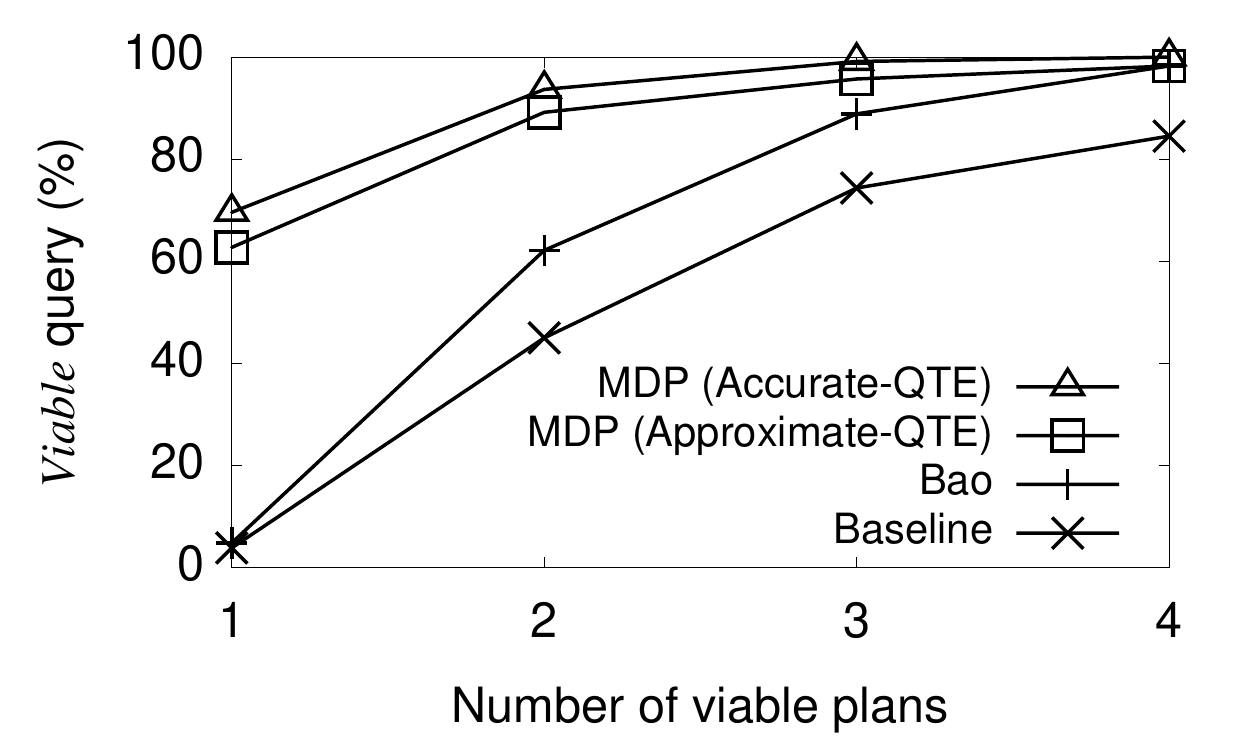}
        \caption{Time budget $\tau = 1.0$ second.}
    \end{subfigure}
    \caption{Viable query percentages using PostgreSQL on {\sf Twitter} dataset for different time budgets. \label{fig:viable-query-percentage-time-budgets}}
\end{figure*}

\begin{figure*}[htb]
    \centering
    \begin{subfigure}[t]{0.33\linewidth}
        \includegraphics[width=\linewidth]{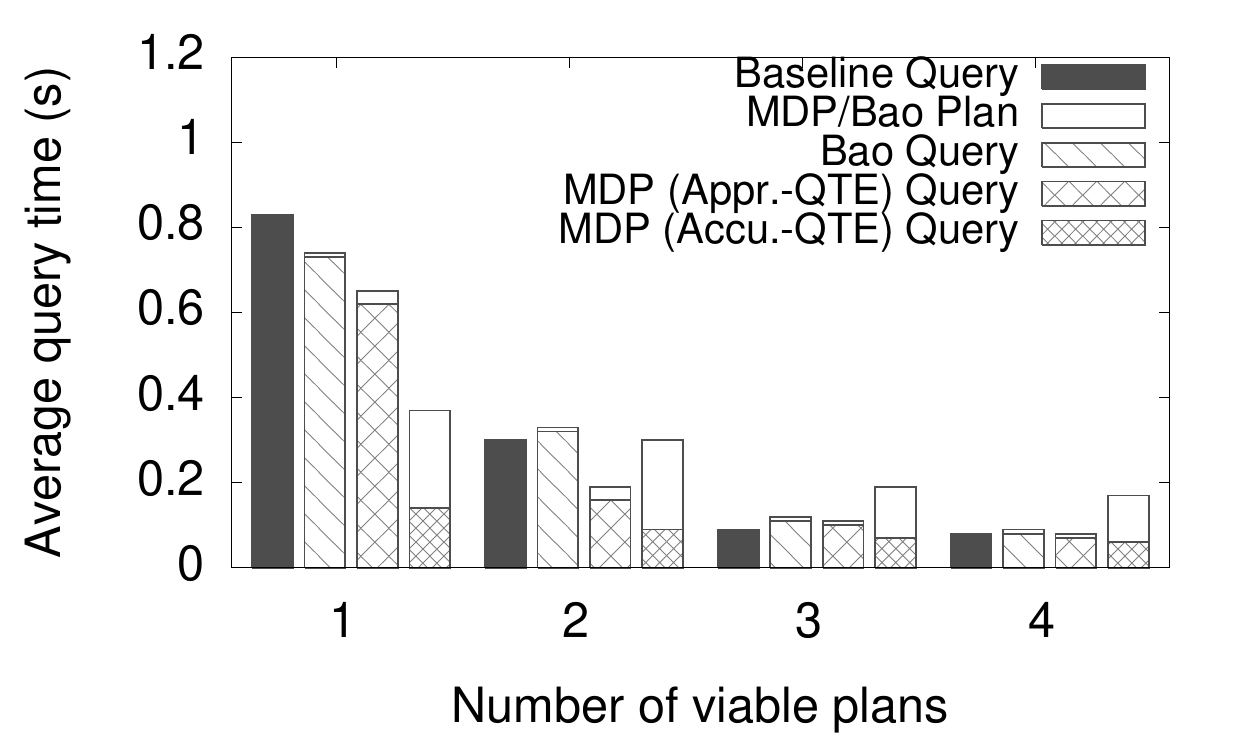}
        \caption{Time budget $\tau = 0.25$ second.}
    \end{subfigure}
    \hfill
    \begin{subfigure}[t]{0.33\linewidth}
        \includegraphics[width=\linewidth]{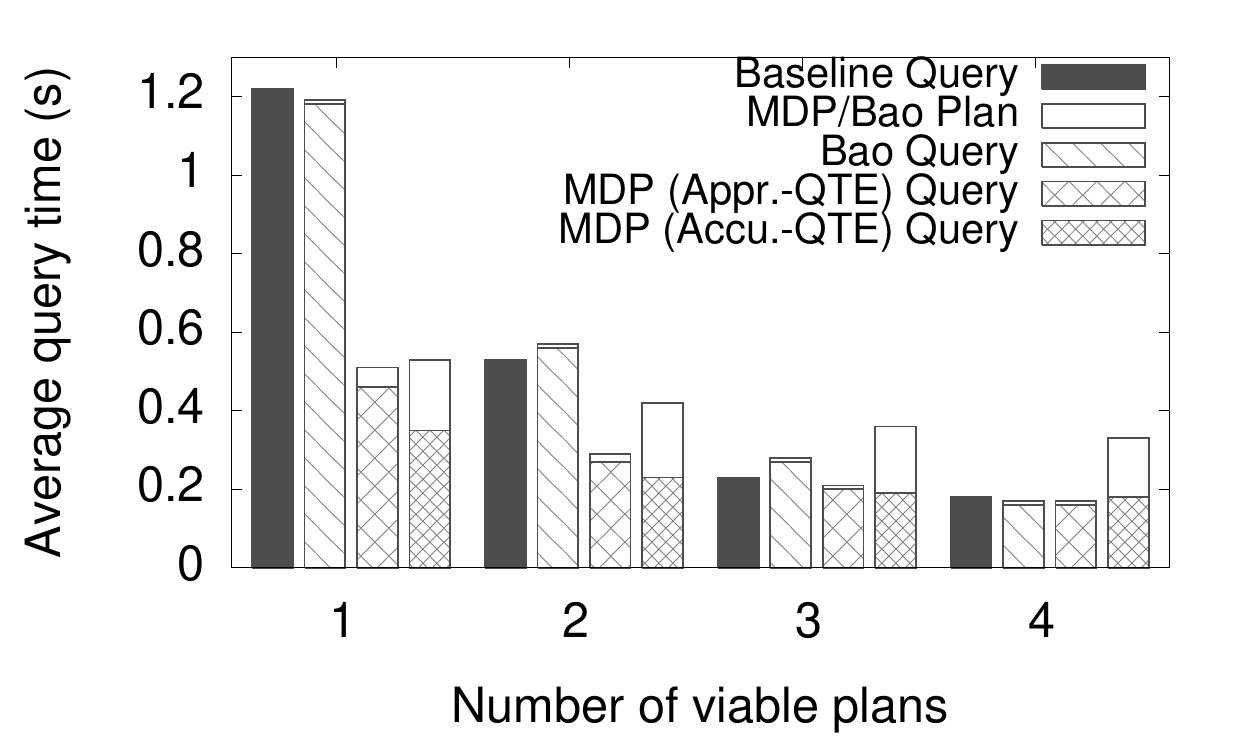}
        \caption{Time budget $\tau = 0.75$ second.}
    \end{subfigure}
    \hfill
    \begin{subfigure}[t]{0.33\linewidth}
        \includegraphics[width=\linewidth]{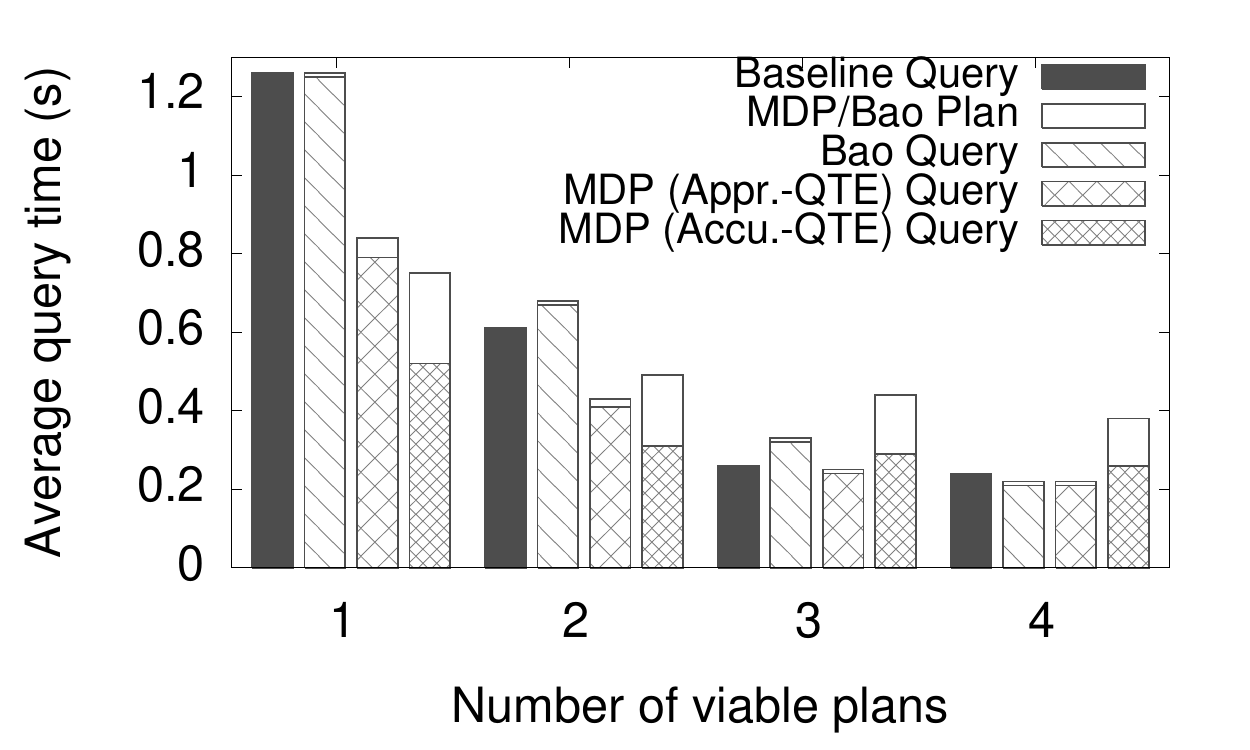}
        \caption{Time budget $\tau = 1.0$ second.}
    \end{subfigure}
    \caption{Average query response time using PostgreSQL on {\sf Twitter} dataset for different time budgets. \label{fig:avg-query-time-time-budgets}}
\end{figure*}

Figure~\ref{fig:avg-query-time-dimensions} shows the results of average query-response time.  Consistent with the VQP results, all MDP-based approaches outperformed both Bao and the baseline approach for hard queries with fewer viable plans.  For example, the MDP-based approach using an Approximate-QTE reduced the average query-response time from the baseline's $1.13$ seconds and Bao's $1.05$ seconds to $0.66$ seconds for workloads with $16$ rewrite options and one or two viable plans.  On the other hand, for queries with more viable plans, the MDP approach using an approximate-QTE had a comparable query-response time to Bao and the baseline approach.  However, the MDP approach using an accurate-QTE always had a lower average query time than all other approaches. That is, the former generated a more efficient plan for most queries than the latter, but the total response time might be longer due to the extra planning time.

\subsection{Effect of Time Budget}

We evaluated the effect of time budget on the performance of different approaches. We varied the time budget to 0.25, 0.75, and 1.0 seconds on the {\sf Twitter} dataset. For each time budget, we used different query workloads with different numbers of viable plans, and collected both the VQP and AQRT metrics for each approach. 

Figure~\ref{fig:viable-query-percentage-time-budgets} shows the results of VQP for different time budgets. The MDP-based approaches outperformed both Bao and the baseline approach significantly for all the budgets. When the time budget was 0.25 second, MDP (Approximate-QTE) outperformed MDP (Accurate-QTE) significantly because the latter was too expensive for planning. On the other hand, when the time budget was 1.0 second, MDP (Accurate-QTE) outperformed MDP (Approximate-QTE) since the agent could afford the expensive estimation cost for more accurate estimations to find better-rewritten queries. These results show that the MDP model is adaptive to QTE's with different costs and accuracies for different time budgets.

Figure~\ref{fig:avg-query-time-time-budgets} shows the results of the average response time for different time budgets. For queries with one or two viable plans, all MDP-based approaches outperformed both Bao and the baseline approach. For queries with three or four viable plans, MDP (Approximate-QTE) had a performance comparable to Bao and the baseline approach.

\subsection{Performance on Join Queries}
\label{subsec:join-experiment}

To evaluate the performance of \sysName on queries with joins, we set up a workload of queries joining the \texttt{tweets} and \texttt{users} tables with filtering conditions on three attributes.  We compared the performance of the MDP-based approaches with Bao and the baseline approach.  For the MDP-based approaches and Bao, we considered $7$ different ways of using or not using indexes on the three attributes and $3$ different join methods (i.e., nest-loop-join, hash-join, and merge-join) between the two tables. Thus we had $21$ query-hint sets in total as the rewrite options.  We varied the evaluation workloads with different groups of viable-plan numbers and collected each approach's VQP and AQRT metrics.

\begin{figure}[htb]
  \centering
  \begin{subfigure}[t]{0.65\linewidth}
    \includegraphics[width=\linewidth]{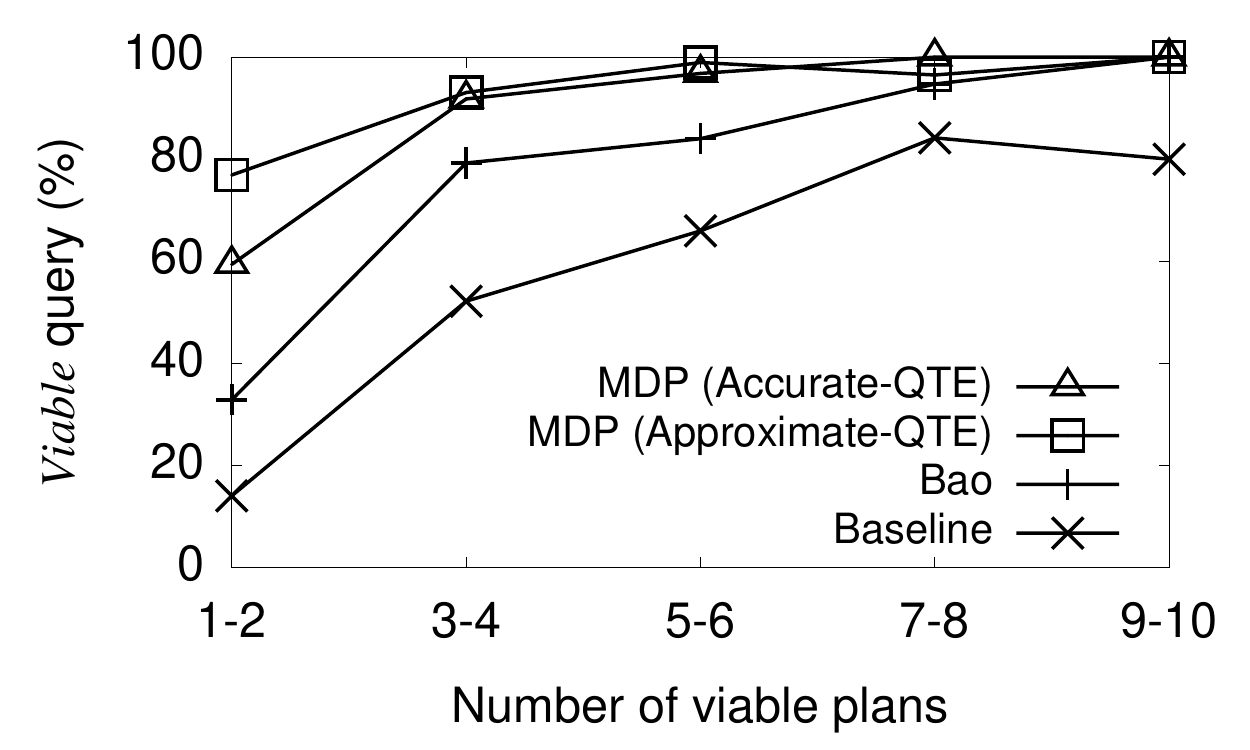}
    \caption{Viable query percentage.}
  \end{subfigure}
  \begin{subfigure}[t]{0.65\linewidth}
    \includegraphics[width=\linewidth]{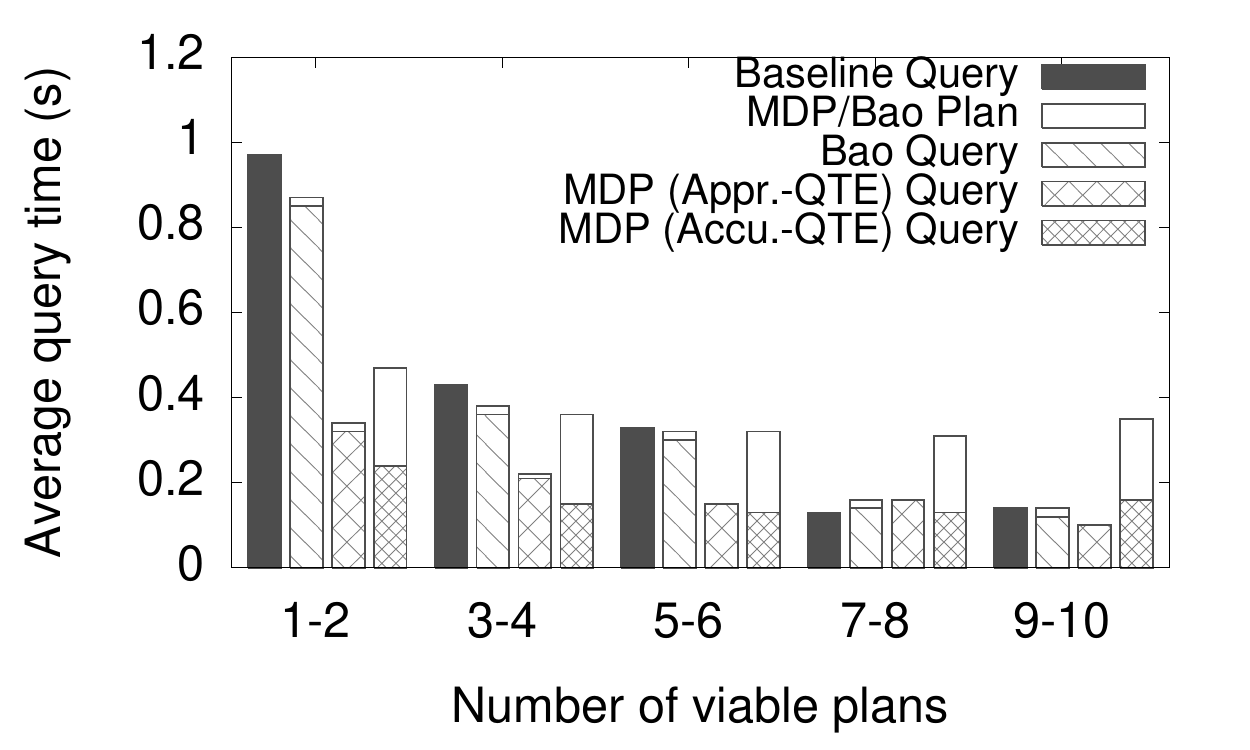}
    \caption{Average query response time.}
  \end{subfigure}
  \caption{Performance on join queries using PostgreSQL on {\sf Twitter} dataset. (Time budget $\tau=500 ms$.) \label{fig:join-queries}}
\end{figure}

Figure~\ref{fig:join-queries}(a) shows that for all workloads of different viable-plan numbers, the MDP-based approaches outperformed Bao. For the queries with only one or two viable plans, MDP (Approximate-QTE) generated more than twice as many viable plans than Bao. Figure~\ref{fig:join-queries}(b) shows that MDP (Approximate-QTE) outperformed Bao in all cases. For queries with one or two viable plans, the MDP-based approach reduced the average query response time from Bao's $0.87$ second to $0.34$ second.  Bao uses the plan tree and operators' cost estimations from the physical plan generated by the underlying PostgreSQL database. Bao's QTE also suffered from estimation errors where the database had wrong cardinality estimations.  \sysName mitigated this problem by applying a more expensive sampling-based approximate-QTE. By judiciously choosing rewritten queries to run the expensive QTE, \sysName generated more efficient queries than Bao.

\boldstart{Maliva vs Bao:}
In general, the MDP-based approaches performed much better than Bao for workloads of queries with textual and spatial filtering conditions (e.g., \sf{Twitter} and \sf{NYC}).  The reason was that Bao's QTE relied on the plan tree and operators' cost estimations from the physical plan generated by PostgreSQL.  It also suffered from the large estimation errors by PostgreSQL for textual and spatial filtering conditions.  \sysName mitigated this problem by applying a sampling-based approximate-QTE that had better estimations for textual and spatial conditions.  By judiciously choosing rewritten queries to run the expensive QTE, \sysName generated more efficient queries than Bao.  We also evaluated the approach that used Bao with the expensive approximate QTE (labeled ``Naive (Approximate-QTE)'' in Figure~\ref{fig:viable-query-percentage-dimensions}(a)), and \sysName still outperformed Bao because of Bao's long time of enumerating all possible rewritten queries.

\begin{figure}[htbp]
  \centering
  \begin{subfigure}[t]{0.65\linewidth}
    \includegraphics[width=\linewidth]{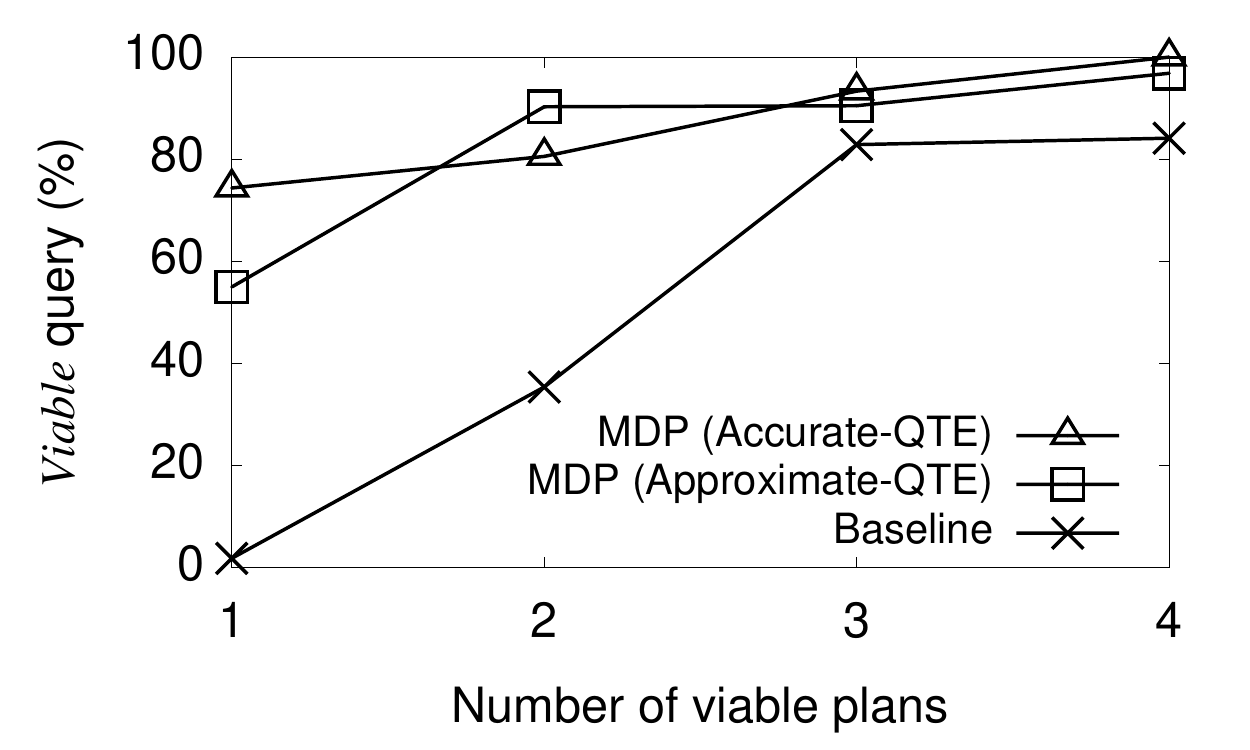}
    \caption{Unseen queries ($\tau=500ms$).}
  \end{subfigure}
  \hfill
  \begin{subfigure}[t]{0.65\linewidth}
    \includegraphics[width=\linewidth]{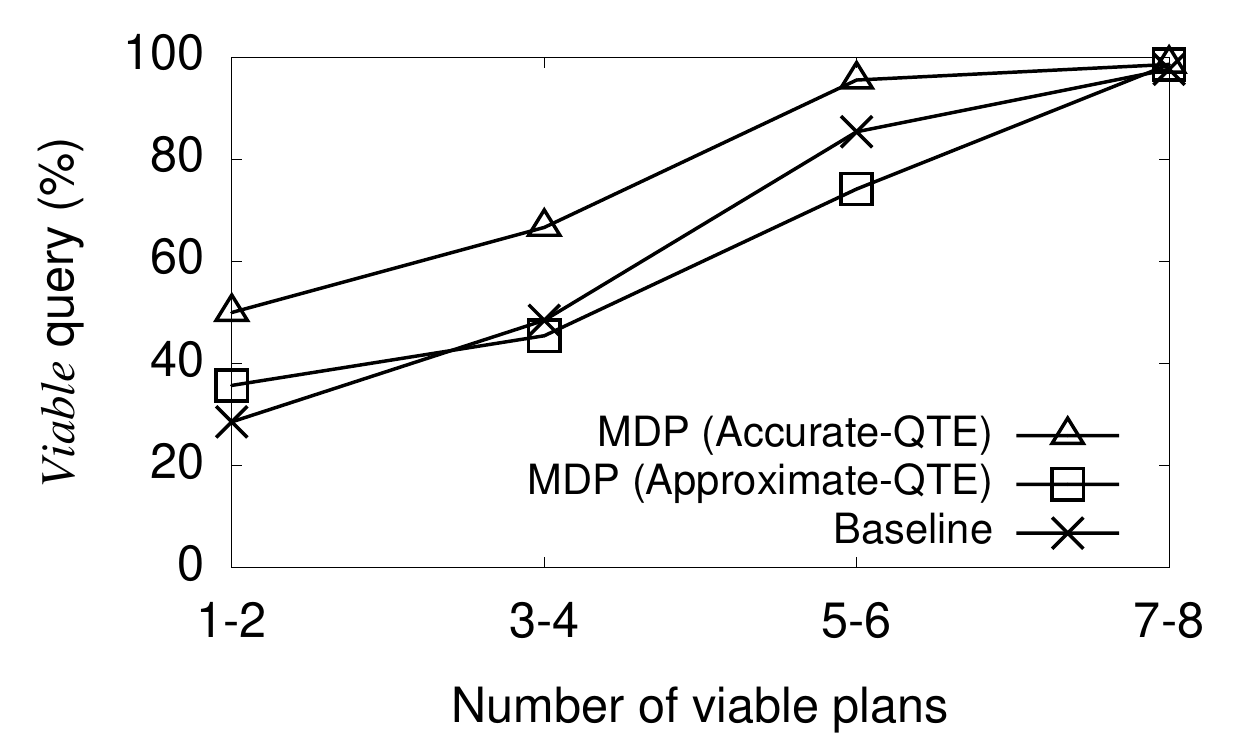}
    \caption{Commercial DB ($\tau=250ms$).}
  \end{subfigure}
  \caption{Generalization to (a) handle unseen queries and (b) use a commercial database.\label{fig:unseen-queries-and-other-databases}}
\end{figure}

\subsection{Unseen Queries and Other Databases}
\label{subsec:unseen-queries-and-other-databases}

To evaluate how well \sysName can be generalized to handle unseen queries, we did experiments on the {\sf Twitter} dataset to train and test the MDP model using two workloads with different query shapes.  The training queries were on a single {\tt tweets} table with three filtering conditions.  In comparison, the testing queries joined the {\tt tweets} table and the {\tt users} table on {\tt user\_id} with three filtering conditions on the former table.  As shown in Figure~\ref{fig:unseen-queries-and-other-databases}(a), the MDP-based approaches outperformed the baseline significantly on the workload with unseen queries.  For example, for queries with a single viable plan, the MDP (Approximate-QTE) approach increased the VQP from the baseline's $2\%$ to $55\%$, and the MDP (Accurate-QTE) approach further increased it to $74\%$.

We also did experiments on the {\sf Twitter} dataset using a commercial database.  We used a smaller table with $10$ million records and thus a smaller time budget ($250ms$).  The result is shown in Figure\ref{fig:unseen-queries-and-other-databases}(b).  Due to the commercial database's complex behaviors, the approximate QTE had a much lower accuracy (two orders of magnitude) than it had on PostgreSQL.  The reason was the approximate QTE only considered predicates' selectivities for estimation, but more factors in the commercial database affected the query time, such as buffering and dynamic execution plan change.  However, MDP (Approximate QTE) still had comparable performance (VQP) to the baseline.   With a more accurate yet more expensive QTE, MDP (Accurate-QTE) outperformed the baseline for all the queries.  For example, for queries with one or two viable plans, the baseline had a VQP of $23\%$, MDP (approximate-QTE) had a VQP of $36\%$, and MDP (Accurate-QTE) increased the VQP to $50\%$.

\begin{figure*}[htb]
    \centering
    \begin{subfigure}[t]{0.33\linewidth}
        \includegraphics[width=\linewidth]{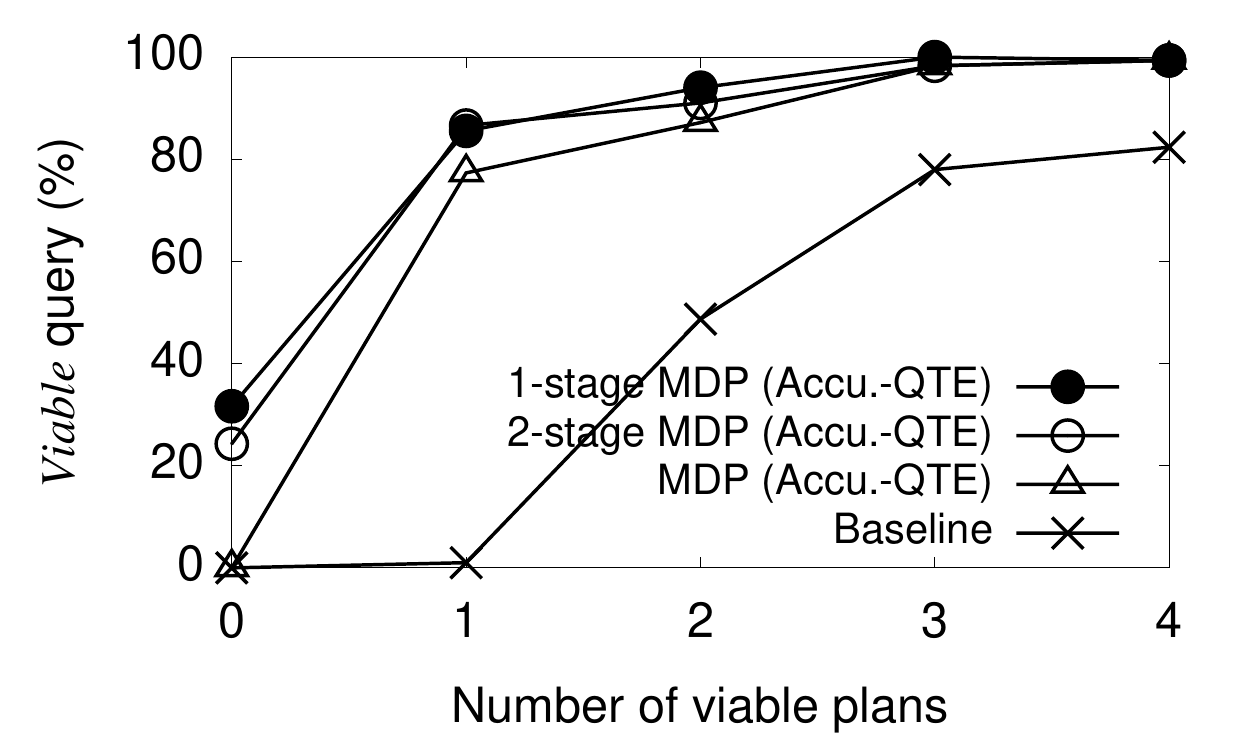}
        \caption{Viable query percentage.}
    \end{subfigure}
    \hfill
    \begin{subfigure}[t]{0.33\linewidth}
        \includegraphics[width=\linewidth]{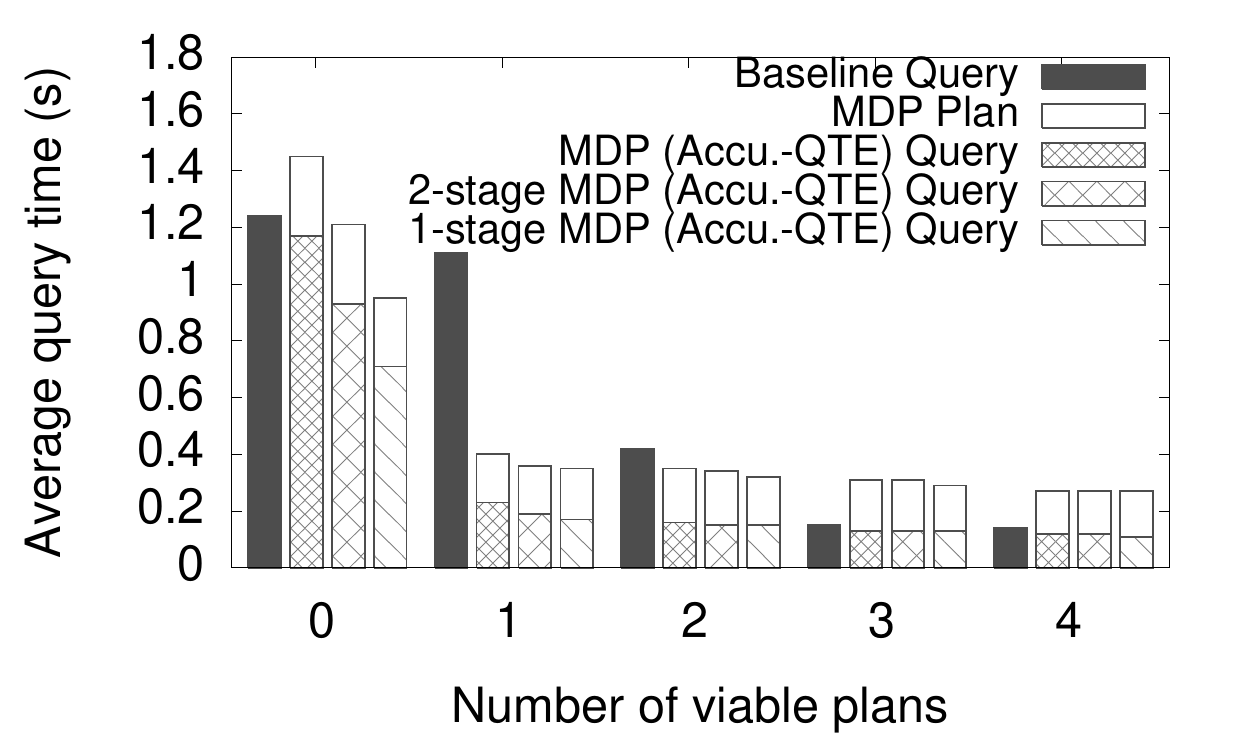}
        \caption{Average query response time.}
    \end{subfigure}
    \hfill
    \begin{subfigure}[t]{0.33\linewidth}
        \includegraphics[width=\linewidth]{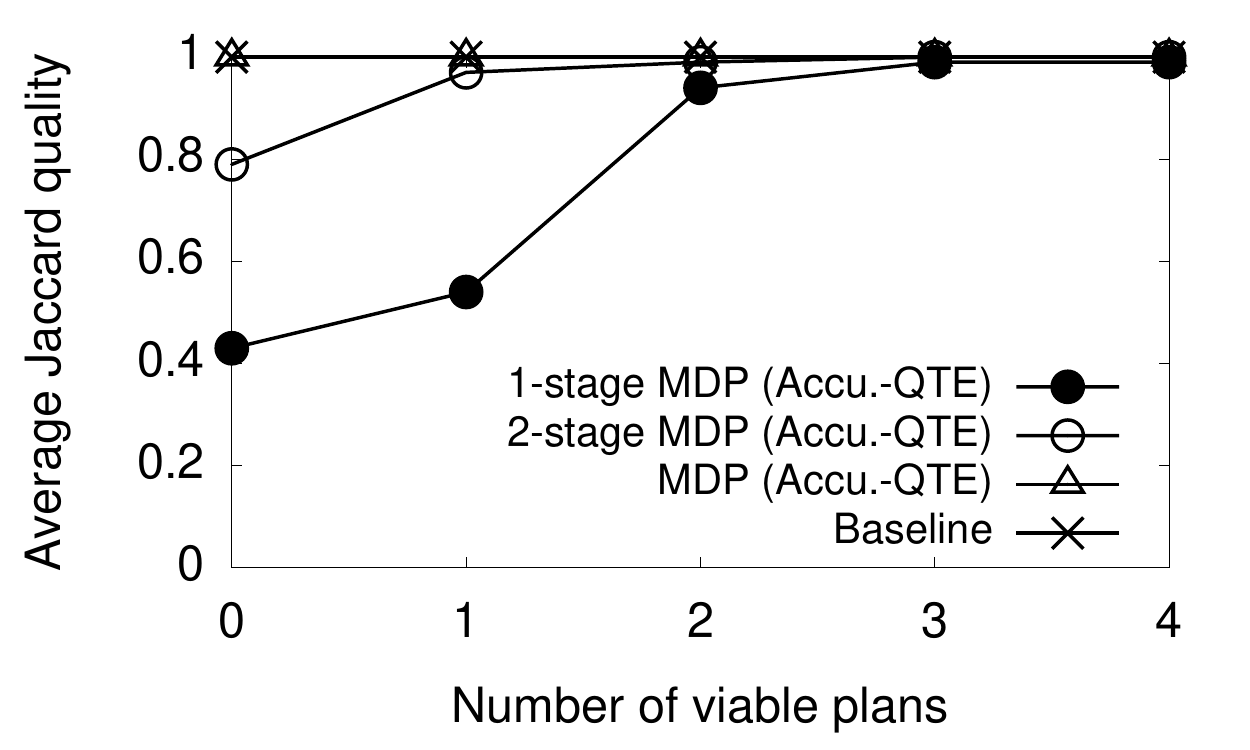}
        \caption{Average Jaccard-based quality.}
    \end{subfigure}
    \caption{Performance of quality-aware rewriting approaches using PostgreSQL on {\sf Twitter} dataset. (Time budget $\tau=500 ms$.) \label{fig:quality-aware}}
\end{figure*}

\begin{figure*}[htb]
    \centering
    \begin{subfigure}[t]{0.33\linewidth}
        \includegraphics[width=\linewidth]{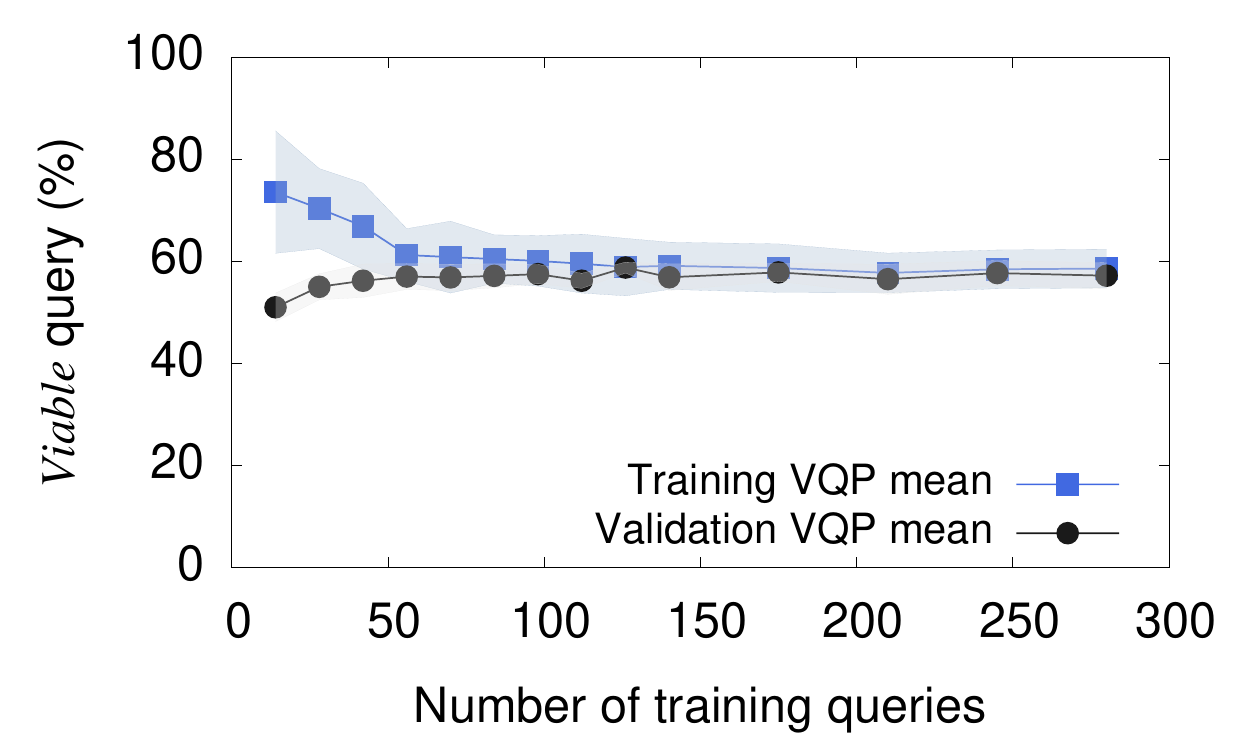}
        \caption{Learning curve for $8$ rewrite options.}
    \end{subfigure}
    \hfill
    \begin{subfigure}[t]{0.33\linewidth}
        \includegraphics[width=\linewidth]{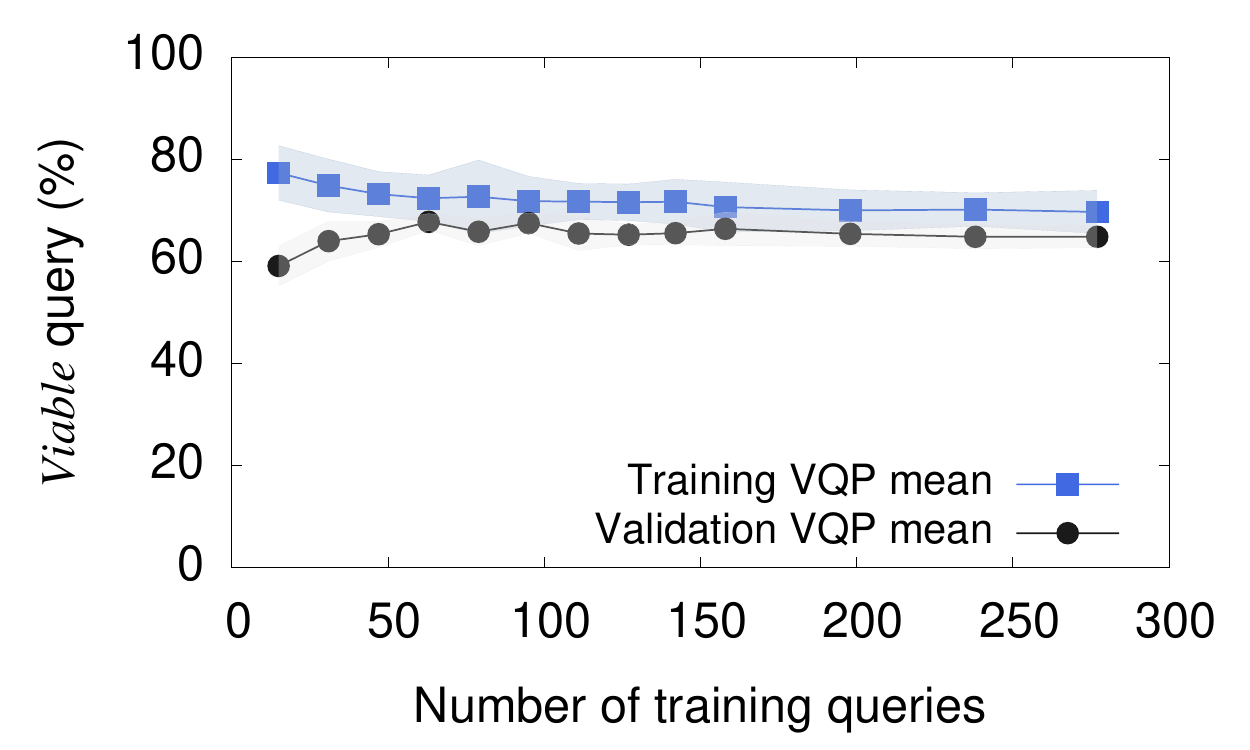}
        \caption{Learning curve for $32$ rewrite options.}
    \end{subfigure}
    \hfill
    \begin{subfigure}[t]{0.33\linewidth}
        \includegraphics[width=\linewidth]{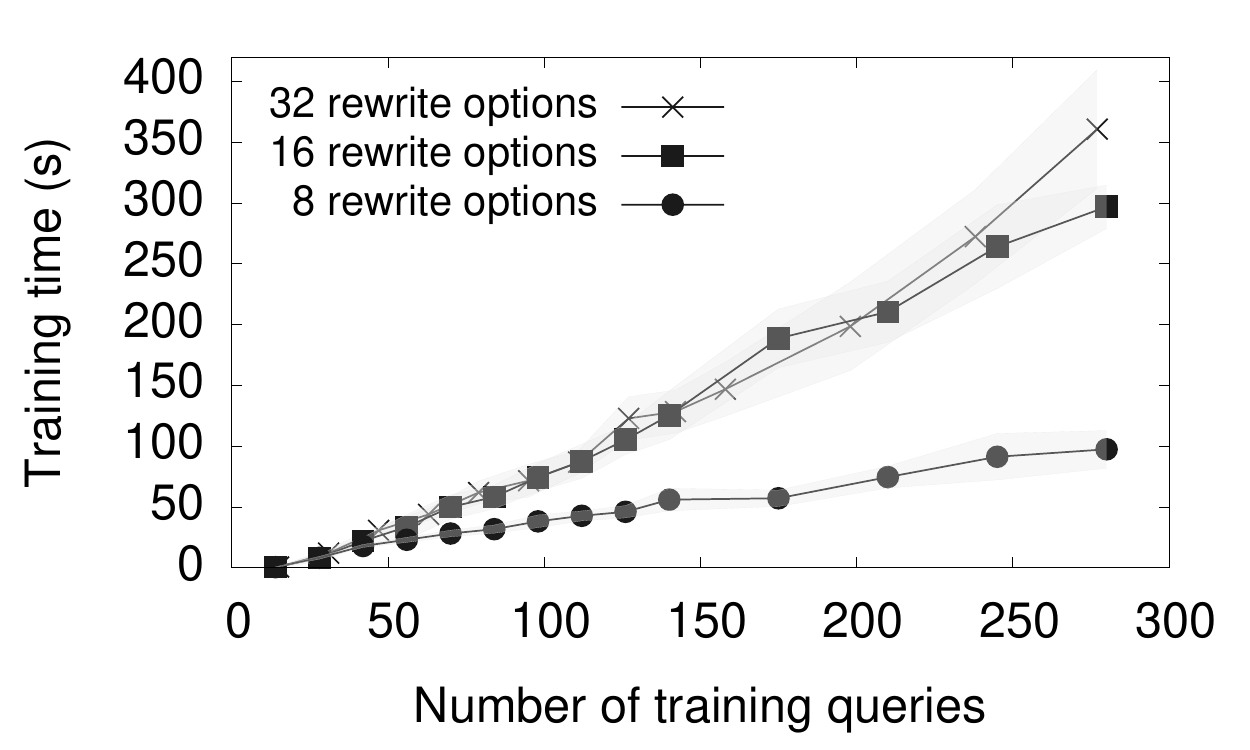}
        \caption{Training time curves for $8$, $16$, and $32$ rewrite options.}
    \end{subfigure}
    \caption{Learning curves and training time curves (varying number of training queries) for $8$ and $32$ rewrite options using PostgreSQL on {\sf Twitter} dataset (shaded area is plotted with (mean $+$ standard deviation) as the upper bound and (mean $-$ standard deviation) as the lower bound). \label{fig:learning-curves}}
\end{figure*}

\subsection{Performance of Quality-Aware Rewriting}

We evaluated the performance of the two quality-aware query rewriting approaches (i.e., one-stage and two-stage) described in Section~\ref{sec:generalization}.  We used them on the same {\sf Twitter} dataset and workload as in Section~\ref{subsec:non-approximate}. We compared them with the baseline approach and the MDP approach without considering approximation rules.  For the quality-aware rewriting approaches, we considered five approximation rules (i.e., adding a \texttt{LIMIT} clause with $0.032\%$, $0.16\%$, $0.8\%$, $4\%$, and $20\%$ of the estimated cardinality of the query) in addition to the eight query-hint sets (i.e., using or not using indexes on the three filtering attributes) considered in Section~\ref{subsec:non-approximate}.  All MDP-based approaches used an accurate-QTE.  Besides the AQP and AQRT metrics, we collected a new metric called {\em Jaccard-based Quality}, which computed the Jaccard similarity between the visualization result of a rewritten query and that of the original query. 

Figure~\ref{fig:quality-aware}(a) shows the VQP of these approaches.  For the group of queries without any viable plan, the MDP-based approach without considering approximation rules and the baseline approach had a zero VQP.  By generating approximate rewritten queries, the two-stage MDP-based approach increased the VQP to $24\%$, and the one-stage MDP approach further increased the VQP to $31\%$.  There were $518$ queries in the $0$-viable-plan workload (Table~\ref{table:workloads}), and the one-stage MDP approach generated more than $35$ viable queries than the two-stage approach.  As shown in Figure~\ref{fig:quality-aware}(b), the one-stage MDP approach also reduced the average query response time of the $0$-viable-plan queries from the two-stage approach's $1.21$ seconds to $0.95$ seconds.  In terms of efficiency, the one-stage MDP approach outperformed the two-stage approach in all cases.  Figure~\ref{fig:quality-aware}(c) shows the average Jaccard-based quality of the rewritten queries generated by different approaches.  Both the baseline and the MDP-based approach without considering approximation rules had no quality loss.  The two-stage MDP approach had a significant advantage over the one-stage approach in terms of quality. For example, The former increased the quality of the $0$-viable-plan queries from the one-stage approach's $0.43$ to $0.79$.

\subsection{Training Performance}

We evaluated the training performance for workloads with eight, sixteen, and thirty-two rewrite options on the {\sf Twitter} dataset. For each workload, we divided a set of about $1,400$ queries into a training set and a validation set. Then we varied the number of training queries and randomly sampled those from the training set without replacement. We then used the sampled queries to train an MDP agent and tested its performance on both the training queries and the validation queries. We repeated the step ten times for each number of training queries and collected the mean and standard deviation of the VQPs. We conducted the experiments on the MDP-based approach using the Accurate-QTE. The unit cost for the three workloads was $100ms$, $60ms$, and $50ms$, respectively, and the time budget was $0.5$ seconds.

Figure~\ref{fig:learning-curves} (a) and Figure~\ref{fig:learning-curves}(b) show the trend when we varied the number of training queries. For the workload with eight rewrite options, the VQP on the validation set got close to the VQP on the training set for about $50$ training queries, i.e., adding more training queries did not further improve the performance. For workloads with sixteen or thirty-two rewrite options, the validation VQP got close to the training VQP at about $80$ and $150$ training queries, respectively. For different sizes of the training set, we also collected the mean and standard deviation of the training time. Figure~\ref{fig:learning-curves}(c) shows the training time of different numbers of rewrite options on the training sizes. For the same number of training queries, more rewrite options resulted in a larger q-network, which took more time to update the weights. For the workload with thirty-two rewrite options, it took about $150$ seconds to train an MDP agent on $150$ training queries.

\boldstart{Remarks:}  The experiments show that \sysName outperformed the baseline approach in terms of both the number of viable queries and average query response time. \sysName generated up to $70\times$ more viable queries than the baseline approach.  The advantages of \sysName were shown in both the real datasets and synthetic dataset, for different numbers of rewriting options, different time budgets and different query workloads. Its offline training overhead was relatively small. By considering approximation rules considered, \sysName generated even more viable queries. The comparison with Bao shows the advantage of \sysName due to the fact these two techniques were designed with different settings and optimization goals.


\section{Conclusions}
In this paper we studied how to rewrite database queries to improve execution performance in middleware-based visualization systems.  We explored two optimization options of adding hints and doing approximation. We developed a novel solution called \sysName, which adopts a Markov Decision Process (MDP) model to rewrite a visualization request under a tight time constraint.  We gave a full specification of the solution, including how to construct an MDP model, how to train an agent, and how to use approximating rewriting options.  Our experiments on both real and synthetic datasets showed that  \sysName performed significantly better than the baseline without no-rewriting options in terms of both the probability of serving a visualization request within a time budget and query execution time.


\bibliographystyle{ACM-Reference-Format}
\bibliography{references, localrefs}


\begin{thebibliography}{73}


\ifx \showCODEN    \undefined \def \showCODEN     #1{\unskip}     \fi
\ifx \showDOI      \undefined \def \showDOI       #1{#1}\fi
\ifx \showISBNx    \undefined \def \showISBNx     #1{\unskip}     \fi
\ifx \showISBNxiii \undefined \def \showISBNxiii  #1{\unskip}     \fi
\ifx \showISSN     \undefined \def \showISSN      #1{\unskip}     \fi
\ifx \showLCCN     \undefined \def \showLCCN      #1{\unskip}     \fi
\ifx \shownote     \undefined \def \shownote      #1{#1}          \fi
\ifx \showarticletitle \undefined \def \showarticletitle #1{#1}   \fi
\ifx \showURL      \undefined \def \showURL       {\relax}        \fi
\providecommand\bibfield[2]{#2}
\providecommand\bibinfo[2]{#2}
\providecommand\natexlab[1]{#1}
\providecommand\showeprint[2][]{arXiv:#2}

\bibitem[\protect\citeauthoryear{ands Ling~Xiao, Gerth, and Hanrahan}{ands
  Ling~Xiao et~al\mbox{.}}{2008}]%
        {conf/ieeevast/ChanXGH08}
\bibfield{author}{\bibinfo{person}{Sye{-}Min~Chan ands Ling~Xiao},
  \bibinfo{person}{John Gerth}, {and} \bibinfo{person}{Pat Hanrahan}.}
  \bibinfo{year}{2008}\natexlab{}.
\newblock \showarticletitle{Maintaining interactivity while exploring massive
  time series}. In \bibinfo{booktitle}{\emph{Proceedings of the {IEEE}
  Symposium on Visual Analytics Science and Technology, {IEEE} {VAST} 2008,
  Columbus, Ohio, USA, 19-24 October 2008}}.
\newblock
\urldef\tempurl%
\url{https://doi.org/10.1109/VAST.2008.4677357}
\showDOI{\tempurl}


\bibitem[\protect\citeauthoryear{{{AsterixDB Query Hints}}}{{{AsterixDB Query
  Hints}}}{[n.d.]}]%
        {asterixdb:query-hints}
{{AsterixDB Query Hints}}.
\newblock
\newblock
\newblock
\shownote{\url{http://asterixdb.apache.org/docs/0.9.6/sqlpp/manual.html##Query_hints}.}


\bibitem[\protect\citeauthoryear{{{BaoForPostgreSQL}}}{{{BaoForPostgreSQL}}}{[n.d.]}]%
        {BaoForPostgreSQL}
{{BaoForPostgreSQL}}.
\newblock
\newblock
\newblock
\shownote{\url{https://github.com/learnedsystems/baoforpostgresql}.}


\bibitem[\protect\citeauthoryear{Battle, Chang, and Stonebraker}{Battle
  et~al\mbox{.}}{2016}]%
        {conf/sigmod/BattleCS16}
\bibfield{author}{\bibinfo{person}{Leilani Battle}, \bibinfo{person}{Remco
  Chang}, {and} \bibinfo{person}{Michael Stonebraker}.}
  \bibinfo{year}{2016}\natexlab{}.
\newblock \showarticletitle{Dynamic Prefetching of Data Tiles for Interactive
  Visualization}. In \bibinfo{booktitle}{\emph{Proceedings of the 2016
  International Conference on Management of Data, {SIGMOD} Conference 2016, San
  Francisco, CA, USA, June 26 - July 01, 2016}},
  \bibfield{editor}{\bibinfo{person}{Fatma {\"{O}}zcan},
  \bibinfo{person}{Georgia Koutrika}, {and} \bibinfo{person}{Sam Madden}}
  (Eds.).
\newblock
\urldef\tempurl%
\url{https://doi.org/10.1145/2882903.2882919}
\showDOI{\tempurl}


\bibitem[\protect\citeauthoryear{Battle, Crouser, Nakeshimana, Montoly, Chang,
  and Stonebraker}{Battle et~al\mbox{.}}{2020}]%
        {journals/tvcg/BattleCNMCS20}
\bibfield{author}{\bibinfo{person}{Leilani Battle}, \bibinfo{person}{R.~Jordan
  Crouser}, \bibinfo{person}{Audace Nakeshimana}, \bibinfo{person}{Ananda
  Montoly}, \bibinfo{person}{Remco Chang}, {and} \bibinfo{person}{Michael
  Stonebraker}.} \bibinfo{year}{2020}\natexlab{}.
\newblock \showarticletitle{The Role of Latency and Task Complexity in
  Predicting Visual Search Behavior}.
\newblock \bibinfo{journal}{\emph{{IEEE} Trans. Vis. Comput. Graph.}}
  \bibinfo{volume}{26}, \bibinfo{number}{1} (\bibinfo{year}{2020}).
\newblock
\urldef\tempurl%
\url{https://doi.org/10.1109/TVCG.2019.2934556}
\showDOI{\tempurl}


\bibitem[\protect\citeauthoryear{Budiu, Gopalan, Suresh, Wieder, Kruiger, and
  Aguilera}{Budiu et~al\mbox{.}}{2019}]%
        {journals/pvldb/BudiuGSWKA19}
\bibfield{author}{\bibinfo{person}{Mihai Budiu}, \bibinfo{person}{Parikshit
  Gopalan}, \bibinfo{person}{Lalith Suresh}, \bibinfo{person}{Udi Wieder},
  \bibinfo{person}{Han Kruiger}, {and} \bibinfo{person}{Marcos~K. Aguilera}.}
  \bibinfo{year}{2019}\natexlab{}.
\newblock \showarticletitle{Hillview: {A} trillion-cell spreadsheet for big
  data}.
\newblock \bibinfo{journal}{\emph{{PVLDB}}} \bibinfo{volume}{12},
  \bibinfo{number}{11} (\bibinfo{year}{2019}).
\newblock
\urldef\tempurl%
\url{http://www.vldb.org/pvldb/vol12/p1442-budiu.pdf}
\showURL{%
\tempurl}


\bibitem[\protect\citeauthoryear{Cheng, Schretlen, Kronenfeld, Bozowsky, and
  Wright}{Cheng et~al\mbox{.}}{2013}]%
        {conf/bigdataconf/ChengSKBW13}
\bibfield{author}{\bibinfo{person}{Daniel Cheng}, \bibinfo{person}{Peter
  Schretlen}, \bibinfo{person}{Nathan Kronenfeld}, \bibinfo{person}{Neil
  Bozowsky}, {and} \bibinfo{person}{William Wright}.}
  \bibinfo{year}{2013}\natexlab{}.
\newblock \showarticletitle{Tile based visual analytics for Twitter big data
  exploratory analysis}. In \bibinfo{booktitle}{\emph{Proceedings of the 2013
  {IEEE} International Conference on Big Data, 6-9 October 2013, Santa Clara,
  CA, {USA}}}.
\newblock
\urldef\tempurl%
\url{https://doi.org/10.1109/BigData.2013.6691787}
\showDOI{\tempurl}


\bibitem[\protect\citeauthoryear{Crotty, Galakatos, Zgraggen, Binnig, and
  Kraska}{Crotty et~al\mbox{.}}{2015}]%
        {journals/pvldb/CrottyGZBK15}
\bibfield{author}{\bibinfo{person}{Andrew Crotty}, \bibinfo{person}{Alex
  Galakatos}, \bibinfo{person}{Emanuel Zgraggen}, \bibinfo{person}{Carsten
  Binnig}, {and} \bibinfo{person}{Tim Kraska}.}
  \bibinfo{year}{2015}\natexlab{}.
\newblock \showarticletitle{Vizdom: Interactive Analytics through Pen and
  Touch}.
\newblock \bibinfo{journal}{\emph{{PVLDB}}} \bibinfo{volume}{8},
  \bibinfo{number}{12} (\bibinfo{year}{2015}).
\newblock
\urldef\tempurl%
\url{https://doi.org/10.14778/2824032.2824127}
\showDOI{\tempurl}


\bibitem[\protect\citeauthoryear{Crotty, Galakatos, Zgraggen, Binnig, and
  Kraska}{Crotty et~al\mbox{.}}{2016}]%
        {conf/sigmod/CrottyGZBK16}
\bibfield{author}{\bibinfo{person}{Andrew Crotty}, \bibinfo{person}{Alex
  Galakatos}, \bibinfo{person}{Emanuel Zgraggen}, \bibinfo{person}{Carsten
  Binnig}, {and} \bibinfo{person}{Tim Kraska}.}
  \bibinfo{year}{2016}\natexlab{}.
\newblock \showarticletitle{The case for interactive data exploration
  accelerators (IDEAs)}. In \bibinfo{booktitle}{\emph{Proceedings of the
  Workshop on Human-In-the-Loop Data Analytics, HILDA@SIGMOD 2016, San
  Francisco, CA, USA, June 26 - July 01, 2016}},
  \bibfield{editor}{\bibinfo{person}{Carsten Binnig}, \bibinfo{person}{Alan
  Fekete}, {and} \bibinfo{person}{Arnab Nandi}} (Eds.).
\newblock
\urldef\tempurl%
\url{https://doi.org/10.1145/2939502.2939513}
\showDOI{\tempurl}


\bibitem[\protect\citeauthoryear{de~Lara~Pahins, Stephens, Scheidegger, and
  Comba}{de~Lara~Pahins et~al\mbox{.}}{2017}]%
        {journals/tvcg/PahinsSSC17}
\bibfield{author}{\bibinfo{person}{Cicero~Augusto de Lara~Pahins},
  \bibinfo{person}{Sean~A. Stephens}, \bibinfo{person}{Carlos Scheidegger},
  {and} \bibinfo{person}{Jo{\~{a}}o Luiz~Dihl Comba}.}
  \bibinfo{year}{2017}\natexlab{}.
\newblock \showarticletitle{Hashedcubes: Simple, Low Memory, Real-Time Visual
  Exploration of Big Data}.
\newblock \bibinfo{journal}{\emph{{IEEE} Trans. Vis. Comput. Graph.}}
  \bibinfo{volume}{23}, \bibinfo{number}{1} (\bibinfo{year}{2017}).
\newblock
\urldef\tempurl%
\url{https://doi.org/10.1109/TVCG.2016.2598624}
\showDOI{\tempurl}


\bibitem[\protect\citeauthoryear{Ding, Huang, Chaudhuri, Chakrabarti, and
  Wang}{Ding et~al\mbox{.}}{2016}]%
        {conf/sigmod/DingHCC016}
\bibfield{author}{\bibinfo{person}{Bolin Ding}, \bibinfo{person}{Silu Huang},
  \bibinfo{person}{Surajit Chaudhuri}, \bibinfo{person}{Kaushik Chakrabarti},
  {and} \bibinfo{person}{Chi Wang}.} \bibinfo{year}{2016}\natexlab{}.
\newblock \showarticletitle{Sample + Seek: Approximating Aggregates with
  Distribution Precision Guarantee}. In \bibinfo{booktitle}{\emph{Proceedings
  of the 2016 International Conference on Management of Data, {SIGMOD}
  Conference 2016, San Francisco, CA, USA, June 26 - July 01, 2016}}.
\newblock
\urldef\tempurl%
\url{https://doi.org/10.1145/2882903.2915249}
\showDOI{\tempurl}


\bibitem[\protect\citeauthoryear{Eldawy, Mokbel, and Jonathan}{Eldawy
  et~al\mbox{.}}{2016}]%
        {conf/icde/EldawyMJ16}
\bibfield{author}{\bibinfo{person}{Ahmed Eldawy}, \bibinfo{person}{Mohamed~F.
  Mokbel}, {and} \bibinfo{person}{Christopher Jonathan}.}
  \bibinfo{year}{2016}\natexlab{}.
\newblock \showarticletitle{HadoopViz: {A} MapReduce framework for extensible
  visualization of big spatial data}. In \bibinfo{booktitle}{\emph{32nd {IEEE}
  International Conference on Data Engineering, {ICDE} 2016, Helsinki, Finland,
  May 16-20, 2016}}.
\newblock
\urldef\tempurl%
\url{https://doi.org/10.1109/ICDE.2016.7498274}
\showDOI{\tempurl}


\bibitem[\protect\citeauthoryear{Fisher, Popov, Drucker, and m.~c.
  schraefel}{Fisher et~al\mbox{.}}{2012}]%
        {conf/chi/FisherPDs12}
\bibfield{author}{\bibinfo{person}{Danyel Fisher}, \bibinfo{person}{Igor~O.
  Popov}, \bibinfo{person}{Steven~M. Drucker}, {and} \bibinfo{person}{m.~c.
  schraefel}.} \bibinfo{year}{2012}\natexlab{}.
\newblock \showarticletitle{Trust me, i'm partially right: incremental
  visualization lets analysts explore large datasets faster}. In
  \bibinfo{booktitle}{\emph{{CHI} Conference on Human Factors in Computing
  Systems, {CHI} '12, Austin, TX, {USA} - May 05 - 10, 2012}}.
\newblock
\urldef\tempurl%
\url{https://doi.org/10.1145/2207676.2208294}
\showDOI{\tempurl}


\bibitem[\protect\citeauthoryear{Godfrey, Gryz, and Lasek}{Godfrey
  et~al\mbox{.}}{2016}]%
        {journals/tkde/GodfreyGL16}
\bibfield{author}{\bibinfo{person}{Parke Godfrey}, \bibinfo{person}{Jarek
  Gryz}, {and} \bibinfo{person}{Piotr Lasek}.} \bibinfo{year}{2016}\natexlab{}.
\newblock \showarticletitle{Interactive Visualization of Large Data Sets}.
\newblock \bibinfo{journal}{\emph{{IEEE} Trans. Knowl. Data Eng.}}
  \bibinfo{volume}{28}, \bibinfo{number}{8} (\bibinfo{year}{2016}).
\newblock
\urldef\tempurl%
\url{https://doi.org/10.1109/TKDE.2016.2557324}
\showDOI{\tempurl}


\bibitem[\protect\citeauthoryear{Guo, Feng, Cong, and Bao}{Guo
  et~al\mbox{.}}{2018}]%
        {conf/sigmod/GuoFCB18}
\bibfield{author}{\bibinfo{person}{Tao Guo}, \bibinfo{person}{Kaiyu Feng},
  \bibinfo{person}{Gao Cong}, {and} \bibinfo{person}{Zhifeng Bao}.}
  \bibinfo{year}{2018}\natexlab{}.
\newblock \showarticletitle{Efficient Selection of Geospatial Data on Maps for
  Interactive and Visualized Exploration}. In
  \bibinfo{booktitle}{\emph{Proceedings of the 2018 International Conference on
  Management of Data, {SIGMOD} Conference 2018, Houston, TX, USA, June 10-15,
  2018}}.
\newblock
\urldef\tempurl%
\url{https://doi.org/10.1145/3183713.3183738}
\showDOI{\tempurl}


\bibitem[\protect\citeauthoryear{Hasan, Thirumuruganathan, Augustine, Koudas,
  and Das}{Hasan et~al\mbox{.}}{2020}]%
        {conf/sigmod/HasanTAK020}
\bibfield{author}{\bibinfo{person}{Shohedul Hasan}, \bibinfo{person}{Saravanan
  Thirumuruganathan}, \bibinfo{person}{Jees Augustine}, \bibinfo{person}{Nick
  Koudas}, {and} \bibinfo{person}{Gautam Das}.}
  \bibinfo{year}{2020}\natexlab{}.
\newblock \showarticletitle{Deep Learning Models for Selectivity Estimation of
  Multi-Attribute Queries}. In \bibinfo{booktitle}{\emph{Proceedings of the
  2020 International Conference on Management of Data, {SIGMOD} Conference
  2020, online conference [Portland, OR, USA], June 14-19, 2020}},
  \bibfield{editor}{\bibinfo{person}{David Maier}, \bibinfo{person}{Rachel
  Pottinger}, \bibinfo{person}{AnHai Doan}, \bibinfo{person}{Wang{-}Chiew Tan},
  \bibinfo{person}{Abdussalam Alawini}, {and} \bibinfo{person}{Hung~Q. Ngo}}
  (Eds.).
\newblock
\urldef\tempurl%
\url{https://doi.org/10.1145/3318464.3389741}
\showDOI{\tempurl}


\bibitem[\protect\citeauthoryear{{{Hint(SQL}}}{{{Hint(SQL}}}{[n.d.]}]%
        {wiki:query-hints}
{{Hint(SQL}}.
\newblock
\newblock
\newblock
\shownote{\url{https://en.wikipedia.org/wiki/Hint_(SQL)}.}


\bibitem[\protect\citeauthoryear{Hu, Bakker, Li, Kraska, and Hidalgo}{Hu
  et~al\mbox{.}}{2019}]%
        {conf/chi/HuBLKH19}
\bibfield{author}{\bibinfo{person}{Kevin~Zeng Hu}, \bibinfo{person}{Michiel~A.
  Bakker}, \bibinfo{person}{Stephen Li}, \bibinfo{person}{Tim Kraska}, {and}
  \bibinfo{person}{C{\'{e}}sar~A. Hidalgo}.} \bibinfo{year}{2019}\natexlab{}.
\newblock \showarticletitle{VizML: {A} Machine Learning Approach to
  Visualization Recommendation}. In \bibinfo{booktitle}{\emph{Proceedings of
  the 2019 {CHI} Conference on Human Factors in Computing Systems, {CHI} 2019,
  Glasgow, Scotland, UK, May 04-09, 2019}},
  \bibfield{editor}{\bibinfo{person}{Stephen~A. Brewster},
  \bibinfo{person}{Geraldine Fitzpatrick}, \bibinfo{person}{Anna~L. Cox}, {and}
  \bibinfo{person}{Vassilis Kostakos}} (Eds.).
\newblock
\urldef\tempurl%
\url{https://doi.org/10.1145/3290605.3300358}
\showDOI{\tempurl}


\bibitem[\protect\citeauthoryear{Im, Villegas, and McGuffin}{Im
  et~al\mbox{.}}{2013}]%
        {conf/bigdataconf/ImVM13}
\bibfield{author}{\bibinfo{person}{Jean{-}Francois Im},
  \bibinfo{person}{Felix~Giguere Villegas}, {and} \bibinfo{person}{Michael~J.
  McGuffin}.} \bibinfo{year}{2013}\natexlab{}.
\newblock \showarticletitle{VisReduce: Fast and responsive incremental
  information visualization of large datasets}. In
  \bibinfo{booktitle}{\emph{Proceedings of the 2013 {IEEE} International
  Conference on Big Data, 6-9 October 2013, Santa Clara, CA, {USA}}},
  \bibfield{editor}{\bibinfo{person}{Xiaohua Hu}, \bibinfo{person}{Tsau~Young
  Lin}, \bibinfo{person}{Vijay~V. Raghavan}, \bibinfo{person}{Benjamin~W. Wah},
  \bibinfo{person}{Ricardo~A. Baeza{-}Yates}, \bibinfo{person}{Geoffrey~C.
  Fox}, \bibinfo{person}{Cyrus Shahabi}, \bibinfo{person}{Matthew Smith},
  \bibinfo{person}{Qiang Yang}, \bibinfo{person}{Rayid Ghani},
  \bibinfo{person}{Wei Fan}, \bibinfo{person}{Ronny Lempel}, {and}
  \bibinfo{person}{Raghunath Nambiar}} (Eds.).
\newblock
\urldef\tempurl%
\url{https://doi.org/10.1109/BigData.2013.6691710}
\showDOI{\tempurl}


\bibitem[\protect\citeauthoryear{{Jia Yu} and Sarwat}{{Jia Yu} and
  Sarwat}{2020}]%
        {conf/icde/YuS20}
\bibfield{author}{\bibinfo{person}{{Jia Yu}} {and} \bibinfo{person}{Mohamed
  Sarwat}.} \bibinfo{year}{2020}\natexlab{}.
\newblock \showarticletitle{Accelerating Spatial Data Visualization Dashboards
  via a Materialized Sampling Approach}. In
  \bibinfo{booktitle}{\emph{Proceedings of the International Conference on Data
  Engineering, ICDE}}.
\newblock


\bibitem[\protect\citeauthoryear{Jiang, Rahman, and Nandi}{Jiang
  et~al\mbox{.}}{2018}]%
        {conf/sigmod/JiangR018}
\bibfield{author}{\bibinfo{person}{Lilong Jiang}, \bibinfo{person}{Protiva
  Rahman}, {and} \bibinfo{person}{Arnab Nandi}.}
  \bibinfo{year}{2018}\natexlab{}.
\newblock \showarticletitle{Evaluating Interactive Data Systems: Workloads,
  Metrics, and Guidelines}. In \bibinfo{booktitle}{\emph{Proceedings of the
  2018 International Conference on Management of Data, {SIGMOD} Conference
  2018, Houston, TX, USA, June 10-15, 2018}}.
\newblock
\urldef\tempurl%
\url{https://doi.org/10.1145/3183713.3197386}
\showDOI{\tempurl}


\bibitem[\protect\citeauthoryear{Kamat, Jayachandran, Tunga, and Nandi}{Kamat
  et~al\mbox{.}}{2014}]%
        {conf/icde/KamatJTN14}
\bibfield{author}{\bibinfo{person}{Niranjan Kamat}, \bibinfo{person}{Prasanth
  Jayachandran}, \bibinfo{person}{Karthik Tunga}, {and} \bibinfo{person}{Arnab
  Nandi}.} \bibinfo{year}{2014}\natexlab{}.
\newblock \showarticletitle{Distributed and interactive cube exploration}. In
  \bibinfo{booktitle}{\emph{{IEEE} 30th International Conference on Data
  Engineering, Chicago, {ICDE} 2014, IL, USA, March 31 - April 4, 2014}},
  \bibfield{editor}{\bibinfo{person}{Isabel~F. Cruz}, \bibinfo{person}{Elena
  Ferrari}, \bibinfo{person}{Yufei Tao}, \bibinfo{person}{Elisa Bertino}, {and}
  \bibinfo{person}{Goce Trajcevski}} (Eds.).
\newblock
\urldef\tempurl%
\url{https://doi.org/10.1109/ICDE.2014.6816674}
\showDOI{\tempurl}


\bibitem[\protect\citeauthoryear{Kraska}{Kraska}{2018}]%
        {journals/pvldb/Kraska18}
\bibfield{author}{\bibinfo{person}{Tim Kraska}.}
  \bibinfo{year}{2018}\natexlab{}.
\newblock \showarticletitle{Northstar: An Interactive Data Science System}.
\newblock \bibinfo{journal}{\emph{{PVLDB}}} \bibinfo{volume}{11},
  \bibinfo{number}{12} (\bibinfo{year}{2018}).
\newblock
\urldef\tempurl%
\url{http://www.vldb.org/pvldb/vol11/p2150-kraska.pdf}
\showURL{%
\tempurl}


\bibitem[\protect\citeauthoryear{Krishnan, Yang, Goldberg, Hellerstein, and
  Stoica}{Krishnan et~al\mbox{.}}{2018}]%
        {journals/corr/abs-1808-03196}
\bibfield{author}{\bibinfo{person}{Sanjay Krishnan}, \bibinfo{person}{Zongheng
  Yang}, \bibinfo{person}{Ken Goldberg}, \bibinfo{person}{Joseph~M.
  Hellerstein}, {and} \bibinfo{person}{Ion Stoica}.}
  \bibinfo{year}{2018}\natexlab{}.
\newblock \showarticletitle{Learning to Optimize Join Queries With Deep
  Reinforcement Learning}.
\newblock \bibinfo{journal}{\emph{CoRR}}  \bibinfo{volume}{abs/1808.03196}
  (\bibinfo{year}{2018}).
\newblock
\showeprint[arXiv]{1808.03196}
\urldef\tempurl%
\url{http://arxiv.org/abs/1808.03196}
\showURL{%
\tempurl}


\bibitem[\protect\citeauthoryear{Lee and Parameswaran}{Lee and
  Parameswaran}{2018}]%
        {journals/debu/LeeP18}
\bibfield{author}{\bibinfo{person}{Doris Jung~Lin Lee} {and}
  \bibinfo{person}{Aditya~G. Parameswaran}.} \bibinfo{year}{2018}\natexlab{}.
\newblock \showarticletitle{The Case for a Visual Discovery Assistant: {A}
  Holistic Solution for Accelerating Visual Data Exploration}.
\newblock \bibinfo{journal}{\emph{{IEEE} Data Eng. Bull.}}
  \bibinfo{volume}{41}, \bibinfo{number}{3} (\bibinfo{year}{2018}).
\newblock
\urldef\tempurl%
\url{http://sites.computer.org/debull/A18sept/p3.pdf}
\showURL{%
\tempurl}


\bibitem[\protect\citeauthoryear{Li and Li}{Li and Li}{2018}]%
        {journals/dase/LiL18}
\bibfield{author}{\bibinfo{person}{Kaiyu Li} {and} \bibinfo{person}{Guoliang
  Li}.} \bibinfo{year}{2018}\natexlab{}.
\newblock \showarticletitle{Approximate Query Processing: What is New and Where
  to Go? - {A} Survey on Approximate Query Processing}.
\newblock \bibinfo{journal}{\emph{Data Science and Engineering}}
  \bibinfo{volume}{3}, \bibinfo{number}{4} (\bibinfo{year}{2018}).
\newblock
\urldef\tempurl%
\url{https://doi.org/10.1007/s41019-018-0074-4}
\showDOI{\tempurl}


\bibitem[\protect\citeauthoryear{Lins, Klosowski, and Scheidegger}{Lins
  et~al\mbox{.}}{2013}]%
        {journals/tvcg/LinsKS13}
\bibfield{author}{\bibinfo{person}{Lauro~Didier Lins},
  \bibinfo{person}{James~T. Klosowski}, {and} \bibinfo{person}{Carlos~Eduardo
  Scheidegger}.} \bibinfo{year}{2013}\natexlab{}.
\newblock \showarticletitle{Nanocubes for Real-Time Exploration of
  Spatiotemporal Datasets}.
\newblock \bibinfo{journal}{\emph{{IEEE} Trans. Vis. Comput. Graph.}}
  \bibinfo{volume}{19}, \bibinfo{number}{12} (\bibinfo{year}{2013}).
\newblock
\urldef\tempurl%
\url{https://doi.org/10.1109/TVCG.2013.179}
\showDOI{\tempurl}


\bibitem[\protect\citeauthoryear{Liu and Heer}{Liu and Heer}{2014}]%
        {journals/tvcg/LiuH14}
\bibfield{author}{\bibinfo{person}{Zhicheng Liu} {and} \bibinfo{person}{Jeffrey
  Heer}.} \bibinfo{year}{2014}\natexlab{}.
\newblock \showarticletitle{The Effects of Interactive Latency on Exploratory
  Visual Analysis}.
\newblock \bibinfo{journal}{\emph{{IEEE} Trans. Vis. Comput. Graph.}}
  \bibinfo{volume}{20}, \bibinfo{number}{12} (\bibinfo{year}{2014}).
\newblock
\urldef\tempurl%
\url{https://doi.org/10.1109/TVCG.2014.2346452}
\showDOI{\tempurl}


\bibitem[\protect\citeauthoryear{Liu, Jiang, and Heer}{Liu
  et~al\mbox{.}}{2013}]%
        {journals/cgf/LiuJH13}
\bibfield{author}{\bibinfo{person}{Zhicheng Liu}, \bibinfo{person}{Biye Jiang},
  {and} \bibinfo{person}{Jeffrey Heer}.} \bibinfo{year}{2013}\natexlab{}.
\newblock \showarticletitle{\emph{imMens}: Real-time Visual Querying of Big
  Data}.
\newblock \bibinfo{journal}{\emph{Comput. Graph. Forum}} \bibinfo{volume}{32},
  \bibinfo{number}{3} (\bibinfo{year}{2013}).
\newblock
\urldef\tempurl%
\url{https://doi.org/10.1111/cgf.12129}
\showDOI{\tempurl}


\bibitem[\protect\citeauthoryear{Lohman}{Lohman}{2014}]%
        {acmblog:is-query-optimization-a-solved-problem}
\bibfield{author}{\bibinfo{person}{G. Lohman}.}
  \bibinfo{year}{2014}\natexlab{}.
\newblock \showarticletitle{Is Query Optimization a ``Solved'' Problem?}
\newblock \bibinfo{journal}{\emph{ACM SIGMOD Blog.}} \bibinfo{volume}{ACM
  Blog}, \bibinfo{number}{14'} (\bibinfo{year}{2014}).
\newblock


\bibitem[\protect\citeauthoryear{Luo, Chai, Qin, Tang, and Li}{Luo
  et~al\mbox{.}}{2020}]%
        {conf/icde/LuoCQ0020}
\bibfield{author}{\bibinfo{person}{Yuyu Luo}, \bibinfo{person}{Chengliang
  Chai}, \bibinfo{person}{Xuedi Qin}, \bibinfo{person}{Nan Tang}, {and}
  \bibinfo{person}{Guoliang Li}.} \bibinfo{year}{2020}\natexlab{}.
\newblock \showarticletitle{Interactive Cleaning for Progressive Visualization
  through Composite Questions}. In \bibinfo{booktitle}{\emph{36th {IEEE}
  International Conference on Data Engineering, {ICDE} 2020, Dallas, TX, USA,
  April 20-24, 2020}}.
\newblock
\urldef\tempurl%
\url{https://doi.org/10.1109/ICDE48307.2020.00069}
\showDOI{\tempurl}


\bibitem[\protect\citeauthoryear{Luo, Qin, Tang, Li, and Wang}{Luo
  et~al\mbox{.}}{2018}]%
        {conf/sigmod/LuoQ00W18}
\bibfield{author}{\bibinfo{person}{Yuyu Luo}, \bibinfo{person}{Xuedi Qin},
  \bibinfo{person}{Nan Tang}, \bibinfo{person}{Guoliang Li}, {and}
  \bibinfo{person}{Xinran Wang}.} \bibinfo{year}{2018}\natexlab{}.
\newblock \showarticletitle{DeepEye: Creating Good Data Visualizations by
  Keyword Search}. In \bibinfo{booktitle}{\emph{Proceedings of the 2018
  International Conference on Management of Data, {SIGMOD} Conference 2018,
  Houston, TX, USA, June 10-15, 2018}},
  \bibfield{editor}{\bibinfo{person}{Gautam Das},
  \bibinfo{person}{Christopher~M. Jermaine}, {and} \bibinfo{person}{Philip~A.
  Bernstein}} (Eds.).
\newblock
\urldef\tempurl%
\url{https://doi.org/10.1145/3183713.3193545}
\showDOI{\tempurl}


\bibitem[\protect\citeauthoryear{Marcus, Negi, Mao, Tatbul, Alizadeh, and
  Kraska}{Marcus et~al\mbox{.}}{2020}]%
        {journals/corr/abs-2004-03814}
\bibfield{author}{\bibinfo{person}{Ryan Marcus}, \bibinfo{person}{Parimarjan
  Negi}, \bibinfo{person}{Hongzi Mao}, \bibinfo{person}{Nesime Tatbul},
  \bibinfo{person}{Mohammad Alizadeh}, {and} \bibinfo{person}{Tim Kraska}.}
  \bibinfo{year}{2020}\natexlab{}.
\newblock \showarticletitle{Bao: Learning to Steer Query Optimizers}.
\newblock \bibinfo{journal}{\emph{CoRR}}  \bibinfo{volume}{abs/2004.03814}
  (\bibinfo{year}{2020}).
\newblock
\showeprint[arxiv]{2004.03814}
\urldef\tempurl%
\url{https://arxiv.org/abs/2004.03814}
\showURL{%
\tempurl}


\bibitem[\protect\citeauthoryear{Marcus, Negi, Mao, Tatbul, Alizadeh, and
  Kraska}{Marcus et~al\mbox{.}}{2021}]%
        {conf/sigmod/MarcusNMTAK21}
\bibfield{author}{\bibinfo{person}{Ryan Marcus}, \bibinfo{person}{Parimarjan
  Negi}, \bibinfo{person}{Hongzi Mao}, \bibinfo{person}{Nesime Tatbul},
  \bibinfo{person}{Mohammad Alizadeh}, {and} \bibinfo{person}{Tim Kraska}.}
  \bibinfo{year}{2021}\natexlab{}.
\newblock \showarticletitle{Bao: Making Learned Query Optimization Practical}.
  In \bibinfo{booktitle}{\emph{{SIGMOD} '21: International Conference on
  Management of Data, Virtual Event, China, June 20-25, 2021}},
  \bibfield{editor}{\bibinfo{person}{Guoliang Li}, \bibinfo{person}{Zhanhuai
  Li}, \bibinfo{person}{Stratos Idreos}, {and} \bibinfo{person}{Divesh
  Srivastava}} (Eds.).
\newblock
\urldef\tempurl%
\url{https://doi.org/10.1145/3448016.3452838}
\showDOI{\tempurl}


\bibitem[\protect\citeauthoryear{Marcus and Papaemmanouil}{Marcus and
  Papaemmanouil}{2018}]%
        {conf/sigmod/MarcusP18}
\bibfield{author}{\bibinfo{person}{Ryan Marcus} {and} \bibinfo{person}{Olga
  Papaemmanouil}.} \bibinfo{year}{2018}\natexlab{}.
\newblock \showarticletitle{Deep Reinforcement Learning for Join Order
  Enumeration}. In \bibinfo{booktitle}{\emph{Proceedings of the First
  International Workshop on Exploiting Artificial Intelligence Techniques for
  Data Management, aiDM@SIGMOD 2018, Houston, TX, USA, June 10, 2018}},
  \bibfield{editor}{\bibinfo{person}{Rajesh Bordawekar} {and}
  \bibinfo{person}{Oded Shmueli}} (Eds.).
\newblock
\urldef\tempurl%
\url{https://doi.org/10.1145/3211954.3211957}
\showDOI{\tempurl}


\bibitem[\protect\citeauthoryear{Marcus, Negi, Mao, Zhang, Alizadeh, Kraska,
  Papaemmanouil, and Tatbul}{Marcus et~al\mbox{.}}{2019}]%
        {journals/pvldb/MarcusNMZAKPT19}
\bibfield{author}{\bibinfo{person}{Ryan~C. Marcus}, \bibinfo{person}{Parimarjan
  Negi}, \bibinfo{person}{Hongzi Mao}, \bibinfo{person}{Chi Zhang},
  \bibinfo{person}{Mohammad Alizadeh}, \bibinfo{person}{Tim Kraska},
  \bibinfo{person}{Olga Papaemmanouil}, {and} \bibinfo{person}{Nesime Tatbul}.}
  \bibinfo{year}{2019}\natexlab{}.
\newblock \showarticletitle{Neo: {A} Learned Query Optimizer}.
\newblock \bibinfo{journal}{\emph{Proc. {VLDB} Endow.}} \bibinfo{volume}{12},
  \bibinfo{number}{11} (\bibinfo{year}{2019}).
\newblock
\urldef\tempurl%
\url{https://doi.org/10.14778/3342263.3342644}
\showDOI{\tempurl}


\bibitem[\protect\citeauthoryear{Mnih, Kavukcuoglu, Silver, Graves, Antonoglou,
  Wierstra, and Riedmiller}{Mnih et~al\mbox{.}}{2013}]%
        {journals/corr/MnihKSGAWR13}
\bibfield{author}{\bibinfo{person}{Volodymyr Mnih}, \bibinfo{person}{Koray
  Kavukcuoglu}, \bibinfo{person}{David Silver}, \bibinfo{person}{Alex Graves},
  \bibinfo{person}{Ioannis Antonoglou}, \bibinfo{person}{Daan Wierstra}, {and}
  \bibinfo{person}{Martin~A. Riedmiller}.} \bibinfo{year}{2013}\natexlab{}.
\newblock \showarticletitle{Playing Atari with Deep Reinforcement Learning}.
\newblock \bibinfo{journal}{\emph{CoRR}}  \bibinfo{volume}{abs/1312.5602}
  (\bibinfo{year}{2013}).
\newblock
\showeprint[arxiv]{1312.5602}
\urldef\tempurl%
\url{http://arxiv.org/abs/1312.5602}
\showURL{%
\tempurl}


\bibitem[\protect\citeauthoryear{Moritz, Fisher, Ding, and Wang}{Moritz
  et~al\mbox{.}}{2017}]%
        {conf/chi/MoritzFD017}
\bibfield{author}{\bibinfo{person}{Dominik Moritz}, \bibinfo{person}{Danyel
  Fisher}, \bibinfo{person}{Bolin Ding}, {and} \bibinfo{person}{Chi Wang}.}
  \bibinfo{year}{2017}\natexlab{}.
\newblock \showarticletitle{Trust, but Verify: Optimistic Visualizations of
  Approximate Queries for Exploring Big Data}. In
  \bibinfo{booktitle}{\emph{Proceedings of the 2017 {CHI} Conference on Human
  Factors in Computing Systems, Denver, CO, USA, May 06-11, 2017.}}
\newblock
\urldef\tempurl%
\url{https://doi.org/10.1145/3025453.3025456}
\showDOI{\tempurl}


\bibitem[\protect\citeauthoryear{Moritz, Howe, and Heer}{Moritz
  et~al\mbox{.}}{2019}]%
        {conf/chi/MoritzHH19}
\bibfield{author}{\bibinfo{person}{Dominik Moritz}, \bibinfo{person}{Bill
  Howe}, {and} \bibinfo{person}{Jeffrey Heer}.}
  \bibinfo{year}{2019}\natexlab{}.
\newblock \showarticletitle{Falcon: Balancing Interactive Latency and
  Resolution Sensitivity for Scalable Linked Visualizations}. In
  \bibinfo{booktitle}{\emph{Proceedings of the 2019 {CHI} Conference on Human
  Factors in Computing Systems, {CHI} 2019, Glasgow, Scotland, UK, May 04-09,
  2019}}.
\newblock
\urldef\tempurl%
\url{https://doi.org/10.1145/3290605.3300924}
\showDOI{\tempurl}


\bibitem[\protect\citeauthoryear{Mozafari, Ramnarayan, Menon, Mahajan,
  Chakraborty, Bhanawat, and Bachhav}{Mozafari et~al\mbox{.}}{2017}]%
        {conf/cidr/MozafariRMMCBB17}
\bibfield{author}{\bibinfo{person}{Barzan Mozafari}, \bibinfo{person}{Jags
  Ramnarayan}, \bibinfo{person}{Sudhir Menon}, \bibinfo{person}{Yogesh
  Mahajan}, \bibinfo{person}{Soubhik Chakraborty}, \bibinfo{person}{Hemant
  Bhanawat}, {and} \bibinfo{person}{Kishor Bachhav}.}
  \bibinfo{year}{2017}\natexlab{}.
\newblock \showarticletitle{SnappyData: {A} Unified Cluster for Streaming,
  Transactions and Interactice Analytics}. In \bibinfo{booktitle}{\emph{{CIDR}
  2017, 8th Biennial Conference on Innovative Data Systems Research, Chaminade,
  CA, USA, January 8-11, 2017, Online Proceedings}}.
\newblock
\urldef\tempurl%
\url{http://cidrdb.org/cidr2017/papers/p28-mozafari-cidr17.pdf}
\showURL{%
\tempurl}


\bibitem[\protect\citeauthoryear{{{MySQL Optimizer Hints}}}{{{MySQL Optimizer
  Hints}}}{[n.d.]}]%
        {mysql:optimizer-hints}
{{MySQL Optimizer Hints}}.
\newblock
\newblock
\newblock
\shownote{\url{https://dev.mysql.com/doc/refman/8.0/en/optimizer-hints.html}.}


\bibitem[\protect\citeauthoryear{{{NYC Taxi Data}}}{{{NYC Taxi
  Data}}}{[n.d.]}]%
        {nyc-taxi-data:website}
{{NYC Taxi Data}}.
\newblock
\newblock
\newblock
\shownote{\url{https://www1.nyc.gov/site/tlc/about/tlc-trip-record-data.page}.}


\bibitem[\protect\citeauthoryear{{{Oracle Using Optimizer Hints}}}{{{Oracle
  Using Optimizer Hints}}}{[n.d.]}]%
        {oracle:optimizer-hints}
{{Oracle Using Optimizer Hints}}.
\newblock
\newblock
\newblock
\shownote{\url{https://docs.oracle.com/cd/B19306_01/server.102/b14211/hintsref.htm##i8327}.}


\bibitem[\protect\citeauthoryear{Park, Cafarella, and Mozafari}{Park
  et~al\mbox{.}}{2016}]%
        {conf/icde/ParkCM16}
\bibfield{author}{\bibinfo{person}{Yongjoo Park}, \bibinfo{person}{Michael~J.
  Cafarella}, {and} \bibinfo{person}{Barzan Mozafari}.}
  \bibinfo{year}{2016}\natexlab{}.
\newblock \showarticletitle{Visualization-aware sampling for very large
  databases}. In \bibinfo{booktitle}{\emph{32nd {IEEE} International Conference
  on Data Engineering, {ICDE} 2016, Helsinki, Finland, May 16-20, 2016}}.
\newblock
\urldef\tempurl%
\url{https://doi.org/10.1109/ICDE.2016.7498287}
\showDOI{\tempurl}


\bibitem[\protect\citeauthoryear{Park, Mozafari, Sorenson, and Wang}{Park
  et~al\mbox{.}}{2018}]%
        {conf/sigmod/ParkMSW18}
\bibfield{author}{\bibinfo{person}{Yongjoo Park}, \bibinfo{person}{Barzan
  Mozafari}, \bibinfo{person}{Joseph Sorenson}, {and} \bibinfo{person}{Junhao
  Wang}.} \bibinfo{year}{2018}\natexlab{}.
\newblock \showarticletitle{VerdictDB: Universalizing Approximate Query
  Processing}. In \bibinfo{booktitle}{\emph{Proceedings of the 2018
  International Conference on Management of Data, {SIGMOD} Conference 2018,
  Houston, TX, USA, June 10-15, 2018}}.
\newblock
\urldef\tempurl%
\url{https://doi.org/10.1145/3183713.3196905}
\showDOI{\tempurl}


\bibitem[\protect\citeauthoryear{Park, Zhong, and Mozafari}{Park
  et~al\mbox{.}}{2020}]%
        {conf/sigmod/ParkZM20}
\bibfield{author}{\bibinfo{person}{Yongjoo Park}, \bibinfo{person}{Shucheng
  Zhong}, {and} \bibinfo{person}{Barzan Mozafari}.}
  \bibinfo{year}{2020}\natexlab{}.
\newblock \showarticletitle{QuickSel: Quick Selectivity Learning with Mixture
  Models}. In \bibinfo{booktitle}{\emph{Proceedings of the 2020 International
  Conference on Management of Data, {SIGMOD} Conference 2020, online conference
  [Portland, OR, USA], June 14-19, 2020}},
  \bibfield{editor}{\bibinfo{person}{David Maier}, \bibinfo{person}{Rachel
  Pottinger}, \bibinfo{person}{AnHai Doan}, \bibinfo{person}{Wang{-}Chiew Tan},
  \bibinfo{person}{Abdussalam Alawini}, {and} \bibinfo{person}{Hung~Q. Ngo}}
  (Eds.).
\newblock
\urldef\tempurl%
\url{https://doi.org/10.1145/3318464.3389727}
\showDOI{\tempurl}


\bibitem[\protect\citeauthoryear{Peng, Zhang, Wang, and Pei}{Peng
  et~al\mbox{.}}{2018}]%
        {conf/sigmod/PengZWP18}
\bibfield{author}{\bibinfo{person}{Jinglin Peng}, \bibinfo{person}{Dongxiang
  Zhang}, \bibinfo{person}{Jiannan Wang}, {and} \bibinfo{person}{Jian Pei}.}
  \bibinfo{year}{2018}\natexlab{}.
\newblock \showarticletitle{{AQP++:} Connecting Approximate Query Processing
  With Aggregate Precomputation for Interactive Analytics}. In
  \bibinfo{booktitle}{\emph{Proceedings of the 2018 International Conference on
  Management of Data, {SIGMOD} Conference 2018, Houston, TX, USA, June 10-15,
  2018}}.
\newblock
\urldef\tempurl%
\url{https://doi.org/10.1145/3183713.3183747}
\showDOI{\tempurl}


\bibitem[\protect\citeauthoryear{{{PostgreSQL Query Hints}}}{{{PostgreSQL Query
  Hints}}}{[n.d.]}]%
        {postgresql:query-hints}
{{PostgreSQL Query Hints}}.
\newblock
\newblock
\newblock
\shownote{\url{https://pghintplan.osdn.jp/pg_hint_plan.html}.}


\bibitem[\protect\citeauthoryear{Psallidas and Wu}{Psallidas and Wu}{2018}]%
        {conf/sigmod/Psallidas018a}
\bibfield{author}{\bibinfo{person}{Fotis Psallidas} {and}
  \bibinfo{person}{Eugene Wu}.} \bibinfo{year}{2018}\natexlab{}.
\newblock \showarticletitle{Provenance for Interactive Visualizations}. In
  \bibinfo{booktitle}{\emph{Proceedings of the Workshop on Human-In-the-Loop
  Data Analytics, HILDA@SIGMOD 2018, Houston, TX, USA, June 10, 2018}}.
\newblock
\urldef\tempurl%
\url{https://doi.org/10.1145/3209900.3209904}
\showDOI{\tempurl}


\bibitem[\protect\citeauthoryear{Qian, Rossi, Du, Kim, Koh, Malik, Lee, and
  Chan}{Qian et~al\mbox{.}}{2020}]%
        {journals/corr/abs-2009-12316}
\bibfield{author}{\bibinfo{person}{Xin Qian}, \bibinfo{person}{Ryan~A. Rossi},
  \bibinfo{person}{Fan Du}, \bibinfo{person}{Sungchul Kim},
  \bibinfo{person}{Eunyee Koh}, \bibinfo{person}{Sana Malik},
  \bibinfo{person}{Tak~Yeon Lee}, {and} \bibinfo{person}{Joel Chan}.}
  \bibinfo{year}{2020}\natexlab{}.
\newblock \showarticletitle{ML-based Visualization Recommendation: Learning to
  Recommend Visualizations from Data}.
\newblock \bibinfo{journal}{\emph{CoRR}}  \bibinfo{volume}{abs/2009.12316}
  (\bibinfo{year}{2020}).
\newblock
\showeprint[arxiv]{2009.12316}
\urldef\tempurl%
\url{https://arxiv.org/abs/2009.12316}
\showURL{%
\tempurl}


\bibitem[\protect\citeauthoryear{Rahman, Aliakbarpour, Kong, Blais, Karahalios,
  Parameswaran, and Rubinfeld}{Rahman et~al\mbox{.}}{2017}]%
        {journals/pvldb/RahmanAKBKPR17}
\bibfield{author}{\bibinfo{person}{Sajjadur Rahman}, \bibinfo{person}{Maryam
  Aliakbarpour}, \bibinfo{person}{Hidy Kong}, \bibinfo{person}{Eric Blais},
  \bibinfo{person}{Karrie Karahalios}, \bibinfo{person}{Aditya~G.
  Parameswaran}, {and} \bibinfo{person}{Ronitt Rubinfeld}.}
  \bibinfo{year}{2017}\natexlab{}.
\newblock \showarticletitle{I've Seen "Enough": Incrementally Improving
  Visualizations to Support Rapid Decision Making}.
\newblock \bibinfo{journal}{\emph{{PVLDB}}} \bibinfo{volume}{10},
  \bibinfo{number}{11} (\bibinfo{year}{2017}).
\newblock
\urldef\tempurl%
\url{https://doi.org/10.14778/3137628.3137637}
\showDOI{\tempurl}


\bibitem[\protect\citeauthoryear{{{Rectifier (neural networks)}}}{{{Rectifier
  (neural networks)}}}{[n.d.]}]%
        {relu}
{{Rectifier (neural networks)}}.
\newblock
\newblock
\newblock
\shownote{\url{https://en.wikipedia.org/wiki/Rectifier_(neural_networks)}.}


\bibitem[\protect\citeauthoryear{Rundensteiner, Ward, Xie, Cui, Wad, Yang, and
  Huang}{Rundensteiner et~al\mbox{.}}{2007}]%
        {conf/sigmod/RundensteinerWXCWYH07}
\bibfield{author}{\bibinfo{person}{Elke~A. Rundensteiner},
  \bibinfo{person}{Matthew~O. Ward}, \bibinfo{person}{Zaixian Xie},
  \bibinfo{person}{Qingguang Cui}, \bibinfo{person}{Charudatta~V. Wad},
  \bibinfo{person}{Di Yang}, {and} \bibinfo{person}{Shiping Huang}.}
  \bibinfo{year}{2007}\natexlab{}.
\newblock \showarticletitle{Xmdvtool\({}^{\mbox{\emph{Q}}}\): : quality-aware
  interactive data exploration}. In \bibinfo{booktitle}{\emph{Proceedings of
  the {ACM} {SIGMOD} International Conference on Management of Data, Beijing,
  China, June 12-14, 2007}}.
\newblock
\urldef\tempurl%
\url{https://doi.org/10.1145/1247480.1247623}
\showDOI{\tempurl}


\bibitem[\protect\citeauthoryear{Sikdar and Jermaine}{Sikdar and
  Jermaine}{2020}]%
        {conf/sigmod/SikdarJ20}
\bibfield{author}{\bibinfo{person}{Sourav Sikdar} {and} \bibinfo{person}{Chris
  Jermaine}.} \bibinfo{year}{2020}\natexlab{}.
\newblock \showarticletitle{{MONSOON:} Multi-Step Optimization and Execution of
  Queries with Partially Obscured Predicates}. In
  \bibinfo{booktitle}{\emph{Proceedings of the 2020 International Conference on
  Management of Data, {SIGMOD} Conference 2020, online conference [Portland,
  OR, USA], June 14-19, 2020}}, \bibfield{editor}{\bibinfo{person}{David
  Maier}, \bibinfo{person}{Rachel Pottinger}, \bibinfo{person}{AnHai Doan},
  \bibinfo{person}{Wang{-}Chiew Tan}, \bibinfo{person}{Abdussalam Alawini},
  {and} \bibinfo{person}{Hung~Q. Ngo}} (Eds.).
\newblock
\urldef\tempurl%
\url{https://doi.org/10.1145/3318464.3389728}
\showDOI{\tempurl}


\bibitem[\protect\citeauthoryear{{{SQL Server Query Hints}}}{{{SQL Server Query
  Hints}}}{[n.d.]}]%
        {sqlserver:query-hints}
{{SQL Server Query Hints}}.
\newblock
\newblock
\newblock
\shownote{\url{https://docs.microsoft.com/en-us/sql/t-sql/queries/hints-transact-sql-query?view=sql-server-ver15}.}


\bibitem[\protect\citeauthoryear{Sun and Li}{Sun and Li}{2019}]%
        {journals/pvldb/SunL19}
\bibfield{author}{\bibinfo{person}{Ji Sun} {and} \bibinfo{person}{Guoliang
  Li}.} \bibinfo{year}{2019}\natexlab{}.
\newblock \showarticletitle{An End-to-End Learning-based Cost Estimator}.
\newblock \bibinfo{journal}{\emph{Proc. {VLDB} Endow.}} \bibinfo{volume}{13},
  \bibinfo{number}{3} (\bibinfo{year}{2019}).
\newblock
\urldef\tempurl%
\url{https://doi.org/10.14778/3368289.3368296}
\showDOI{\tempurl}


\bibitem[\protect\citeauthoryear{Sutton and Barto}{Sutton and Barto}{1998}]%
        {books/lib/SuttonB98}
\bibfield{author}{\bibinfo{person}{Richard~S. Sutton} {and}
  \bibinfo{person}{Andrew~G. Barto}.} \bibinfo{year}{1998}\natexlab{}.
\newblock \bibinfo{booktitle}{\emph{Reinforcement learning - an introduction}}.
\newblock \bibinfo{publisher}{{MIT} Press}.
\newblock
\showISBNx{978-0-262-19398-6}
\urldef\tempurl%
\url{https://www.worldcat.org/oclc/37293240}
\showURL{%
\tempurl}


\bibitem[\protect\citeauthoryear{Tao, Liu, Wang, Battle, Demiralp, Chang, and
  Stonebraker}{Tao et~al\mbox{.}}{2019}]%
        {journals/cgf/TaoLWBDCS19}
\bibfield{author}{\bibinfo{person}{Wenbo Tao}, \bibinfo{person}{Xiaoyu Liu},
  \bibinfo{person}{Yedi Wang}, \bibinfo{person}{Leilani Battle},
  \bibinfo{person}{{\c{C}}agatay Demiralp}, \bibinfo{person}{Remco Chang},
  {and} \bibinfo{person}{Michael Stonebraker}.}
  \bibinfo{year}{2019}\natexlab{}.
\newblock \showarticletitle{Kyrix: Interactive Pan/Zoom Visualizations at
  Scale}.
\newblock \bibinfo{journal}{\emph{Comput. Graph. Forum}} \bibinfo{volume}{38},
  \bibinfo{number}{3} (\bibinfo{year}{2019}).
\newblock
\urldef\tempurl%
\url{https://doi.org/10.1111/cgf.13708}
\showDOI{\tempurl}


\bibitem[\protect\citeauthoryear{Tokic}{Tokic}{2010}]%
        {conf/ki/Tokic10}
\bibfield{author}{\bibinfo{person}{Michel Tokic}.}
  \bibinfo{year}{2010}\natexlab{}.
\newblock \showarticletitle{Adaptive epsilon-Greedy Exploration in
  Reinforcement Learning Based on Value Difference}. In
  \bibinfo{booktitle}{\emph{{KI} 2010: Advances in Artificial Intelligence,
  33rd Annual German Conference on AI, Karlsruhe, Germany, September 21-24,
  2010. Proceedings}} \emph{(\bibinfo{series}{Lecture Notes in Computer
  Science})}, \bibfield{editor}{\bibinfo{person}{R{\"{u}}diger Dillmann},
  \bibinfo{person}{J{\"{u}}rgen Beyerer}, \bibinfo{person}{Uwe~D. Hanebeck},
  {and} \bibinfo{person}{Tanja Schultz}} (Eds.), Vol.~\bibinfo{volume}{6359}.
\newblock
\urldef\tempurl%
\url{https://doi.org/10.1007/978-3-642-16111-7\_23}
\showDOI{\tempurl}


\bibitem[\protect\citeauthoryear{tpch}{tpch}{[n.d.]}]%
        {TPC-H}
tpch.
\newblock
\newblock
\newblock
\shownote{TPC-H Website, \url{http://www.tpc.org/tpch/}.}


\bibitem[\protect\citeauthoryear{Trummer, Wang, Maram, Moseley, Jo, and
  Antonakakis}{Trummer et~al\mbox{.}}{2019}]%
        {conf/sigmod/TrummerWMMJA19}
\bibfield{author}{\bibinfo{person}{Immanuel Trummer}, \bibinfo{person}{Junxiong
  Wang}, \bibinfo{person}{Deepak Maram}, \bibinfo{person}{Samuel Moseley},
  \bibinfo{person}{Saehan Jo}, {and} \bibinfo{person}{Joseph Antonakakis}.}
  \bibinfo{year}{2019}\natexlab{}.
\newblock \showarticletitle{SkinnerDB: Regret-Bounded Query Evaluation via
  Reinforcement Learning}. In \bibinfo{booktitle}{\emph{Proceedings of the 2019
  International Conference on Management of Data, {SIGMOD} Conference 2019,
  Amsterdam, The Netherlands, June 30 - July 5, 2019}},
  \bibfield{editor}{\bibinfo{person}{Peter~A. Boncz}, \bibinfo{person}{Stefan
  Manegold}, \bibinfo{person}{Anastasia Ailamaki}, \bibinfo{person}{Amol
  Deshpande}, {and} \bibinfo{person}{Tim Kraska}} (Eds.).
\newblock
\urldef\tempurl%
\url{https://doi.org/10.1145/3299869.3300088}
\showDOI{\tempurl}


\bibitem[\protect\citeauthoryear{Wang, Christensen, Li, and Yi}{Wang
  et~al\mbox{.}}{2015}]%
        {journals/pvldb/WangCLY15}
\bibfield{author}{\bibinfo{person}{Lu Wang}, \bibinfo{person}{Robert
  Christensen}, \bibinfo{person}{Feifei Li}, {and} \bibinfo{person}{Ke Yi}.}
  \bibinfo{year}{2015}\natexlab{}.
\newblock \showarticletitle{Spatial Online Sampling and Aggregation}.
\newblock \bibinfo{journal}{\emph{{PVLDB}}} \bibinfo{volume}{9},
  \bibinfo{number}{3} (\bibinfo{year}{2015}).
\newblock
\urldef\tempurl%
\url{https://doi.org/10.14778/2850583.2850584}
\showDOI{\tempurl}


\bibitem[\protect\citeauthoryear{Wang, Chen, Wang, and Qu}{Wang
  et~al\mbox{.}}{2020}]%
        {journals/corr/abs-2012-00467}
\bibfield{author}{\bibinfo{person}{Qianwen Wang}, \bibinfo{person}{Zhutian
  Chen}, \bibinfo{person}{Yong Wang}, {and} \bibinfo{person}{Huamin Qu}.}
  \bibinfo{year}{2020}\natexlab{}.
\newblock \showarticletitle{Applying Machine Learning Advances to Data
  Visualization: {A} Survey on {ML4VIS}}.
\newblock \bibinfo{journal}{\emph{CoRR}}  \bibinfo{volume}{abs/2012.00467}
  (\bibinfo{year}{2020}).
\newblock
\showeprint[arxiv]{2012.00467}
\urldef\tempurl%
\url{https://arxiv.org/abs/2012.00467}
\showURL{%
\tempurl}


\bibitem[\protect\citeauthoryear{Wang, Feng, Chu, Zhang, Fu, Sedlmair, Yu, and
  Chen}{Wang et~al\mbox{.}}{2018}]%
        {journals/tvcg/WangFCZFSYC18}
\bibfield{author}{\bibinfo{person}{Yunhai Wang}, \bibinfo{person}{Kang Feng},
  \bibinfo{person}{Xiaowei Chu}, \bibinfo{person}{Jian Zhang},
  \bibinfo{person}{Chi{-}Wing Fu}, \bibinfo{person}{Michael Sedlmair},
  \bibinfo{person}{Xiaohui Yu}, {and} \bibinfo{person}{Baoquan Chen}.}
  \bibinfo{year}{2018}\natexlab{}.
\newblock \showarticletitle{A Perception-Driven Approach to Supervised
  Dimensionality Reduction for Visualization}.
\newblock \bibinfo{journal}{\emph{{IEEE} Trans. Vis. Comput. Graph.}}
  \bibinfo{volume}{24}, \bibinfo{number}{5} (\bibinfo{year}{2018}).
\newblock
\urldef\tempurl%
\url{https://doi.org/10.1109/TVCG.2017.2701829}
\showDOI{\tempurl}


\bibitem[\protect\citeauthoryear{Wang, Ferreira, Wei, Bhaskar, and
  Scheidegger}{Wang et~al\mbox{.}}{2017}]%
        {journals/tvcg/WangFWBS17}
\bibfield{author}{\bibinfo{person}{Zhe Wang}, \bibinfo{person}{Nivan Ferreira},
  \bibinfo{person}{Youhao Wei}, \bibinfo{person}{Aarthy~Sankari Bhaskar}, {and}
  \bibinfo{person}{Carlos Scheidegger}.} \bibinfo{year}{2017}\natexlab{}.
\newblock \showarticletitle{Gaussian Cubes: Real-Time Modeling for Visual
  Exploration of Large Multidimensional Datasets}.
\newblock \bibinfo{journal}{\emph{{IEEE} Trans. Vis. Comput. Graph.}}
  \bibinfo{volume}{23}, \bibinfo{number}{1} (\bibinfo{year}{2017}).
\newblock
\urldef\tempurl%
\url{https://doi.org/10.1109/TVCG.2016.2598694}
\showDOI{\tempurl}


\bibitem[\protect\citeauthoryear{Watkins and Dayan}{Watkins and Dayan}{1992}]%
        {journals/ml/WatkinsD92}
\bibfield{author}{\bibinfo{person}{Christopher J. C.~H. Watkins} {and}
  \bibinfo{person}{Peter Dayan}.} \bibinfo{year}{1992}\natexlab{}.
\newblock \showarticletitle{Technical Note Q-Learning}.
\newblock \bibinfo{journal}{\emph{Mach. Learn.}}  \bibinfo{volume}{8}
  (\bibinfo{year}{1992}).
\newblock
\urldef\tempurl%
\url{https://doi.org/10.1007/BF00992698}
\showDOI{\tempurl}


\bibitem[\protect\citeauthoryear{Wu, Chi, Zhu, Tatemura, Hacig{\"{u}}m{\"{u}}s,
  and Naughton}{Wu et~al\mbox{.}}{2013}]%
        {conf/icde/WuCZTHN13}
\bibfield{author}{\bibinfo{person}{Wentao Wu}, \bibinfo{person}{Yun Chi},
  \bibinfo{person}{Shenghuo Zhu}, \bibinfo{person}{Jun'ichi Tatemura},
  \bibinfo{person}{Hakan Hacig{\"{u}}m{\"{u}}s}, {and}
  \bibinfo{person}{Jeffrey~F. Naughton}.} \bibinfo{year}{2013}\natexlab{}.
\newblock \showarticletitle{Predicting query execution time: Are optimizer cost
  models really unusable?}. In \bibinfo{booktitle}{\emph{29th {IEEE}
  International Conference on Data Engineering, {ICDE} 2013, Brisbane,
  Australia, April 8-12, 2013}},
  \bibfield{editor}{\bibinfo{person}{Christian~S. Jensen},
  \bibinfo{person}{Christopher~M. Jermaine}, {and} \bibinfo{person}{Xiaofang
  Zhou}} (Eds.).
\newblock
\urldef\tempurl%
\url{https://doi.org/10.1109/ICDE.2013.6544899}
\showDOI{\tempurl}


\bibitem[\protect\citeauthoryear{Wu, Wu, Hacig{\"{u}}m{\"{u}}s, and
  Naughton}{Wu et~al\mbox{.}}{2014}]%
        {journals/pvldb/WuWHN14}
\bibfield{author}{\bibinfo{person}{Wentao Wu}, \bibinfo{person}{Xi Wu},
  \bibinfo{person}{Hakan Hacig{\"{u}}m{\"{u}}s}, {and}
  \bibinfo{person}{Jeffrey~F. Naughton}.} \bibinfo{year}{2014}\natexlab{}.
\newblock \showarticletitle{{Uncertainty Aware Query Execution Time
  Prediction}}.
\newblock \bibinfo{journal}{\emph{{PVLDB}}} \bibinfo{volume}{7},
  \bibinfo{number}{14} (\bibinfo{year}{2014}).
\newblock


\bibitem[\protect\citeauthoryear{Yu, Moraffah, and Sarwat}{Yu
  et~al\mbox{.}}{2017}]%
        {conf/icde/YuMS17}
\bibfield{author}{\bibinfo{person}{Jia Yu}, \bibinfo{person}{Raha Moraffah},
  {and} \bibinfo{person}{Mohamed Sarwat}.} \bibinfo{year}{2017}\natexlab{}.
\newblock \showarticletitle{Hippo in Action: Scalable Indexing of a Billion New
  York City Taxi Trips and Beyond}. In \bibinfo{booktitle}{\emph{33rd {IEEE}
  International Conference on Data Engineering, {ICDE} 2017, San Diego, CA,
  USA, April 19-22, 2017}}.
\newblock
\urldef\tempurl%
\url{https://doi.org/10.1109/ICDE.2017.201}
\showDOI{\tempurl}


\bibitem[\protect\citeauthoryear{Yu, Zhang, and Sarwat}{Yu
  et~al\mbox{.}}{2018}]%
        {conf/ssdbm/YuZS18}
\bibfield{author}{\bibinfo{person}{Jia Yu}, \bibinfo{person}{Zongsi Zhang},
  {and} \bibinfo{person}{Mohamed Sarwat}.} \bibinfo{year}{2018}\natexlab{}.
\newblock \showarticletitle{GeoSparkViz: a scalable geospatial data
  visualization framework in the apache spark ecosystem}. In
  \bibinfo{booktitle}{\emph{Proceedings of the 30th International Conference on
  Scientific and Statistical Database Management, {SSDBM} 2018, Bozen-Bolzano,
  Italy, July 09-11, 2018}}.
\newblock
\urldef\tempurl%
\url{https://doi.org/10.1145/3221269.3223040}
\showDOI{\tempurl}


\bibitem[\protect\citeauthoryear{Yu, Li, Chai, and Tang}{Yu
  et~al\mbox{.}}{2020}]%
        {conf/icde/Yu0C020}
\bibfield{author}{\bibinfo{person}{Xiang Yu}, \bibinfo{person}{Guoliang Li},
  \bibinfo{person}{Chengliang Chai}, {and} \bibinfo{person}{Nan Tang}.}
  \bibinfo{year}{2020}\natexlab{}.
\newblock \showarticletitle{Reinforcement Learning with Tree-LSTM for Join
  Order Selection}. In \bibinfo{booktitle}{\emph{36th {IEEE} International
  Conference on Data Engineering, {ICDE} 2020, Dallas, TX, USA, April 20-24,
  2020}}.
\newblock
\urldef\tempurl%
\url{https://doi.org/10.1109/ICDE48307.2020.00116}
\showDOI{\tempurl}


\bibitem[\protect\citeauthoryear{Zeng, Agarwal, Dave, Armbrust, and
  Stoica}{Zeng et~al\mbox{.}}{2015}]%
        {conf/sigmod/ZengADAS15}
\bibfield{author}{\bibinfo{person}{Kai Zeng}, \bibinfo{person}{Sameer Agarwal},
  \bibinfo{person}{Ankur Dave}, \bibinfo{person}{Michael Armbrust}, {and}
  \bibinfo{person}{Ion Stoica}.} \bibinfo{year}{2015}\natexlab{}.
\newblock \showarticletitle{{G-OLA:} Generalized On-Line Aggregation for
  Interactive Analysis on Big Data}. In \bibinfo{booktitle}{\emph{Proceedings
  of the 2015 {ACM} {SIGMOD} International Conference on Management of Data,
  Melbourne, Victoria, Australia, May 31 - June 4, 2015}}.
\newblock
\urldef\tempurl%
\url{https://doi.org/10.1145/2723372.2735381}
\showDOI{\tempurl}


\bibitem[\protect\citeauthoryear{Zhang, Wang, Yin, and Ji}{Zhang
  et~al\mbox{.}}{2016}]%
        {journals/pvldb/ZhangWYJ16}
\bibfield{author}{\bibinfo{person}{Xuhong Zhang}, \bibinfo{person}{Jun Wang},
  \bibinfo{person}{Jiangling Yin}, {and} \bibinfo{person}{Shouling Ji}.}
  \bibinfo{year}{2016}\natexlab{}.
\newblock \showarticletitle{Sapprox: Enabling Efficient and Accurate
  Approximations on Sub-datasets with Distribution-aware Online Sampling}.
\newblock \bibinfo{journal}{\emph{{PVLDB}}} \bibinfo{volume}{10},
  \bibinfo{number}{3} (\bibinfo{year}{2016}).
\newblock
\urldef\tempurl%
\url{https://doi.org/10.14778/3021924.3021928}
\showDOI{\tempurl}


\end{thebibliography}

%
% ...

\end{document}